\documentclass[useAMS,usenatbib]{mn2e}

\usepackage[latin1]{inputenc}
\usepackage{graphicx}
\usepackage{epstopdf}

\usepackage{subfigure}
\usepackage{rotate}
\usepackage{amsmath, float}
\usepackage{txfonts}
\usepackage{fleqn}
\usepackage{color}
\usepackage{comment}

\usepackage{mathrsfs}  
\usepackage{booktabs}
\usepackage{colortbl}

\newcolumntype{G}{>{\columncolor[gray]{0.8}}l} 

\newcommand{\be}{\begin{equation}}
\newcommand{\ee}{\end{equation}}
\newcommand{\bdm}{\begin{displaymath}}
\newcommand{\edm}{\end{displaymath}}
\newcommand{\bea}{\begin{multline}}
\newcommand{\eea}{\end{multline}}
\newcommand{\ba}{\begin{align}}
\newcommand{\ea}{\end{align}}

\newcommand{\G}{G }

\renewcommand{\sigma}{\varpi^2}

\newcommand{\Ebar}{{\hat E}}
\newcommand{\Sbar}{{\hat S}}

\newcommand{\Rcirc}{R_{\tiny{\mbox{circ}}}}
\newcommand{\Bmax}{B_{\tiny{\mbox{max}}}}

\def\simlt{\mathrel{\hbox{\rlap{\hbox{\lower4pt\hbox{$\sim$}}}\hbox{$<$}}}}
\def\simgt{\mathrel{\hbox{\rlap{\hbox{\lower4pt\hbox{$\sim$}}}\hbox{$>$}}}}

\title[Axisymmetric equilibrium models for magnetized neutron stars in General Relativity]
{Axisymmetric equilibrium models for magnetized neutron stars in
  General Relativity under the Conformally Flat Condition}
\author[A.~G. Pili, N. Bucciantini, L. Del Zanna]{
A.~G. Pili$^{1,2,3}$\thanks{E-mail: pili@arcetri.astro.it}, N. Bucciantini$^{2,3}$\thanks{E-mail: niccolo@arcetri.astro.it}, L. Del Zanna$^{1,2,3}$ \\
$^{1}$Dipartimento di Fisica e Astronomia, Universit\`a degli Studi di Firenze, Via G. Sansone 1, 
I-50019 Sesto F.~no  (Firenze), Italy\\
$^{2}$INAF - Osservatorio Astrofisico di Arcetri, Largo E. Fermi 5, I-50125 Firenze, Italy\\
$^{3}$INFN - Sezione di Firenze, Via G. Sansone 1, I-50019 Sesto F.~no  (Firenze), Italy}

\begin{document}
 
\date{Accepted / Received}

\maketitle

\label{firstpage}

\begin{abstract}

Extremely magnetized neutron stars with magnetic fields as strong as $\sim 10^{15-16}$~G, 
or \emph{magnetars}, have received considerable attention in the last decade due
to their identification as a plausible source for Soft Gamma Repeaters and Anomalous
X-ray Pulsars. Moreover, this class of compact objects has been proposed as
a possible engine capable of powering both Long and Short Gamma-Ray Bursts,
if the rotation period in their formation stage is short enough ($\sim 1$~ms).
Such strong fields are expected to induce substantial deformations 
of the star and thus to produce the emission of gravitational waves.
Here we investigate, by means of numerical modeling, axisymmetric static equilibria
of polytropic and strongly magnetized stars in full general relativity, within the ideal
magneto-hydrodynamic regime.
The \emph{eXtended Conformally Flat Condition} (XCFC) for the metric is assumed, 
allowing us to employ the techniques introduced for the X-ECHO code 
[\emph{Bucciantini \& Del Zanna, 2011, Astron. Astrophys. 528, A101}], 
proven to be accurate, efficient, and stable. The updated XNS code for 
magnetized neutron star equilibria is made publicly available for the community
(see \texttt{www.arcetri.astro.it/science/ahead/XNS}).
Several sequences of models are here retrieved, from the purely toroidal (resolving a
controversy in the literature) or poloidal cases, to the so-called \emph{twisted torus} 
mixed configurations, expected to be dynamically stable, which are solved for the first time 
in the non-perturbative regime. 

\end{abstract}


\section{Introduction}

Neutron Stars (NSs) are the most compact objects in the universe
endowed with an internal structure. Proposed originally by \citet{Baade_Zwicky34a}
in the context of supernova explosions, they were discovered only in
1967 by \citet{Hewish_Bell+68a} as radio pulsars. Today, NSs are among the most
studied objects in high-energy astrophysics because they are known to
power many astrophysical sources of high energy emission. The extreme
conditions characterizing their interior make them also interesting
objects from the point of view of nuclear and condense matter
physics, and future combined observations of both mass and radius of such compact
objects may finally discriminate on the different equations of state (EoS) 
so far proposed \citep{Feroci_Stella+12a}.

It was immediately evident that NSs can also harbour very high magnetic fields,
usually inferred to be in the range $10^{8-12}$ G for normal pulsars. It is indeed
this very strong magnetic field  that is responsible for most of their
phenomenology and emission. The amplification of magnetic fields form the
initial values prior to collapse to those enhanced values is believed 
to take place during the formation of the compact object itself: surely
due to the compression associated with the collapse of the core
of the progenitor star \citep{Spruit09a}, it can be further increased by
differential rotation in the core leading to the twisting of fieldlines \citep{Burrows_Dessart+07a},
and to possible dynamo effects \citep{Bonanno_Rezzolla+03a,Rheinhardt_Geppert05a}. 
In principle there is a large store of free energy available during and immediately
following the collapse of the core and the formation of a proto-NS,
such that a magnetic field as high as $10^{17-18}$~G could be
even reached. 

The {\it magnetar} model for Anomalous X-Ray Pulsars and Soft Gamma
Repeaters \citep{Thompson_Duncan96a,Mereghetti08a} suggests that the magnetic field can 
reach at least values close to $10^{16}$ G at the surface of NSs. Accounting also for the effects of
dissipative processes \citep{Vigano_Rea+13a}, given the typical ages of known
magnetars ($\sim 10^4$ yr), it is not unreasonable to expect that
younger magnetars with even higher magnetic fields might exist, and more so
immediately after collapse and formation, due to the processes discussed above.

Magnetars could be fundamental also to explain another class of objects
typical of high-energy astrophysics, namely Gamma Ray Bursts (GRBs).
The combination of a rapid millisecond-like rotation of a compact NS
with a magnetic field of typical magnetar strength, can easily drive a
relativistic outflow with energetics of the order of $\sim 10^{49-50}$~erg~s$^{-1}$, 
enough to power a classical \emph{Long} GRB. 
\emph{Short} GRBs have been instead usually associated to merger events, 
rather than to core collapse of stellar objects, leading to the formation of a rotating Black Hole (BH),
similarly to the \emph{collapsar} scenario for Long GRBs \citep{Woosley93a,MacFadyen_Woosley99a}. 
However, the recent  discovery, on the one hand of extended emission and
flaring activity (pointing to a long-lived engine) \citep{Rowlinson_OBrien+10a,Norris_Bonnel06a},  
and  on the other of a NS of mass $2.1\,M_\odot$ \citep{Romani_Filippenko+12a}, suggests that it is not
unreasonable to expect a high-mass NS, rather than a BH, to form from the merger of two low-mass NSs.
Indeed, these assumptions are in part at the base of the so-called {\it millisecond magnetar} 
models for Long and Short GRBs
\citep{Bucciantini_Metzger+12a,Metzger_Gioannios+11a,Bucciantini_Quataert+09a}. 

These extremely strong magnetic fields  will inevitably introduce deformations of
the neutron stars \citep[i.e.][and references therein]{Haskell_Samuellsson+08a,Mastrano_Larsky+13a}. 
A purely toroidal field is known to make the star prolate, 
while a poloidal field will tend to make it oblate. 
Also the distribution of matter in the interior will be affected, depending
on the softness or stiffness of the EOS describing the nuclear matter.
Deformations could even be revealed: if the system is rotating a natural consequence 
will be the emission of Gravitational Waves (GWs), and the new generations
of detectors could search for the emission by these objects.
\citet{Mastrano_Melatos+11a,Gualtieri_Ciolfi+11a,Cutler02a,DallOsso_Stella07a} 
have all estimated the losses of energy due to GWs for newly formed NSs, a process that
will compete with the emission of relativistic outflows. More recently
an upper limit to the magnetic field inside the Crab Pulsar of $7\times  10^{16}$~G has been
set  from the non-detection of GWs \citep{Mastrano_Melatos+11a}.

A newly born proto-NS with magnetic field of the order of $10^{15-16}$~G
is expected to rapidly settle into an equilibrium configuration, given that
the corresponding Alfv\'en crossing time is much smaller than the
typical Kelvin-Helmholz timescale \citep{Pons_Reddy+99a}.  
Theoretical models for equilibria of classical magnetized stars have a long tradition,
dating back to \cite{Chandrasekar_Fermi53a} 
\citep[also][]{Monaghan66a,Ostriker_Hartwick68a,Miketinac75a,Monaghan65a,
Woltjer60a,Chandrasekar_Fermi53a,Ferraro54a,Prendergast56a,Roxburgh66a,Roberts55a},
up to more recent developments \citep{Tomimura_Eriguchi05a,Yoshida_Yoshida+06a}. 
Models for stars endowed with strong magnetic fields in General Relativity (GR)
have started to appear only in the last years, due to the additional complexity of the equations.
Many of these models focus on simple configurations of either a purely toroidal
\citep{Kiuchi_Yoshida08a,Kiuchi_Kotake+09a,Frieben_Rezzolla12a} or
a purely poloidal magnetic field \citep{Bocquet_Bonazzola+95a,Konno01a,Yazadjiev12a}. 
However, as originally suggested by \cite{Prendergast56a}, such configurations are 
expected to be unstable \citep{Wright73a,Tayler73a,Tayler74a,Markey_Tayler73a}. 
More recently \citet{Braithwaite_Norlund06a,Braithwaite_Spruit06a,Braithwaite09a} 
have shown, via numerical simulation, that such
instability can rapidly rearrange the magnetic configuration of the stars. 
It is found that, if the magnetic helicity is finite, the
magnetic field relaxes to a mixed configuration of toroidal and
poloidal field, which is roughly axisymmetric. In these configurations the toroidal field is confined
in a ring-like region, immediately below the stellar surface, while the
poloidal field smoothly extends outwards. Such configurations are
usually referred as {\it Twisted Torus} (TT), and these models have been
presented so far either in Newtonian regime 
\citep{Lander_Jones09a,Lander_Jones12a,Glampedakis_Andersson+12a,Fujisawa_Yoshida+12a}, 
or within GR metrics following a perturbative approach
\citep{Ciolfi_Ferrari+09a,Ciolfi_Ferrari+10a,Ciolfi_Rezzolla13a},
where either the metric or the field are only developed
considering first order deviations. In all cases, until very recently 
\citep{Ciolfi_Rezzolla13a}, it was difficult to investigate toroidally dominated
configurations (precisely those more likely to result from the rearrangement of the field). 

As we will show, convergence of the models in the extreme cases of very strong
magnetic field often requires higher order corrections, even for the simplest configurations.
For purely toroidal fields, for example,
the validity of the results in \citet{Kiuchi_Yoshida08a} (KY08
hereafter) has been recently questioned
by \citet{Frieben_Rezzolla12a} (FR12 hereafter), where different models have been found for the same
set of parameters. On the other hand, purely poloidal configurations have been
presented only by \citet{Bocquet_Bonazzola+95a} (BB95 hereafter) and \citet{Konno01a}, and a study of both the
parameter space and the role of the distribution of internal currents
have not been fully carried out yet.

The main difficulty in solving for magnetized equilibrium models in GR
is due to the non-linear nature of Einstein equations for the metric. 
In particular for TT configurations and if rotation is included, 
as we will show in the next section, many metric terms must be retained 
and a large set of coupled elliptic partial differential equations
has to be solved by means of numerical methods. 
However, it is well known that non-linear elliptical equations can be numerically unstable, 
depending on the way the non-linear terms are cast. This might in part explain the 
discrepancies sometimes present in the literature. 

We present here a novel approach to compute magnetized equilibrium models for
NSs. Instead of looking for an exact solution of Einstein equations, we
make the simplifying assumption that the metric is conformally flat, imposing
the so-called \emph{Conformally Flat Condition} (CFC) by \cite{Wilson_Mathews03a,Wilson_Mathews+96a}.
This allows us to greatly simplify the equations to be solved, and
to cast them in a form that is numerically stable
\citep{Cordero-Carrion_Cerda-Duran+09a,Bucciantini_Del-Zanna11a}.
Moreover, this approach improves upon previous works
\citep{Ciolfi_Ferrari+09a,Ciolfi_Ferrari+10a,Ciolfi_Rezzolla13a}
where the metric was assumed to be spherically symmetric. By approximating
the metric, we are able to solve for equilibrium without resorting to
perturbative approaches. This allows us, on the one hand to investigate
cases with a higher magnetic field, on the other to capture strong deformations 
of the stellar shape. Interestingly, where a comparison was
possible, we have verified that the assumption of a conformally flat
metric leads to results that are indistinguishable, within the
accuracy of the numerical scheme, from those obtained in the correct regime. 
This suggests that the
simplification of our approach does not compromise the accuracy of the
results, while greatly simplifying their computation.

This paper is structured in the following way. In Sect.~\ref{sec:formalism}, the
general formalism, the CFC approximation , and the model equations
describing the structure and geometry of the magnetic field and
related currents are presented. In Sect~\ref{sec:numerical} we briefly describe
our numerical scheme and its accuracy. In Sect.~\ref{sec:results} we
illustrate our results, for various magnetic configurations, and
compare them with existing ones. Finally
we conclude in Sect.~\ref{sec:conclusions}.

In the following we assume a signature $(-,+,+,+)$ for the spacetime
metric and we use Greek letters $\mu, \nu, \lambda,\ldots$
(running from 0 to 3) for 4D space-time tensor components, while Latin
letters $i, j, k, \ldots$ (running from 1 to 3) will be employed for
3D spatial tensor components. Moreover, we set $c=G =1$ and
all $\sqrt{4\pi}$ factors will be absorbed in the definition of the electromagnetic fields.

\section{General formalism and model equations}
\label{sec:formalism}

In this section we will introduce the general formalism we have
adopted to construct equilibrium models. We will firstly present and
justify our assumptions on the symmetries and form of the spacetime,
that we have chosen. We will show how, under those assumptions, given a
distribution of momentum-energy, one can solve Einstein's equations,
and determine the associated metric. Then we will illustrate how to
determine an equilibrium configuration, for the matter and the fields,
on a given metric.

\subsection{The $3+1$ formalism and  \emph{Conformal Flatness}}

Numerical relativity codes for the evolution of Einstein's equations, or for the evolution
of fluid/MHD quantities within a fixed or evolving spacetime, are nowadays built
on top of the so-called $3 + 1$ formalism \citep[e.g.][]{Alcubierre08a,Gourgoulhon12a}.
Any generic spacetime endowed with a metric tensor $g_{\mu\nu}$ 
can be split into spacelike hypersurfaces $\Sigma_t$, with
a timelike unit normal $n_\mu$ (the velocity of the \emph{Eulerian observer}). 
The induced 3-metric on each hypersurface is 
$\gamma_{\mu\nu} := g_{\mu\nu} +n_\mu n_\nu$.
If $x^\mu := (t, x^i)$ are the spacetime coordinates adapted to the foliation 
introduced above, the generic line element is 
\be
ds^2 = -\alpha^2dt^2+ \gamma_{ij}(dx^i+\beta^i dt)(dx^j+\beta^j dt),
\ee
where the \emph{lapse} function $\alpha$ and the \emph{shift vector}  $\beta^i$
(a purely spatial vector) are free \emph{gauge} functions. When $\beta^i=0$ the
spacetime is said to be \emph{static}.

Consider now spherical-like coordinates $x^\mu = (t, r, \theta, \phi)$ and assume
that our spacetime is  \emph{stationary} and \emph{axisymmetric}. 
This implies the existence of two commuting \emph{Killing vectors}, $t^\mu :=(\partial_t)^\mu$ 
(timelike) and $\phi^\mu :=(\partial_\phi)^\mu$ (spacelike) \citep{Carter70a,Carter73a}, spanning
the timelike 2-plane $\Pi := \mathrm{Vect}(t^\mu, \phi^\mu)$. Any
vector  $V^\mu$ is said to be {\it toroidal} 
if $V^\mu \in \Pi \Rightarrow V^\mu= c_t t^\mu + c_\phi \phi^\mu$ (with $c_\phi\neq 0$), and 
\emph{poloidal} (or \emph{meridional}) if it lies in the spacelike 2-plane perpendicular to $\Pi$.
Additional properties are valid for the subset of \emph{circular} spacetimes,
for which the coordinates $(r,\theta)$ span the 2-surfaces orthogonal to $\Pi$,
leading to the simplification
$g_{tr} = g_{t\theta} = g_{r\phi} = g_{\theta\phi} = 0$, where 
all remaining metric tensor components depend on $r$ and $\theta$ alone.
This type of metric is generated by configurations of matter-energy for which
the momentum-energy tensor $T^{\mu\nu}$ is also circular, and this happens when
\be
t_\mu T^{\mu [ \nu} t^\kappa \phi^{\lambda ]} =0 , \quad 
\phi_\mu T^{\mu [ \nu} t^\kappa \phi^{\lambda ]} =0,
\ee
where square brackets indicates antisymmetrization with respect to
enclosed indexes. 

Consider now the case of rotating, magnetized compact objects to be described
as equilibrium solutions of the GRMHD system. The stress-energy tensor reads
\be
T^{\mu\nu} = (e+p+b^2) u^\mu u^\nu - b^\mu b^\nu + (p + \tfrac{1}{2}b^2) g^{\mu\nu},
\label{eq:grmhd}
\ee
where $e$ is the total energy density, $p$ is the pressure, $u^\mu$ is the 4-velocity
of the fluid, and $b^\mu := F^{*\mu\nu}u_\nu$ is the magnetic field as measured
in the comoving frame, and $F^{\mu\nu}$ is the Faraday tensor (the asterisk indicates
the dual). Notice that the ideal MHD condition is, for a perfect
conductor, $e^\mu := F^{\mu\nu}u_\nu = 0$, thus the comoving electric field must vanish.
For more general forms of Ohm's law see \cite{Bucciantini_Del-Zanna13a}.
When applied to the above form of the momentum-energy tensor, the
circularity condition holds provided the 4-velocity is toroidal, that is
$u^\mu \in \Pi \Rightarrow u^\mu := u^t ( t^\mu + \Omega \phi^\mu)$,
due to $t_\mu u^\mu \neq 0$, where $\Omega := u^\phi/u^t = d\phi/dt$ is the 
fluid angular momentum as measured by an observer at rest at spatial infinity.
If one looks for magnetic configurations independent of the flow structure, 
in the limit of ideal MHD, circularity requires that the comoving magnetic field 
must be either purely toroidal, $b^\mu \in \Pi$, with 
$b_\mu u ^\mu = 0 \Rightarrow b_t = - \Omega b_\phi$, 
or purely poloidal, that is $b^\mu t_\mu=b^\mu\phi_\mu=0$. In the latter case, stationarity
requires solid body rotation $u^\phi / u^t = \mathrm{const}$ \citep{Oron02a},
or $\Omega$ must be a constant on \emph{magnetic surfaces} \citep{Gourgoulhon_Markakis+11a}. 
For mixed (\emph{twisted torus}) configurations circularity does not
hold.

In the case of circular spacetimes and spherical-like coordinates, a common
choice is to assume $g_{r\theta}=0$ and $g_{\theta\theta}=r^2 g_{rr}$
(a two metric is always conformally flat), leading
to the \emph{quasi-isotropic form}, than can be written as
\be
ds^2 = - \alpha^2 dt^2 + \psi^4 (dr^2+ r^2 d\theta^2) + R_{\rm q}^2 \, (d\phi +\beta^\phi dt)^2,
\label{eq:qi}
\ee
where $\alpha(r,\theta)$, $\psi(r,\theta)$ (the so called {\it
  conformal-factor}), $R_{\rm q}(r,\theta)$ (the {\it quasi-isotropic radius}), and $\beta^\phi(r,\theta)$ are the
metric terms to be derived from Einstein's equations. 
Models of stationary and axisymmetric equilibria of  rotating NSs are generally built on top
of this metric \cite[e.g.][]{Gourgoulhon10a}, even in the magnetized
case  (KY08,FR12,BB95)
for either purely poloidal or purely toroidal fields. However, in the mixed case, even if the above
form of the metric is no longer appropriate, sensible deviations are expected to arise only
for unrealistically large values of the magnetic field of $\sim 10^{19}$~G \citep{Oron02a}.
Moreover, it is known that even for highly deformed objects, i.e. for rotating NSs at the
\emph{mass shedding limit}, the difference $\psi^4 r^2 \sin^2 \! \theta-R_{\rm q}^2$ is of order $10^{-4}$, and the metric
can be further simplified to
\be
ds^2 = - \alpha^2 dt^2 + \psi^4 [dr^2+ r^2 d\theta^2+r^2 \sin^2 \! \theta\, (d\phi +\beta^\phi dt)^2].
\label{eq:iso}
\ee
Under this latter assumption, the spatial three-metric is {\it
  conformally flat}, and the spherical coordinates can be identified
with the canonical {\it isotropic coordinates}. This form is better suitable to numerical solution, as it is described
below. 

\subsection{Solving Einstein's equations in the \emph{Conformally Flat
    Condition} (CFC)}

The $3 + 1$ formalism introduced in the previous section, allows us to
recast Einstein's equations, in a form that is particularly well suited for
numerical solutions. The first step in this direction is  to perform a $3+1$ decomposition of the
energy-momentum tensor, on the same foliation.
The  $T^{\mu\nu}$ for the GRMHD system in Eq.~(\ref{eq:grmhd}) splits as
\be
E := n_\mu n_\nu T^{\mu\nu} \! = \!
 (e \! + \! p) \Gamma^2 - p + \tfrac{1}{2}(E^2+B^2),
 \label{eq:E}
 \ee
 \be
S^i := - n_{\,\mu}\gamma^i_\nu T^{\mu\nu} \! = \!
(e \! + \! p) \Gamma^2 \varv^i  + \epsilon^{ijk}E_j B_k,
\label{eq:S^i}
\ee
\be
S^{ij}  := \gamma^i_{\,\mu}\gamma^j_{\,\nu}T^{\mu\nu} \! = \! 
(e \! + \! p) \Gamma^2 \varv^i\varv^j \! - \!  E^i E^j \! - \!  B^i B^j \! + \!  [ p + \tfrac{1}{2}(E^2+B^2) ] \gamma^{ij} \! ,
\label{eq:S^ij}
\ee
where $E^\mu := F^{\mu\nu}n_\nu$ and $B^\mu := F^{*\mu\nu}n_\nu$ are the
spatial electric and magnetic fields, respectively, as measured by the Eulerian observer, that now we have written explicitly.
In the $3+1$ formalism, the ideal MHD assumption of a vanishing electric field
in the comoving frame becomes the usual relation
\be
E_i = \epsilon_{ijk}\varv^jB^k,
\label{eq:ohm}
\ee
where $\epsilon_{ijk}=\sqrt{\gamma}[ijk]$ is the 3D Levi-Civita tensor.

These quantities act as sources for Einstein's equations. 
Einstein's equations are generally written in the so-called ADM form
\citep{Arnowitt_Deser+59a} as a system of evolutionary equations, and
constrained equations. The evolutionary equations
for the 12 unknowns $\gamma_{ij}$ and $K_{ij}$ (the \emph{extrinsic curvature}),
in the case of a stationary metric, as for the GRMHD equilibria we are
looking for, turn into a condition for
the extrinsic curvature, which relates it directly to the spatial derivatives of the shift vector
\be
2\alpha K_{ij} = D_i \beta_j + D_j \beta_i,
\ee
where $D_i$ is the connection for $\gamma_{ij}$ ($D_k\gamma_{ij}\equiv 0$)
and $D_i\beta^i=0$.  The constrained equations, known as
\emph{Hamiltonian} and \emph{momentum} constrains, take the form
\be
R + K^2 - K_{ij}K^{ij} = 16\pi E,
\ee
\be
D_j (K^{ij} - K \gamma^{ij}) = 8\pi S^i,
\ee
where $R$ is the Ricci scalar associated to the 3-metric and $K=K^i_{\,i}$.

Let us now introduce the two final assumptions. First, we shall seek \emph{static}
(non-rotating) configurations with $\varv^i=0$, then from Eq.~(\ref{eq:ohm}) $E_i=0$
and also $S^i=0$ due to Eq.~(\ref{eq:S^i}). In this case also the spacetime is static
with $\beta^i=K_{ij}=0$ and we have a condition of \emph{maximum slicing} ($K=0$).
Second, as we anticipated we will assume that the 3-metric is \emph{conformally flat}
\be
\gamma_{ij} = \psi^4 f_{ij}, \quad
f_{ij} = \mathrm{diag}(1,r^2, r^2 \sin^2\!\theta),
\label{eq:conformal}
\ee
where $f_{ij}$ is the 3-metric of asymptotic flat space,
so that also $\sqrt{\gamma} = \psi^6 r^2 \sin\theta$ (in this case the
coordinates are said to be {\it isotropic}). It is known that such an approximation
is strictly applicable only for spherically symmetric distributions, however
this form of the metric is commonly used also for generic evolving spacetimes \citep{Wilson_Mathews+96a}, especially for perturbations of
quasi-spherical equilibria or even collapses.

Under the above assumptions, Einstein's equations turn into two
Poisson-like elliptic equations for the unknowns $\psi$ and $\alpha$
(CFC equations), of the form
\be
\Delta u = s u^q,
\label{eq:poiss}
\ee
where $\Delta := f^{ij} \nabla_i \nabla_i$ and $\nabla_i$ are, respectively, 
the usual 3D Laplacian and the \emph{nabla} operator of flat space
(in spherical coordinates), $u$ is the generic variable ($\psi$ or $\alpha\psi$),
$s$ is the corresponding source term, and $q$ provides the exponent of the 
non-linearity ($q=0$ for a canonical Poisson equation). However, it can be demonstrated
that only the condition $sq \geq 0$ insures that the solution
$u$ is \emph{locally unique}.  Then the CFC equations are conveniently recast into a
form that guarantees this property, which is of paramount importance in view of
numerical integration of the system. This form is the following
\be
\Delta \psi = [ - 2\pi \hat{E} ] \psi^{-1} ,
\label{eq:xcfc_psi}
\ee
\be
\Delta (\alpha\psi) = [ 2\pi (\hat{E}+2\hat{S})\psi^{-2} ]  (\alpha\psi),
\label{eq:xcfc_alpha}
\ee
where we have introduced rescaled fluid source terms of the form
\be
\Ebar:=\psi^6 E,\quad \Sbar:=\psi^6 S,
\ee
and $S=S^i_{\,i}$. In the case of static GRMHD equilibria, we have
\be
E = e + \tfrac{1}{2}B^2, \quad S^{ij} = - B^iB^j + (p+\tfrac{1}{2}B^2)\gamma^{ij}, \quad S = 3p + \tfrac{1}{2}B^2. 
\ee
Equations (\ref{eq:xcfc_psi}-\ref{eq:xcfc_alpha}) are the system of equations for the metric
that will be solved here. Notice that this is a subset of the XCFC (\emph{eXtended
Conformally Flat Condition}) system, in the static case. This has been first
presented by \cite{Cordero-Carrion_Cerda-Duran+09a}, and extensively validated in \cite{Bucciantini_Del-Zanna11a},
where the metric evolution was solved either as a initial data problem (the XNS code
for polytropic NSs with toroidal velocity and magnetic field),
or combined to the GRMHD equations within the ECHO code \citep{Del-Zanna_Zanotti+07a}.

\subsection{The Bernoulli integral and the Grad-Shafranov equation}

Consider now the equations for static GRMHD equilibria in a stationary and axisymmetric
metric in $3+1$ form, also assuming $\beta^i$ and $K^{ij}=0$ as in the CFC approximation
of the previous sub-section. Let us start from the case where a poloidal magnetic field
is present, for which a formulation based on the so-called
\emph{Grad-Shafranov equation} [see e.g. \citet{Del_Zanna-Chiuderi96a}]
for the toroidal component of the vector potential is more convenient. The most general
formulation of this kind for GRMHD stationary and axisymmetric equilibria, not necessarily
in a circular spacetime, can be found in \citet{Gourgoulhon_Markakis+11a}, 
to which the reader is referred also for additional references.

The first equation to consider is the divergence-free condition for the
magnetic field $D_i B^i=\gamma^{-1/2}\partial_i (\gamma^{1/2}B^i) = 0$,
that under the assumption of a conformally flat metric leads to
\be
B^r = \frac{ \partial_\theta A_\phi}{\psi^6 r^2\sin\theta}, \quad
B^\theta = -  \frac{ \partial_r A_\phi}{\psi^6 r^2\sin\theta},
\label{eq:aphi}
\ee
where we have used the definition $B^i=\epsilon^{ijk}\partial_jA_k$ to rewrite the 
poloidal components as derivatives of $A_\phi$, as anticipated above.
The surfaces with $A_\phi=\mathrm{const}$ are known as \emph{magnetic surfaces},
and they contain the magnetic poloidal fieldlines. The potential
$A_\phi$ is also known as \emph{magnetic flux function}. Any scalar function $\mathcal{S}$ for which
$B^i\partial_i\mathcal{S}= 0$ must necessarily satisfy $\mathcal{S}=\mathcal{S}(A_\phi)$, 
then must be also constant on magnetic surfaces.
The only other non-vanishing equation of the static GRMHD system 
is the Euler equation in the presence of an external electromagnetic field
\be
\partial_i p + (e \! + \! p) \,\partial_i \ln\alpha = L_i := \epsilon_{ijk} J^j B^k,
\label{eq:lorentz}
\ee
where $L_i$ is the Lorentz force and $J^i = \alpha^{-1}\epsilon^{ijk}\partial_j (\alpha B_k)$
is the conduction current (we recall that due to the ideal MHD assumption
the electric field and the displacement current vanish for $\varv^i=0$). 

If we assume, as it is often done for NS equilibria, a \emph{barotropic} EOS, for
instance $e=e(\rho),\, p=p(\rho)$, then also the specific enthalpy $h:= (e+p)/\rho$,
where $\rho$ is the rest mass density,  can be written as a function of one of the 
previous thermodynamical quantities and the above equation becomes
\be
\partial_i \ln h + \partial_i \ln\alpha  = \frac{L_i}{\rho h}.
\ee
Now, since the curl of the left-hand side vanishes, also the right-hand side must do so and,
in particular, it can be written as a gradient of a scalar function. 
Moreover, since $B^iL_i=\epsilon_{ijk}J^jB^kB^i\equiv 0$,
this must be a (free) function of the potential alone, constant on the magnetic surfaces
as previously discussed.
The poloidal component of the Lorentz force can be then obtained through this novel
\emph{magnetization} function $\mathcal{M}(A_\phi)$ as
\be
L_i = \rho h\, \partial_i \mathcal{M} = \rho h\, \frac{d\mathcal{M}}{d A_\phi}\partial_i A_\phi,
\label{eq:lorentz2}
\ee
and Eq.~(\ref{eq:lorentz}) can be integrated providing the \emph{Bernoulli integral}
\be
\ln{\left( \frac{h}{h_c}\right)} + \ln{\left( \frac{\alpha}{\alpha_c}\right)} - \mathcal{M} = 0,
\label{eq:bernoulli}
\ee
which, once the functional form  $\mathcal{M}(A_\phi)$ has been chosen and
$A_\phi(r,\theta)$ has been found, relates the enthalpy at each point to the conditions set 
in the centre(labeled $c$), where we assume $\mathcal{M}_c=0$.

Consider now the $\phi$ component of the Lorentz force,
which must vanish due to axisymmetry. Thanks to Eq.~(\ref{eq:aphi}) we then find
$0 = L_\phi = \alpha^{-1} B^i \partial_i (\alpha B_\phi)$, thus
\be
B_\phi = \alpha^{-1} \mathcal{I}(A_\phi),
\ee
where $\mathcal{I}(A_\phi)$ is another free function and it is constant on the magnetic surfaces.
This function is also strictly related to the poloidal current, since we have
\be
J^r = \alpha^{-1} B^r \frac{d\mathcal{I}}{d A_\phi}, \quad
J^\theta = \alpha^{-1} B^\theta \frac{d\mathcal{I}}{d A_\phi}.
\label{eq:curpol}
\ee
The toroidal current can be retrieved from the poloidal component of the Lorentz force
in Eq.~(\ref{eq:lorentz2}). Using also the original definition $L_i=\epsilon_{ijk}J^jB^k$
we arrive at the expression
\be
J^\phi = \rho h \, \frac{d\mathcal{M}}{d A_\phi} + 
\frac{\mathcal{I}}{\sigma}\frac{d\mathcal{I}}{d A_\phi},
\label{eq:curtor}
\ee
where we have defined $\sigma := \alpha^2 \psi^4 r^2\sin^2\!\theta$. 
If, instead, derivatives of the poloidal magnetic field
components are worked out, one finds
\be
J^\phi=-\frac{1}{\psi^8 r^2 \sin^2\!\theta}
\left[\Delta_*A_\phi+ \partial A_\phi \partial\ln (\alpha\psi^{-2}) \right],
\ee
where the following operators have been introduced
\be
\Delta_* :=  \partial^2_r+\frac{1}{r^2}\partial_\theta^2-\frac{1}{r^2\tan{\theta}}\partial_\theta,
\ee
\be
\partial f\partial g := \partial_r f \partial_r g+\frac{1}{r^2}\partial_\theta f \partial_\theta g .
\ee
Finally, equating the two above expressions for $J^\phi$, and introducing the new
variable $\tilde{A}_\phi := A_\phi / (r\sin\theta)$ and the new operator 
\be
\tilde{\Delta}_3  \! := \!  \Delta - \frac{1}{r^2\sin^2\!\theta}  \! = \! 
\partial^2_r + \frac{2}{r}\partial_r+\frac{1}{r^2}\partial_\theta^2
+ \frac{1}{r^2\tan{\theta}}\partial_\theta - \frac{1}{r^2\sin^2\!\theta},
\ee
for which $\tilde{\Delta}_3 \tilde{A}_\phi = \Delta_* A_\phi / (r\sin\theta)$
(it coincides with the $\phi$ component of the \emph{vector laplacian} in spherical coordinates),
we retrieve the \emph{Grad-Shafranov} equation for the magnetic flux function $A_\phi$
\be
\tilde{\Delta}_3 \tilde{A}_\phi
+  \frac{\partial A_\phi \partial\ln (\alpha\psi^{-2})}{r \sin\theta}
+ \psi^8 r \sin\!\theta \left( \rho h \frac{d \mathcal{M}}{d  A_\phi}
+ \frac{\mathcal{I}}{\sigma}\frac{d\mathcal{I}}{dA_\phi} \right) = 0.
\label{eq:gs}
\ee
Provided the metric is known (the functions $\alpha$ and $\psi$ in CFC),
the solution procedure is the following: after a choice for the free functions
$\mathcal{M}$ and $\mathcal{I}$ is made, Eq.~(\ref{eq:gs}) is solved over
the whole domain (with appropriate boundary conditions), so that the magnetic
field and current components can be worked out. As anticipated, the thermodynamical
quantities are instead provided from the Bernoulli equation Eq.~(\ref{eq:bernoulli}).
In the remainder, we shall provide the choices of the free functions for
the various magnetic configurations we are interested in.

\subsection{Choice for poloidal and twisted torus configurations}

When $A_\phi\neq 0$, for which the whole body of the previous section applies,
we need to specify the free functions $\mathcal{M}$ and $\mathcal{I}$, as discussed
just above, in a way  appropriate for NS modeling.
In analogy with \cite{Ciolfi_Ferrari+09a} we choose here a second-order polynomial functional
form for $\mathcal{M}$, namely
\be
\mathcal{M}(A_\phi)=k_{\rm pol} (A_\phi +\xi \tfrac{1}{2}A^2_\phi),
\label{eq:mbern}
\ee
where $k_{\rm pol}$ is the \emph{poloidal magnetization constant}, and
$\xi$ is the \emph{non-linear poloidal term}. On the other hand,
the functional form for $\mathcal{I}$ is chosen as 
\be
\mathcal{I}(A_\phi)=\frac{a}{\zeta+1}\Theta [A_\phi-A_\phi^{\rm max}] (A_\phi-A_\phi^{\rm max})^{\zeta+1},
\label{eq:fbern}
\ee
where $\Theta [.]$ is the Heaviside function, $A_\phi^{\rm max}$ is
the maximum value the $\phi$ component of the vector potential reaches
on the stellar surface, $a$ is the \emph{twisted torus magnetization constant} and
$\zeta$ is the \emph{twisted torus magnetization index}.

From Eqs.~(\ref{eq:curpol}-\ref{eq:curtor}) the poloidal components of
the conduction current are, for the assumed choices of the free functions
\begin{align}
&J^r = \alpha^{-1} B^r \, a\Theta[A_\phi-A_\phi^{\rm max}](A_\phi-A_\phi^{\rm max})^{\zeta},
\nonumber \\
&J^\theta  = \alpha^{-1} B^\theta \, a\Theta[A_\phi-A_\phi^{\rm max}](A_\phi-A_\phi^{\rm max})^{\zeta},
\end{align}
whereas the toroidal component is
\be
J^\phi = \rho h \, k_{\rm pol} (1+ \xi A_\phi) + 
\frac{a^2}{(\zeta+1)\sigma} \Theta[A_\phi-A_\phi^{\rm max}](A_\phi-A_\phi^{\rm max})^{2\zeta+1}.
\ee

The above choice of $\mathcal{M}(A_\phi)$ and $\mathcal{I}(A_\phi)$
guarantees that the currents are all confined within the star. In the
purely poloidal case $a=0$, the linear term $\propto A_\phi$ in Eq.~(\ref{eq:mbern}) 
always leads to magnetic field configurations which are dominated by a
dipolar component.  Only the non-linear term $\propto A_\phi^2$ can in principle 
lead to currents that produce higher order multipolar magnetic field configurations. 
However, as it will be discussed later, this kind of configuration can only be realized
numerically under special conditions. With our choice, the toroidal
component of the magnetic field differs from zero only in a rope
inside the star, from which the name of \emph{twisted torus} configuration. 

\subsection{Choice for purely toroidal configurations}
\label{subsec:toroidal}

In the case of a purely toroidal field, most of the formalism leading to the Grad-Shafranov 
equation does not apply, since $A_\phi=0$ and we cannot define the usual free functions
on magnetic surfaces. However, Eq.~(\ref{eq:lorentz}) is still valid and we can still
look for a scalar function $\mathcal{M}$ (though no longer a function of $A_\phi$)
such that $L_i=\rho h \partial_i\mathcal{M}$ and leading to the usual Bernoulli equation
Eq.~(\ref{eq:bernoulli}). The Lorentz force is conveniently written in terms of 
$\alpha B_\phi$, and the Euler equation, for the usual assumptions of a barotropic EOS 
and conformal metric, becomes
\be
\partial_i \ln h + \partial_i \ln \alpha + 
\frac{\alpha B_\phi\partial_i ( \alpha B_\phi ) }{\rho h \, \sigma} = 0.
\ee
The above equation is integrable if also the last term can be written as a
gradient of a scalar function.
If we now define the new variable, related to the enthalpy per unit volume $\rho h$, namely
\be
\G := \rho h \, \sigma = \rho h \, \alpha^2 \psi^4r^2\sin^2\!\theta,
\ee
this is possible provided
\be
B_\phi = \alpha^{-1} \mathcal{I}(\G), \quad
\mathcal{M}(\G) = - \int \frac{\mathcal{I}}{\G}\frac{d\mathcal{I}}{d \G}d\G,
\ee
basically as in the previous case but with a change of dependency, where
the magnetization function is to be plugged into Eq.~(\ref{eq:bernoulli}).

A common assumption (KY08,FR12)
 is to choose a barotropic-type
expression for $\mathcal{I}$ too, for example
\be
\mathcal{I}(\G) = K_m \G^m, \quad \mathcal{M}(\G) = - \frac{m
  K^2_m}{2m-1} \G^{2m-1},
\label{eq:bernoullitor}
\ee
where $K_m$ is the \emph{toroidal magnetization constant}, and $m\ge 1$ is
the \emph{toroidal magnetization index}.
Once the CFC metric has been provided (the functions $\alpha$ and $\psi$), the 
equilibrium is then found by first solving the Bernoulli equation for the specific enthalpy $h$
\be
\ln{\left( \frac{h}{h_c}\right)} + \ln{\left( \frac{\alpha}{\alpha_c}\right)} + 
\frac{m K^2_m}{2m-1} (\rho h \, \sigma)^{2m-1} = 0,
\label{eq:bernoulli2}
\ee
providing also $\rho$, $e$ and $p$ through the assumed EOS, while the magnetic field is
\be
B_\phi = \alpha^{-1}K_m (\rho h \, \sigma)^m.
\ee
When applied to the modeling of magnetized NSs, such choice of the free function
$\mathcal{I}$ (and consequently of $\mathcal{M}$) insures that the field is fully confined 
within the star, and that it is symmetric with respect to the equatorial plane.  

\section{Numerical scheme}
\label{sec:numerical}

The non-linear Poisson-like equations
Eqs.~(\ref{eq:xcfc_psi}-\ref{eq:xcfc_alpha}), are a subset of those found in
the XCFC formalism, and for this reason  we employ the same
numerical algorithm described in \citet{Bucciantini_Del-Zanna13a}, to
which the reader is referred for a complete description. Let us here
briefly summarize it for convenience. Solutions, for the
scalar quantities of interest ($\psi$ and $\alpha\psi$), are searched
in terms of a series of spherical harmonics $Y_l(\theta)$
\be
u(r,\theta):=\sum_{l=0}^{\infty}[A_l(r)Y_l(\theta)].
\ee
The Laplacian can then be reduced to a series of radial 2nd order
boundary value ODEs for the
coefficients $A_l(r)$ of each harmonic, which are then solved using
tridiagonal matrix inversion, on the same radial grid where the
solution is discretized.  Given that the equations are
non-linear this procedure is repeated until convergence, using in the
source term the value of the solution computed at the previous
iteration.

If a poloidal field is present, also the Grad-Shafranov, equation
Eq.~(\ref{eq:gs}), needs to be solved. Interestingly, this can be reduced
to the solution of a non-linear vector Poisson equation, which is
formally equivalent to the equation for the shift-vector (to be more
precise its $\phi$ component) in the XCFC
approximation. $\tilde{A}_\phi$ is searched
in terms of a series of vector-spherical harmonics
\be
\tilde{A}_\phi
(r,\theta):=\sum_{l=0}^{\infty}[C_l(r)Y^\prime_l(\theta)].
\label{eq:harmonics}
\ee
 The only difference is that now the source term is
non-linear. Again we can use the same algorithm, with a combination of
vector spherical harmonics decomposition for the angular part, and
matrix inversion for the radial part 
\citep{Bucciantini_Del-Zanna13a}. Now, this is iterated until
convergence, because of the non-linearity of the source terms.

The use of spherical harmonics allows us to preserve the correct
behaviour on the axis, the correct parity at the center, and the correct
asymptotic trend at the other radius, without the need to use a
compactified domain.

Solutions are discretized on a grid in spherical coordinates in the
domain  $r=[0,25]$, $\theta=[0,\pi]$. For purely toroidal or purely
poloidal cases we use 250 points in the radial
direction and 100 points in the angular one. For TT configurations we
instead used  500 points in the radial
direction and 200 points in the angular one. The radial domain has
been chosen such that its outer boundary is far enough from the
stellar surface, so that
higher order multipoles in the various quantities (i.e. in the metric
terms) become negligible. The boundary conditions at the inner radial
boundary at  $r=0$ are chosen such that  each radial coefficients
$A_l(r)$, $C_l(r)$ goes to 0 with parity $(-1)^l$. Note that this is
different from imposing that they go to 0 as $r\,^l$. This latter
choice is only justified in vacuum, for a flat spacetime, while in
all our cases, the source terms (including terms that contains the
vector potential itself) extend all the way to the centre.
The outer boundary of the computational
domain is always located outside the stellar surface,  which is
defined as the place where the density drops below a fiducial small
value (usually $10^{-5} - 10^{-4}$ times the value of the central
density). This implies that at the outer boundary both the equations
for the metric coefficients $\alpha$ and $\phi$ and the equation for the
vector potential reduce to the equations in vacuum. At the outer
radius we impose that each coefficient  $A_l(r)$, $ C_l(r)$ goes to 0 as
$r^{-(l+1)}$. 

Note that, unlike in previous works \citep{Lander_Jones09a,Lander_Jones12a,Glampedakis_Andersson+12a,Ciolfi_Ferrari+09a,Ciolfi_Ferrari+10a,Ciolfi_Rezzolla13a,Tomimura_Eriguchi05a} we do not solve separately the Maxwell and Einstein equations
inside the star and outside it and then match them at the surface. We
instead solve these equations in the full domain, including both the
star (where the source term are confined) and the outside ``vacuum''. This automatically
guarantees that solutions are continuous and smooth at the stellar
surface. It also allows the stellar surface to adjust freely, and not
to any imposed shape. We have verified that the solution we obtain are
independent of the location of the outer radius. Our previous results
(X-ECHO) for the metric solver indicate that this global approach,
where solutions of non-linear elliptic equations are searched over the
entire domain, at once, gives correct results, without the need to
introduce matching conditions, at often undefined surfaces. In fact,
while in a perturbative approach one can safely assume the stellar
surface to be spherical,  this cannot be done for strong fields, and
the shape of the NS surface is itself unknown. The correct behaviour
on the axis is instead automatically guaranteed by the properties of
spherical harmonics. 

We have verified that at this resolution, the
{\it discretization errors} of our solutions are $\simlt 10^{-3}$, and
at most reach $10^{-2}$ for the most extreme Twisted Torus
configurations. This is likely due to the fact that in the latter
case, the toroidal field is concentrated in a narrow torus-like region at
the edge of the star, while for purely poloidal and purely toroidal
cases, all the quantities are smoothly distributed in the domain. 

In models with purely toroidal or purely poloidal field we have used
20 spherical harmonics. For TT configurations we have used about
40 harmonics.
We have also verified that, increasing the number of spherical harmonics,
does not improve significantly the results. Again the twisted torus
configurations are the ones requiring in general a higher number of
spherical harmonics. We found that 10 are already sufficient to provide results
with an accuracy of the order of $10^{-3}$ both for the purely poloidal or
purely toroidal cases. Instead for the most extreme TT cases we used up
to 50 harmonics. A more detailed discussion of the number of spherical
harmonics needed to get convergent results of the Grad-Shafranov
equation alone, Eq.~(\ref{eq:gs}), is presented in
Appendix~\ref{appendixB}.

\section{Results}
\label{sec:results}

In this section we present a study of various equilibrium
configurations. In particular we analyze how the various global
quantities that parametrize the resulting models change, not only as a
function of the magnetic field strength, but also for different choices
of the field structure (the  distribution of currents) and geometry.

Given that our work focus on the role of magnetic field only, we have
adopted a simple polytropic EoS $p = K_{a} \rho^{\gamma_a}$, with an
adiabatic index $\gamma_{a}=2$ and a
polytropic constant $K_{a}=110$ (in geometrized units\footnote{
  This corresponds to $K_{a}=1.6\times 10^5$ cm$^5$g$^{-1}$s$^{-2}$}). These values are
commonly used in literature and allow us a straightforward comparison
with previous results (KY08, FR12, BB95). In the unmagnetized case, for a central density $\rho_c=8.576
\times 10 ^{14} \, \mbox{g} \, \mbox{cm}^{-3}$,
this EoS gives an equilibrium configuration characterized by a
baryonic mass $M_0=1.680 M_{\sun}$, a gravitational 
mass $M=1.551 M_{\sun}$, 
and a circumferential radius $R_{\rm circ}=14.19 \, \mbox{km}$ (see
Tab.~\ref{tab:tor1}). This will be our reference model for comparison to
magnetized cases.

A detailed description of all the global quantities that can be
defined, and that can be used to parametrize each equilibrium model,
can be found in Appendix \ref{appendix}.

\subsection{Purely Toroidal Field}
\label{subsec:ptf}

\begin{figure*}
	\centering
	\includegraphics[width=.35\textwidth,bb=8 0 415 440, clip]{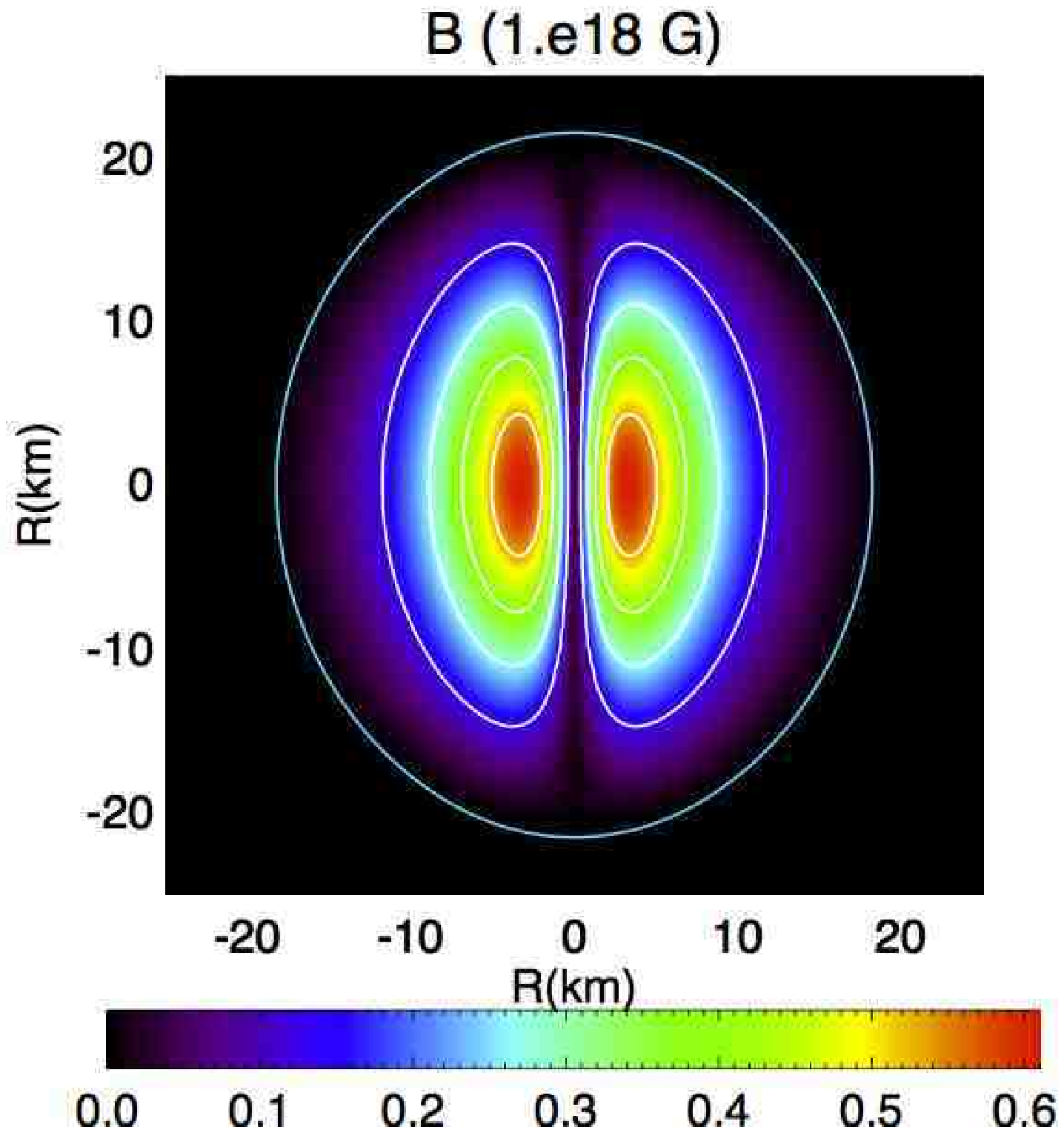} 
	\includegraphics[width=.35\textwidth,bb=8 0 415 440, clip]{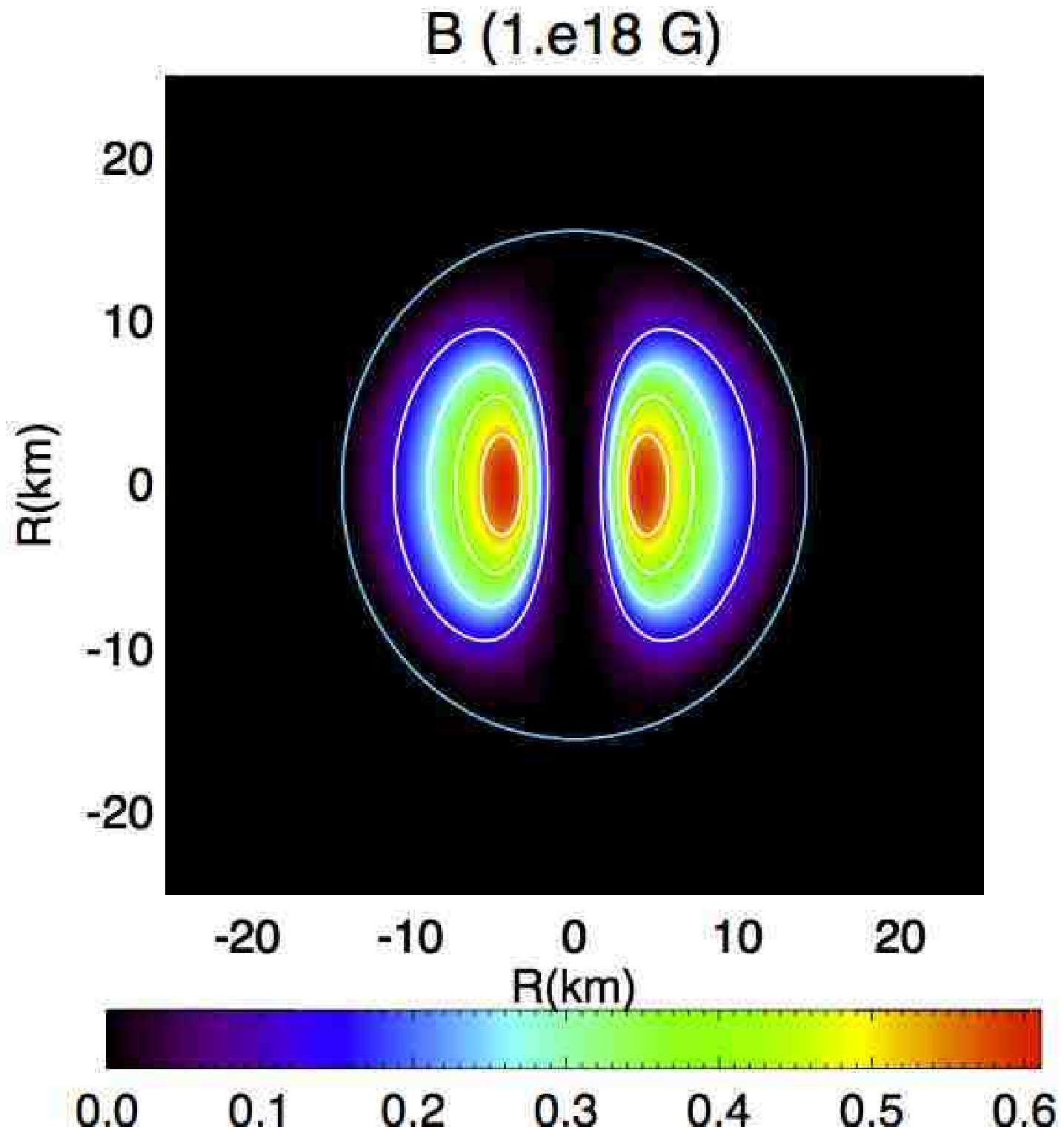} \\
	\includegraphics[width=.35\textwidth,bb=8 0 415 440, clip]{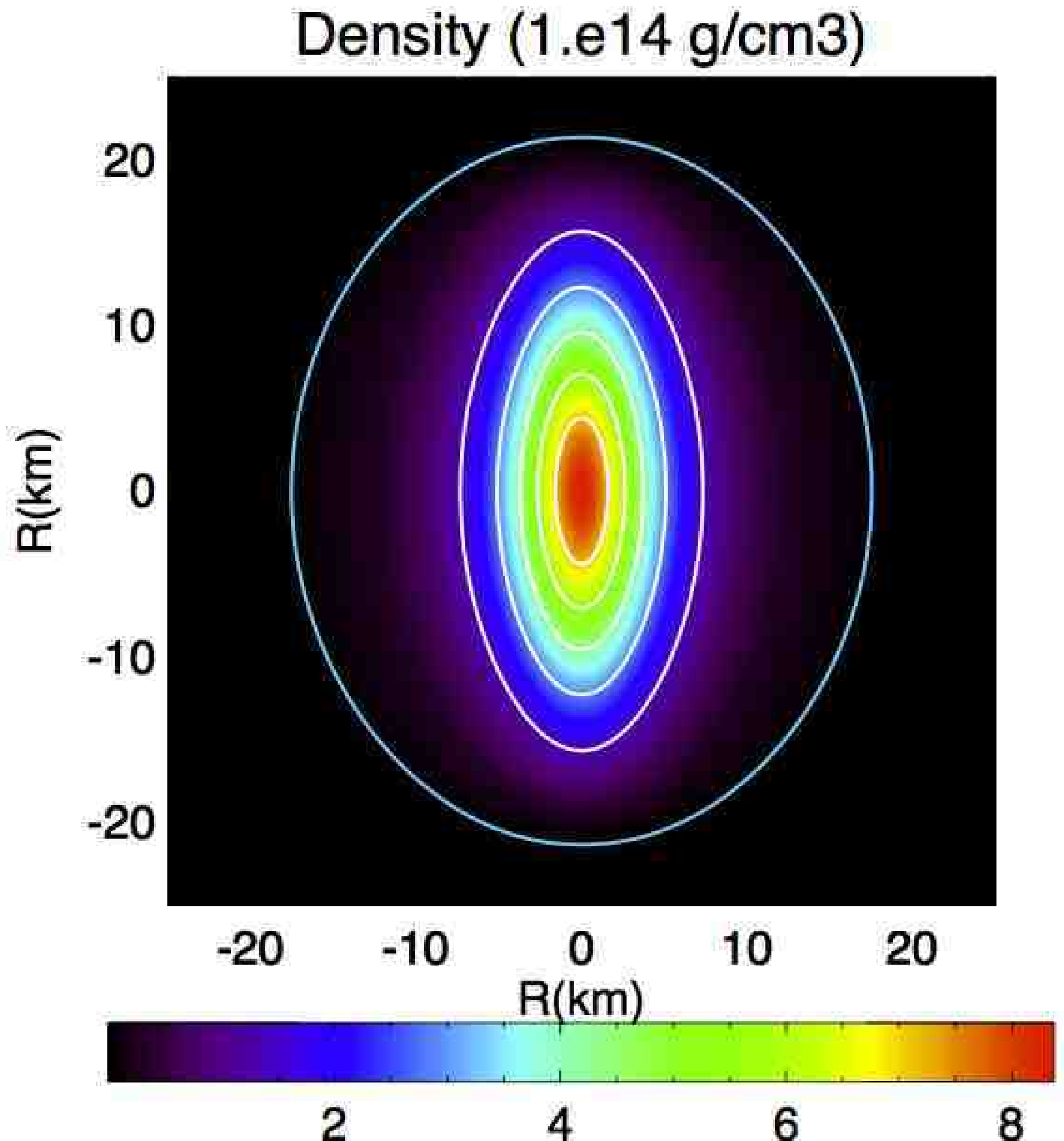}
	\includegraphics[width=.35\textwidth,bb=8 0 415 440, clip]{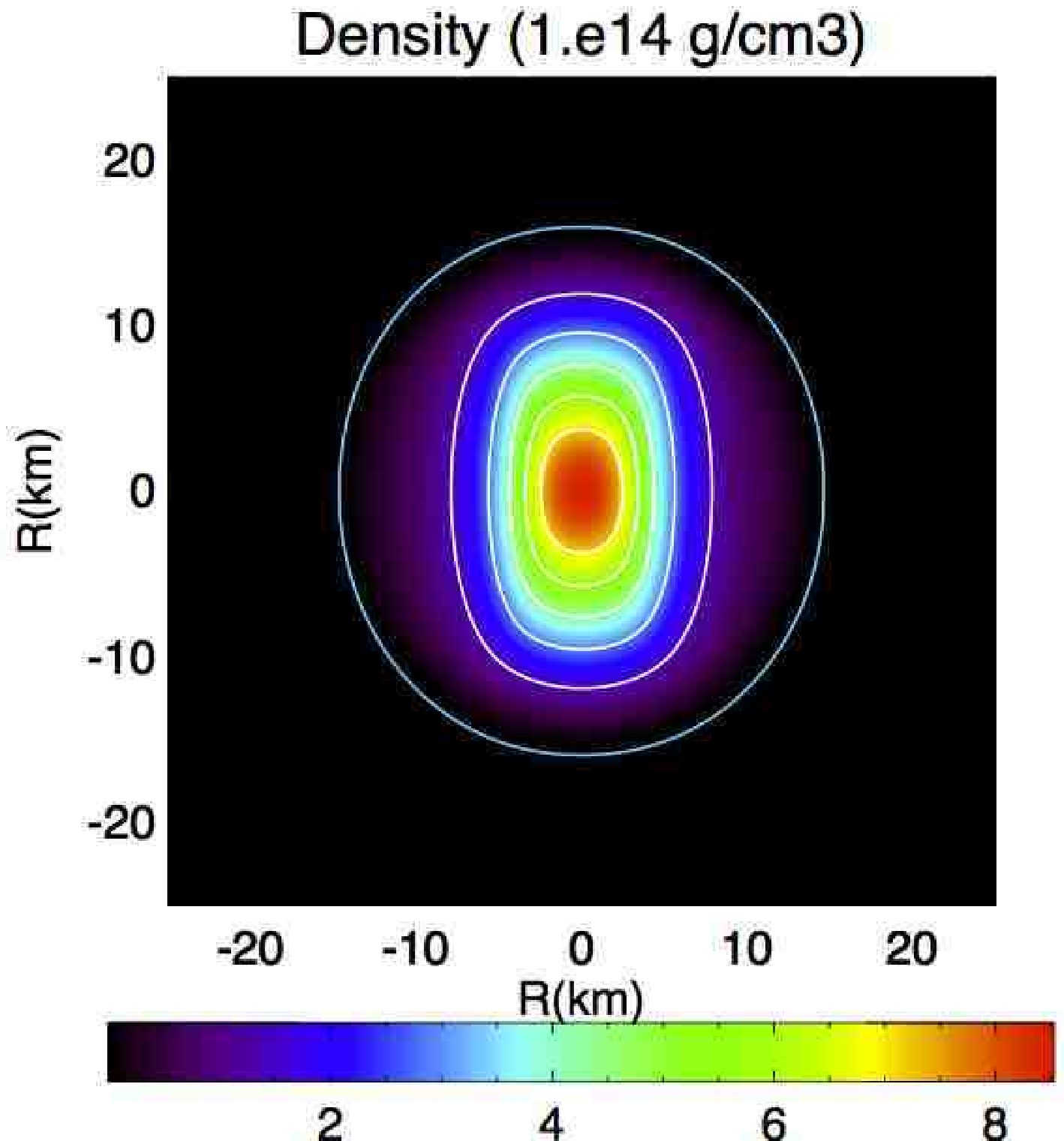} 
	\caption{ Meridional distribution and isocontours of the
          magnetic field strength $B=\sqrt{B^\phi B_\phi}$ (top) 
	and of the baryonic density (bottom) for models with baryonic mass $M_0=1.68 M_{\sun}$, 
	maximum magnetic field strength $\Bmax=6.134\times 10^{17}\,\mbox{G}$, with magnetic
	index $m=1$ (left) and $m=2$ (right). Blue curves represent the surface of the star. Other global 
	quantities related to these configurations are listed in Table~\ref{tab:tor1}.}
	\label{fig:toroidal}
\end{figure*}

Configurations with a purely toroidal magnetic field are obtained with the barotropic-type
expression for $\mathcal{M}(\G)$ in Eq.~(\ref{eq:bernoullitor}).  Let us first discuss
 the role played by the magnetic exponent $m$.
In Fig.~\ref{fig:toroidal} we show the strength of the magnetic field
 and  the distribution of the  baryonic density for two equilibrium configurations 
characterized by the same baryonic mass $M_0=1.68 
M_{\sun}$, the same maximum value of the internal magnetic field strength 
$\Bmax=6.134\times 10^{17}\,\mbox{G}$ but with different values 
of the toroidal magnetization index: $m=1$ and $m=2$
respectively. In Tab.~\ref{tab:tor1} we characterize these models.
Concerning the distribution of magnetic field, they look qualitatively
very similar: as expected for a toroidal
field, in both cases the magnetic field vanishes on the axis of symmetry,
reaches a maximum deep inside the star and then decreases moving
toward the surface where it vanishes. Quantitatively, however, there are
significative differences. In the case $m=1$ the magnetic field
strength goes to zero on the axis as $ r\sin{\theta}$, while the
ratio $B^2/p$, a monotonically increasing function of radius, tends to a constant at the stellar surface (the magnetic
field decreases as fast as the pressure). On the other hand in the case $m=2$ the magnetic field
strength goes to zero on axis $\propto (r\sin{\theta})^3$, while the
ratio $B^2/p$ reaches a maximum inside the star, and then goes to zero at the stellar surface.

\begin{table*}
\caption{\label{tab:tor1}
Global physical quantities of the equilibrium models displayed 
in Fig.~\ref{fig:toroidal} with baryonic mass $M_0=1.68 M_{\sun}$ and maximum 
magnetic field strength $\Bmax=6.134\times 10^{17}\,\mbox{G}$. For the
definition of the various quantities see Appendix \ref{appendix}.}
\begin{tabular}{l*{8}{c}}
\toprule
\toprule
Model &$\rho_c$ &  $M$   & $r_e$  & $r_p/r_e$ & $R_{\rm circ}$ & $\mathscr{H}/\mathscr{W}$ & $\bar{e}$ 
																							 & $\Phi$ \\
 & [$10^{14}\mbox{g}\,{\mbox{cm}}^{-3}$] & [$M_{\odot}$] & [km] & & [km] &   [$10^{-1}$]  & [$10^{-1}$]  &                                                                               
                                                                      [$10^{30}\,\mbox{G}\,\mbox{cm}^2$] \\
\midrule
$m=0$ &  8.576   &  1.551  &  12.08  &  1.000  &  14.19  &  0.000  &  0.000  & 0.000  \\
$m=1$ &  8.430   &  1.596  &  18.10  &  1.139  & 20.15  &  2.013  & -8.130  & 1.538  \\
$m=2$ &  8.588   &  1.577  &  14.01  &  1.104  & 15.92  &  1.246  & -3.730  & 0.862  \\
\bottomrule 
\end{tabular}
\end{table*}

Similar considerations hold for the  distribution
of the baryonic density (Fig.~\ref{fig:toroidal}). 
In both cases the magnetic stresses lead 
to a prolate deformation of the star. This affects the internal layers
even more than the outer ones. Indeed, the typical prolateness of the
 iso-density surfaces in the core is larger  than the deformation
of the stellar surface, and the external low-density
layers. Interestingly, to this axial compression of the internal
layers corresponds an expansion of the outer part of the star to
larger radii, due to the extra pressure support provided by the
magnetic field. There are
two noticeable differences between the $m=1$ and $m=2$ cases, in this
respect. For $m=1$ the iso-density surfaces are, to a good
approximation, prolate ellipsoids, while in the $m=2$ case they tend to
be more barrel-shaped. More important, despite the internal maximum
magnetic field being the same, the $m=2$ case shows a much smaller
deformation. This can be explained recalling that 
the action of the magnetic tension, responsible for the anisotropy, is
$\propto B^2/R$ ($R$ is now
the radius of curvature of the magnetic  field line). For higher values
of $m$ the magnetic field reaches its maximum at increasingly larger
radii, resulting in a relatively smaller tension. Based on our results
it is evident that a magnetic field concentrated at larger radii
will produce smaller effects, than the same magnetic field, buried
deeper inside. This can be rephrased in terms of currents, suggesting
that currents in the outer layers have minor effects with respect to
those residing in the deeper interior.

\begin{figure*}
	\centering
	\includegraphics[width=.4\textwidth]{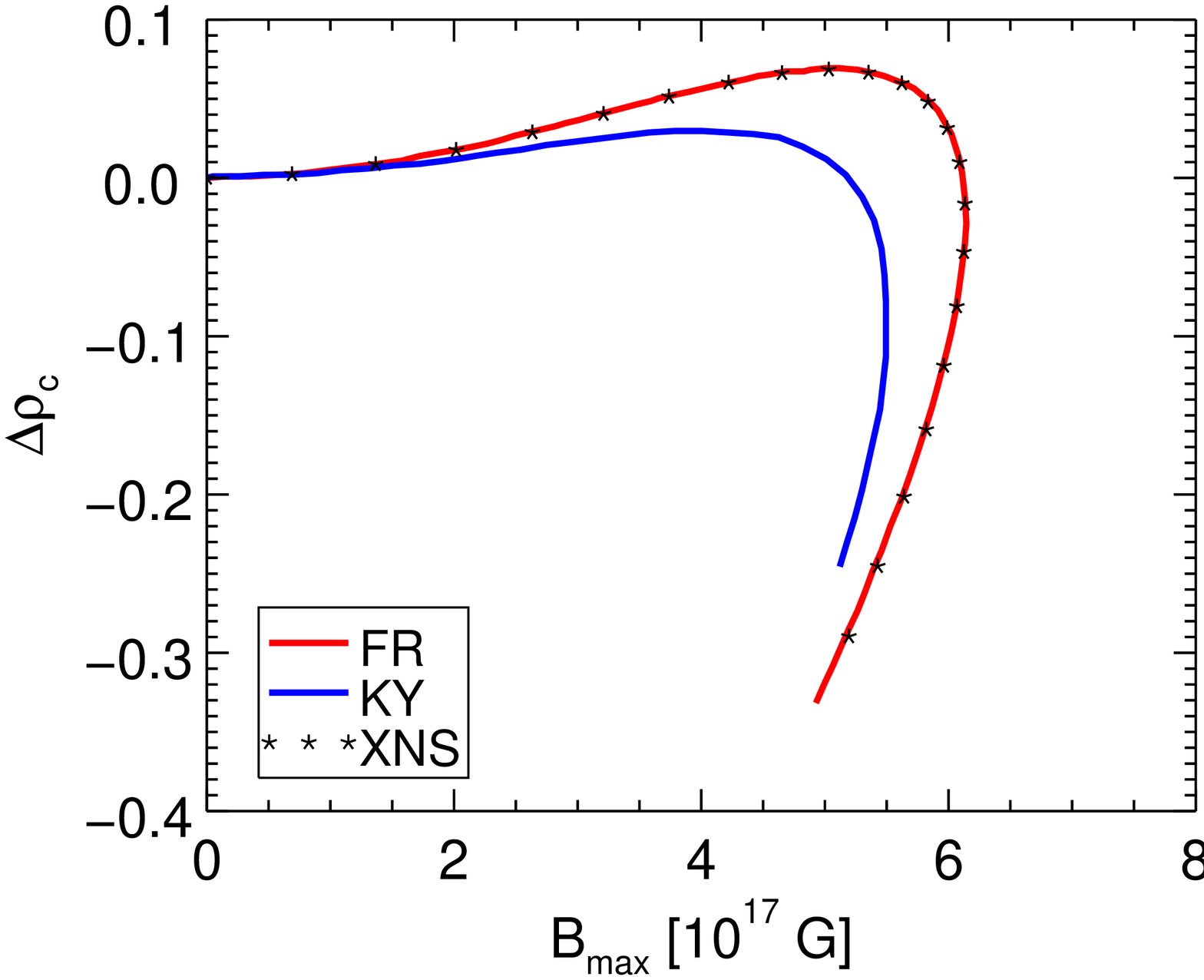} 
	\includegraphics[width=.4\textwidth]{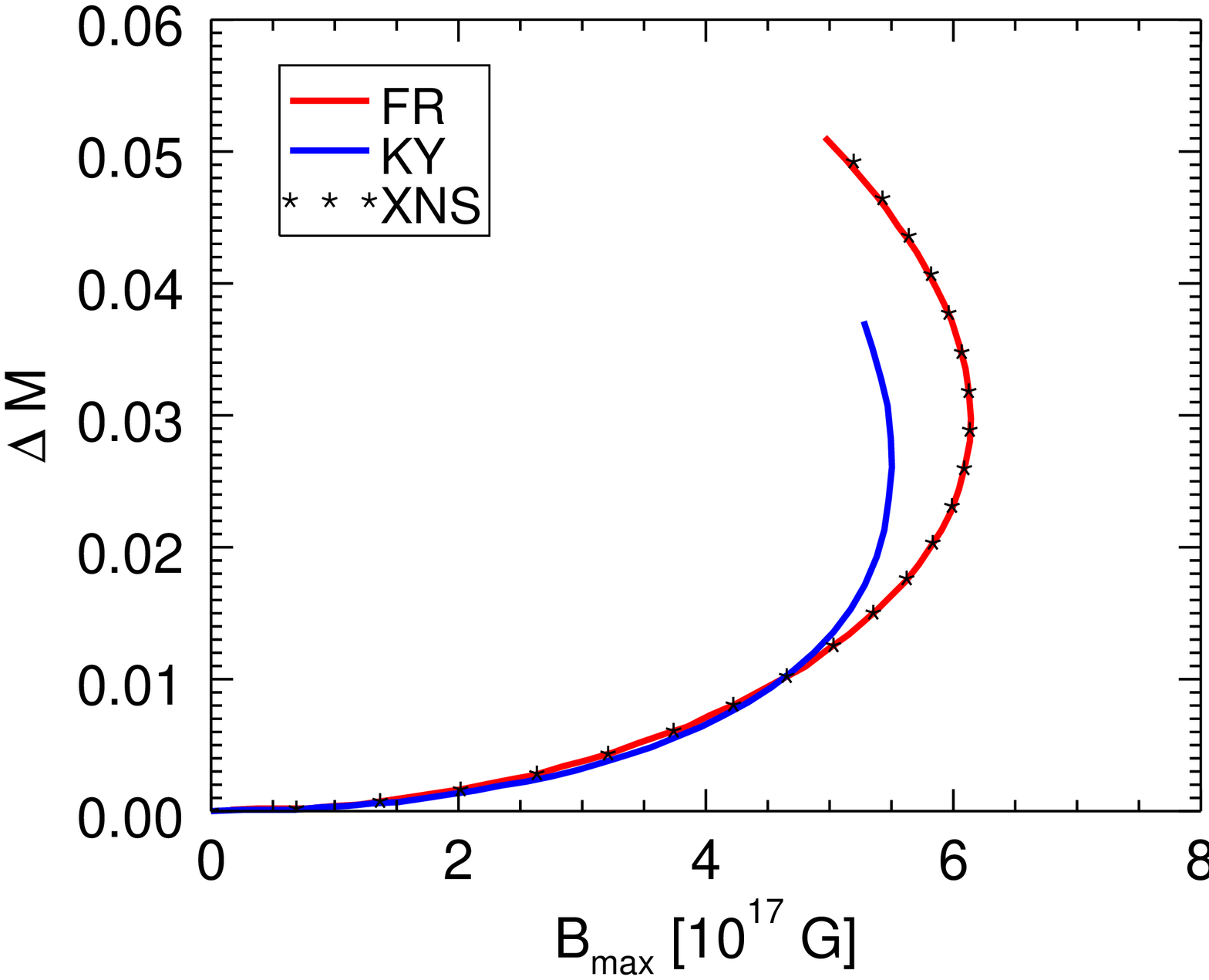} \\
	\includegraphics[width=.4\textwidth]{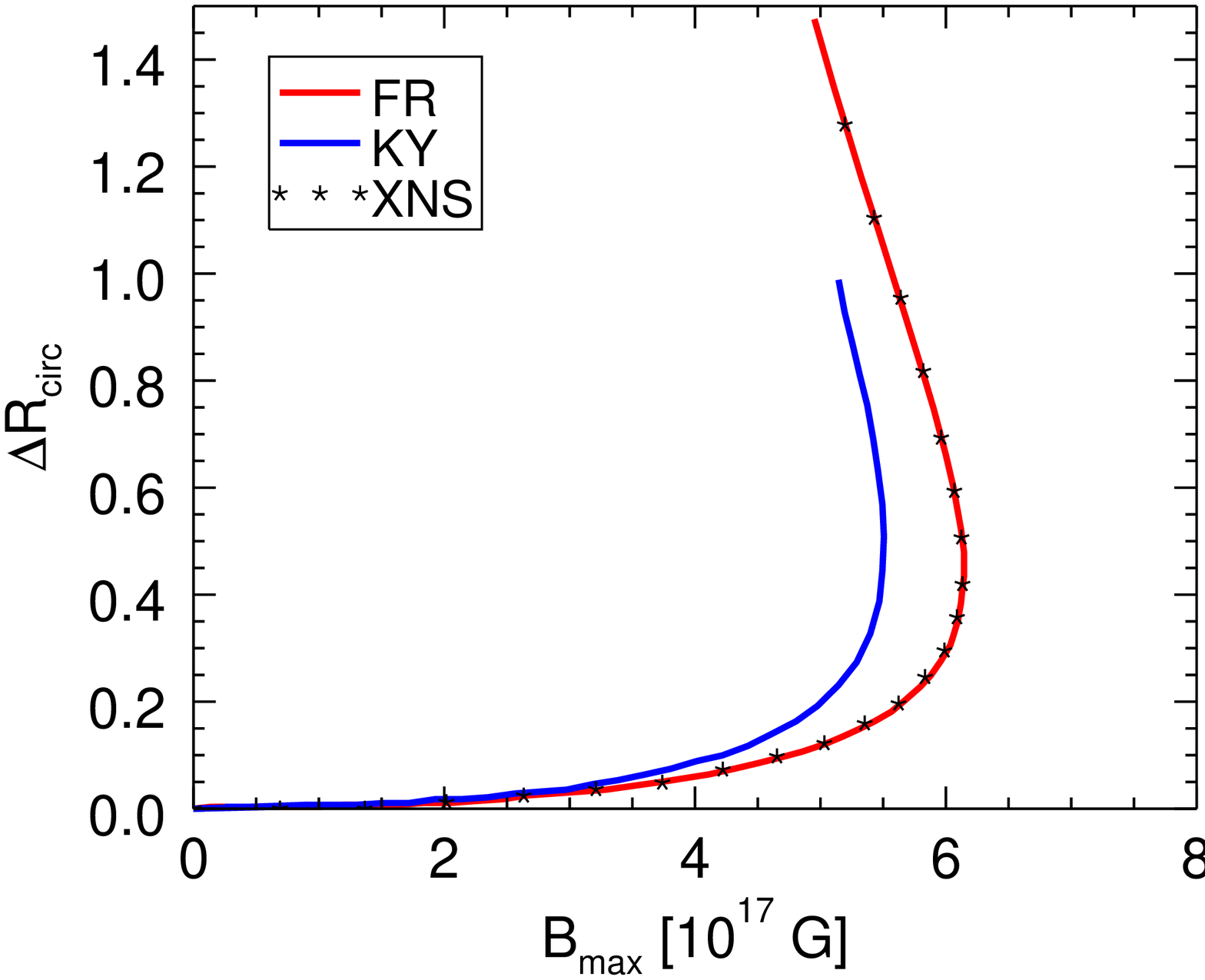} 
	\includegraphics[width=.4\textwidth]{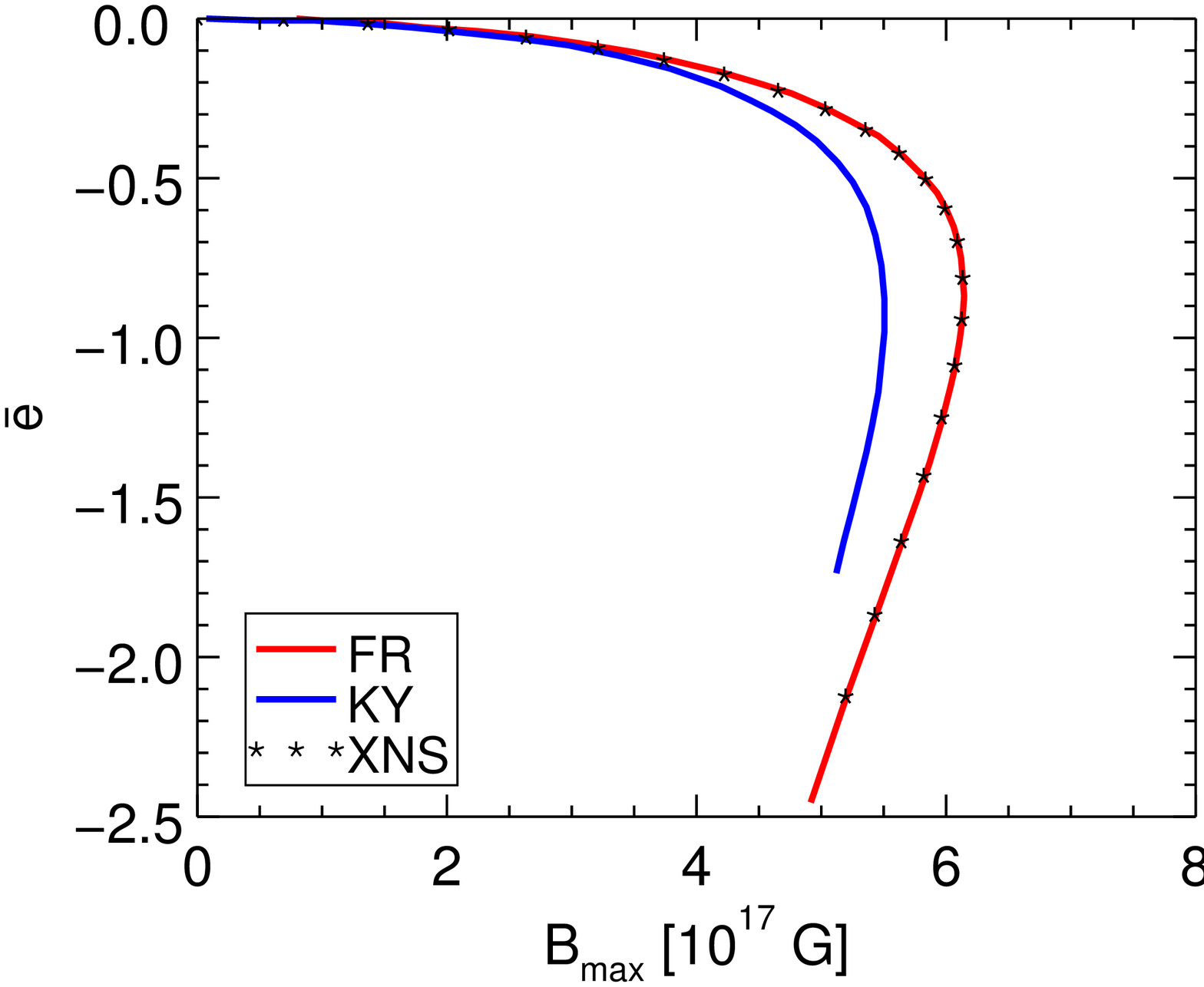}
	\caption{ Variation, with respect to the unmagnetized
          equilibrium model,  of the central baryon density $\rho_c$, 
	of the gravitational mass $M$, of the circumferential radius $\Rcirc$ and of
	the mean deformation rate $\bar e$ along the equilibrium sequence of magnetized 
	 configuration  with constant $ M_0 = 1.68 M_{\sun}$ and $m=1$.
	Lines represent the results by KY08 and FR12, points are our results.}
	\label{fig:cfrKYFR}
\end{figure*}

\begin{figure*}
	\centering
	\includegraphics[width=.4\textwidth]{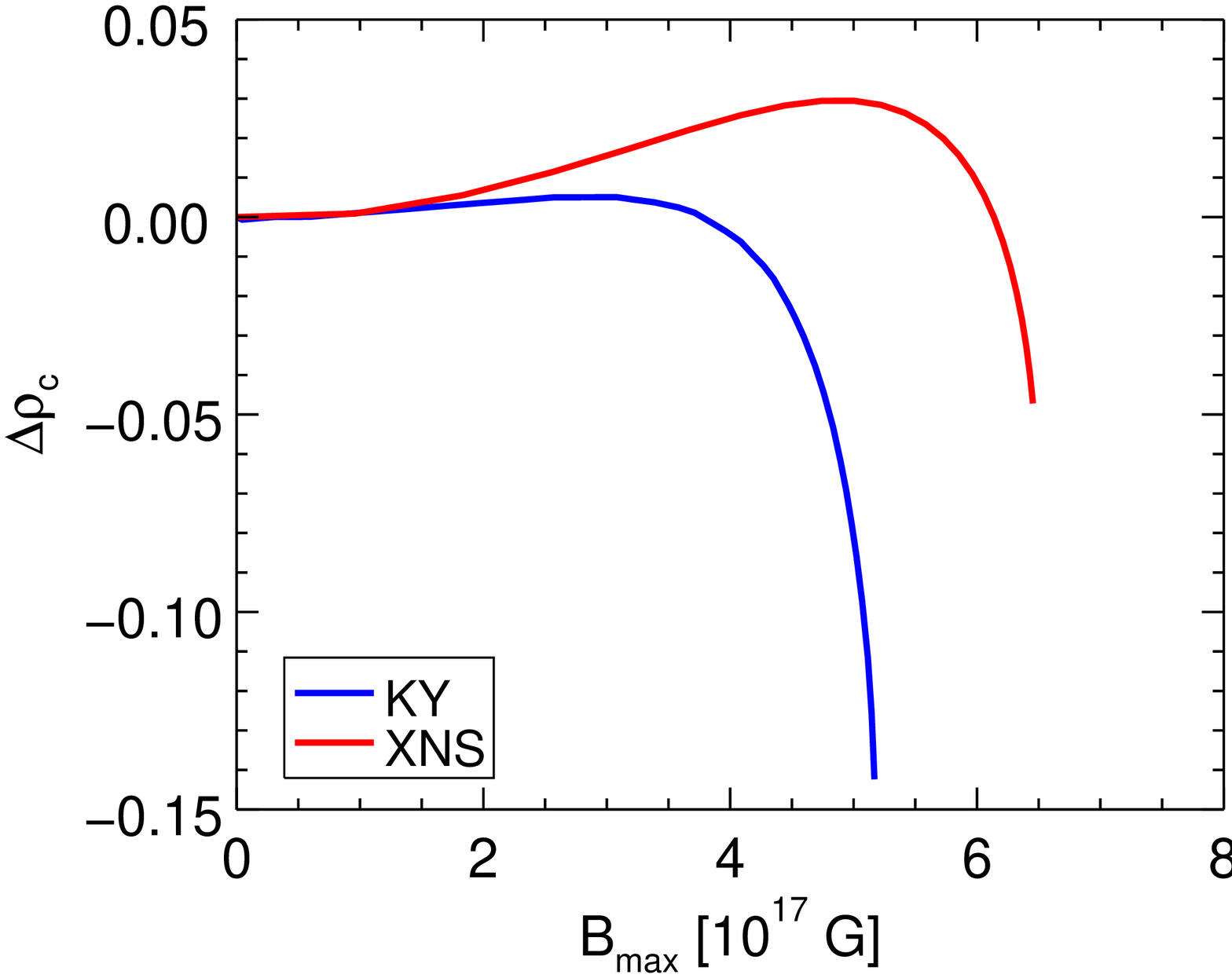} 
	\includegraphics[width=.4\textwidth]{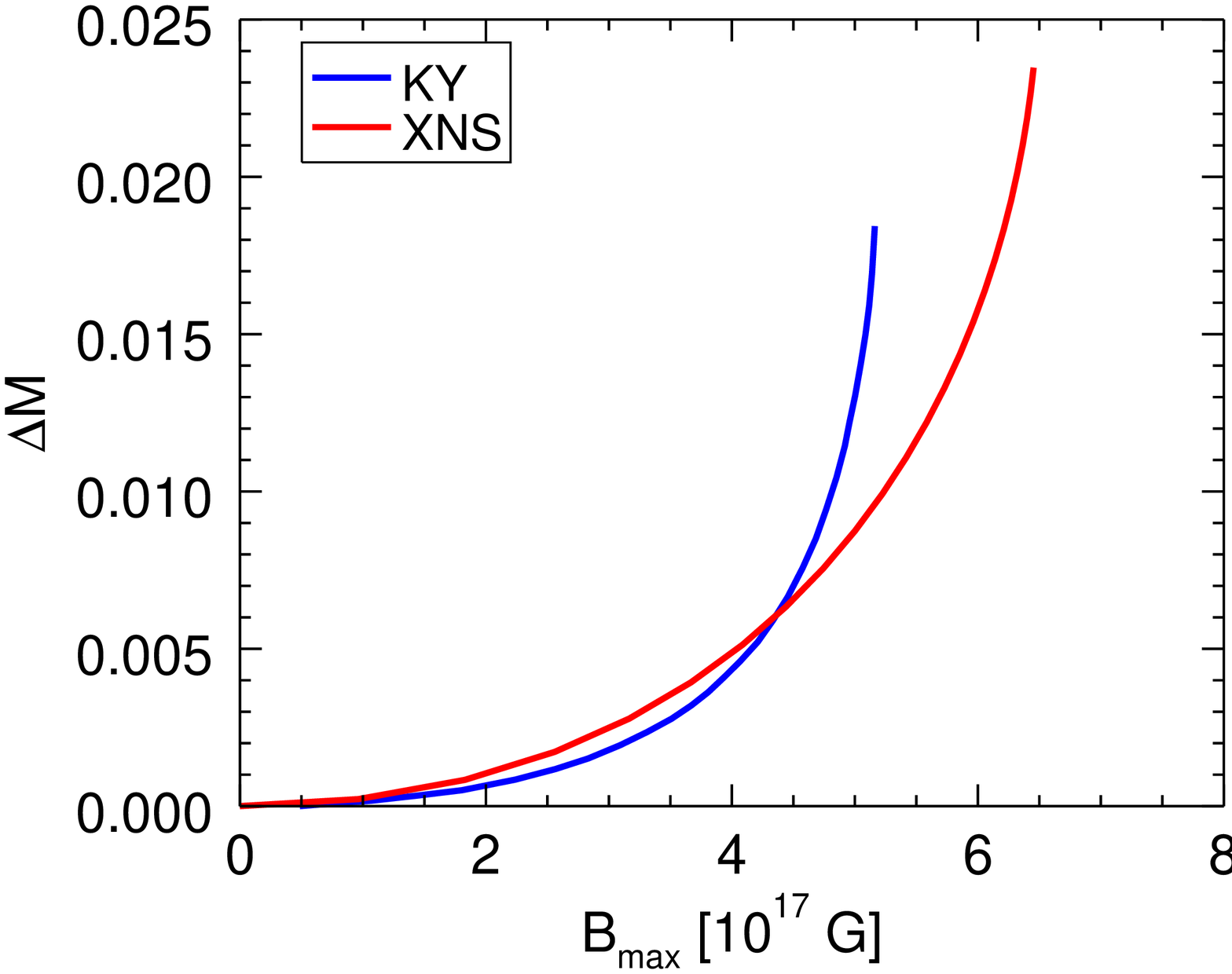} \\
	\includegraphics[width=.4\textwidth]{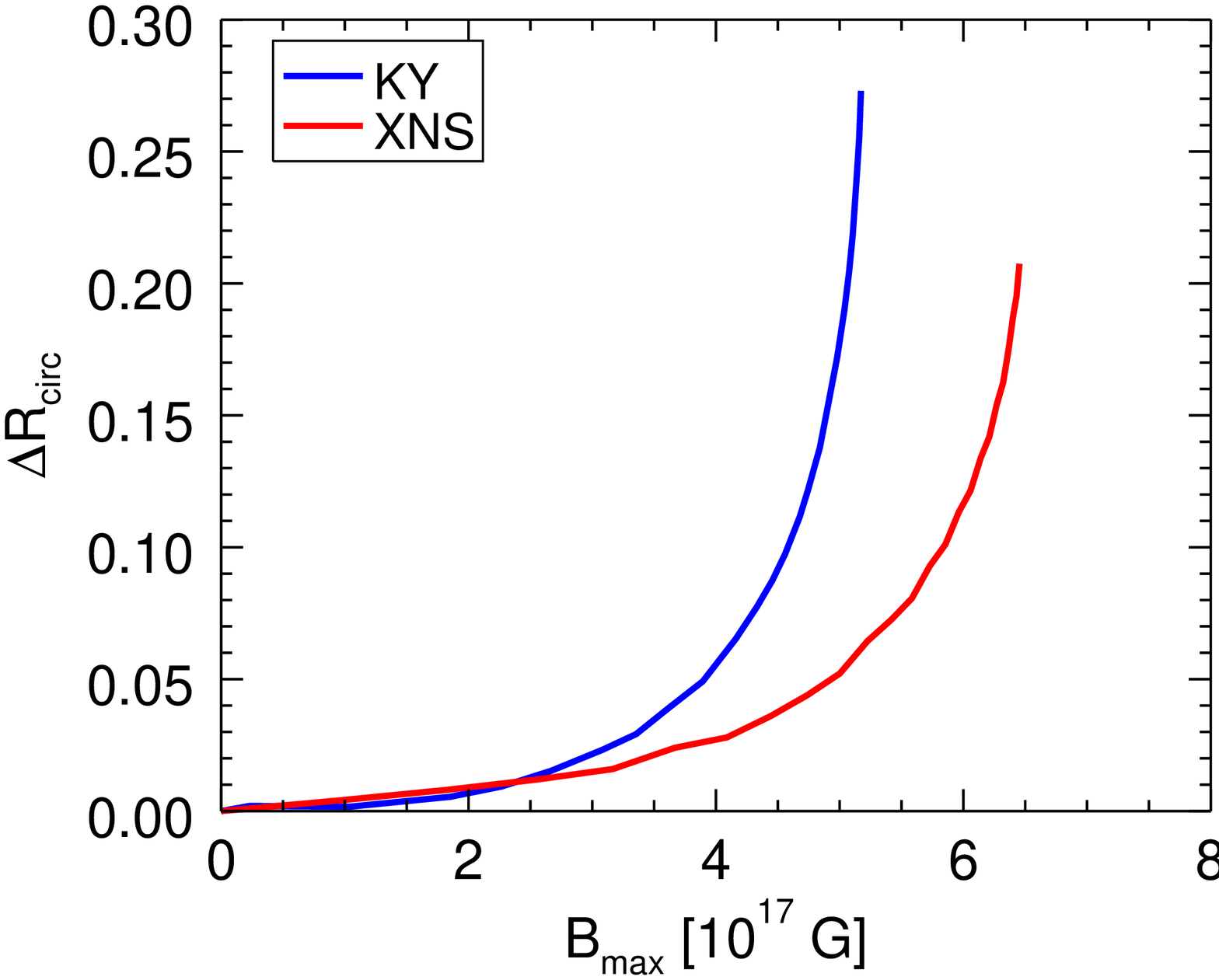}
	\includegraphics[width=.4\textwidth]{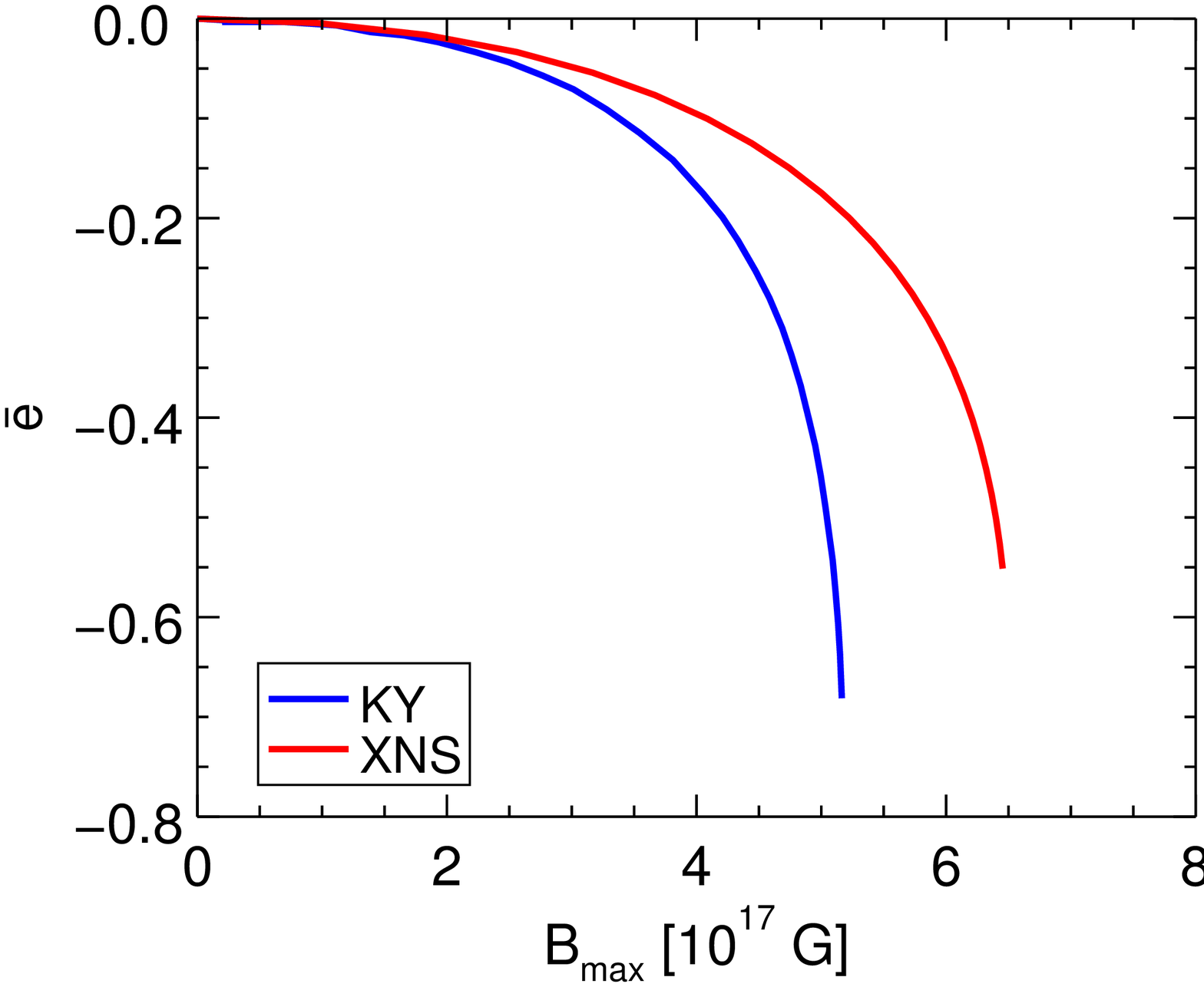}
	\caption{ 
	Same comparison as the one shown in as in
        Fig.~\ref{fig:cfrKYFR} but for the $m=2$ case.}
	\label{fig:cfrKY}
\end{figure*}

Apart from a qualitative  analysis of the structure and configuration
of these equilibrium models, it is possible to investigate in detail
the available parameter space, and how the various quantities are
related. This will allow us also to compare our results with other
previously presented in literature, in particular the results by KY08 and
FR12, for a purely toroidal magnetic field. KY08 and FR12 both solve for
equilibrium in the correct regime for the space-time metric, described
by a quasi-isotropic form. Despite this, the results are significatively
different. 
In Fig.~\ref{fig:cfrKYFR} we compare our results with KY08 and
FR12 (for the case $m=1$). We plot the deviation of four quantities with respect to the
unmagnetized case, as a function of the maximum value of the magnetic
field strength inside the star. The deviation of a quantity $Q$ is
here defined as:
\be
\Delta Q=\frac{[Q(\Bmax,M_0)-Q(0,M_0)]}{Q(0,M_0)}.
\ee
The sequence refers to a set of  
equilibrium models, characterized by a constant baryonic mass  
$M_0=1.68 M_{\sun}$,  as a function of the maximum field strength
$\Bmax$. Following KY08 we show:
the mean deformation rate $\bar e$, the deviation of 
the gravitational mass $\Delta M$, of the circumferential radius
$\Delta R_{\rm circ}$ and of 
the central baryonic density $\Delta \rho_c$. Our models are in
complete agreement with FR12 and confirm the latter results against
KY08. The first thing to notice is that $\Bmax$ is not a monotonic
function of the magnetization constant $K_m$. On the contrary $\Bmax$ initially increases
with $K_m$, till it reaches a maximum value, and then for higher
values of $K_m$ it drops. This is due to the expansion of the
star. For small values of $K_m$, the stellar radius is marginally
affected, and an increase in $K_m$ leads to a higher field. However at
higher values of $K_m$ the radius of the star is largely inflated and
a further increase in $K_m$ translates into an expansion of the star,
and a consequent
reduction of the maximum internal field. If $\Delta M$,
$\Delta \Rcirc$,  or $\bar e$ are plotted
against the total magnetic energy, we find that there appears to be a
monotonic trend, at least in the range covered by our models. A similar effect shows up in the behaviour of the
central density. For small values of $K_m$ the magnetic tension tends to
compress the matter in the core, increasing its density. However as
soon as the magnetic field becomes strong enough to cause the outer
layer of the star to expand, the central density begins to drop
(recall that the sequence is for a fixed baryonic mass). The same
comparison with KY08 in the $m=2$ (FR12 present only the $m=1$ case)  is shown in Fig.~\ref{fig:cfrKY}.


\begin{figure*}
    \includegraphics[width=.49\textwidth]{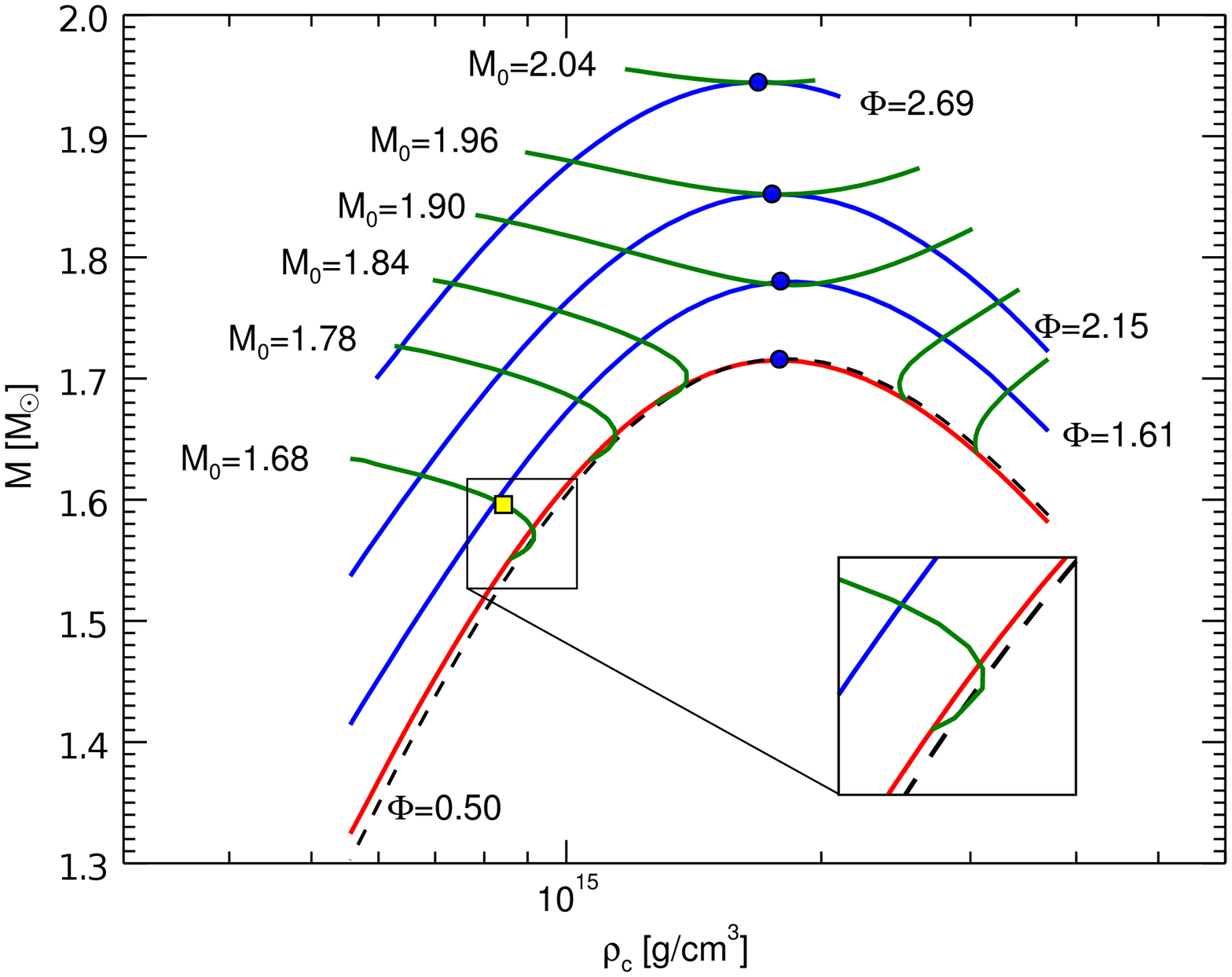}
    \includegraphics[width=.49\textwidth]{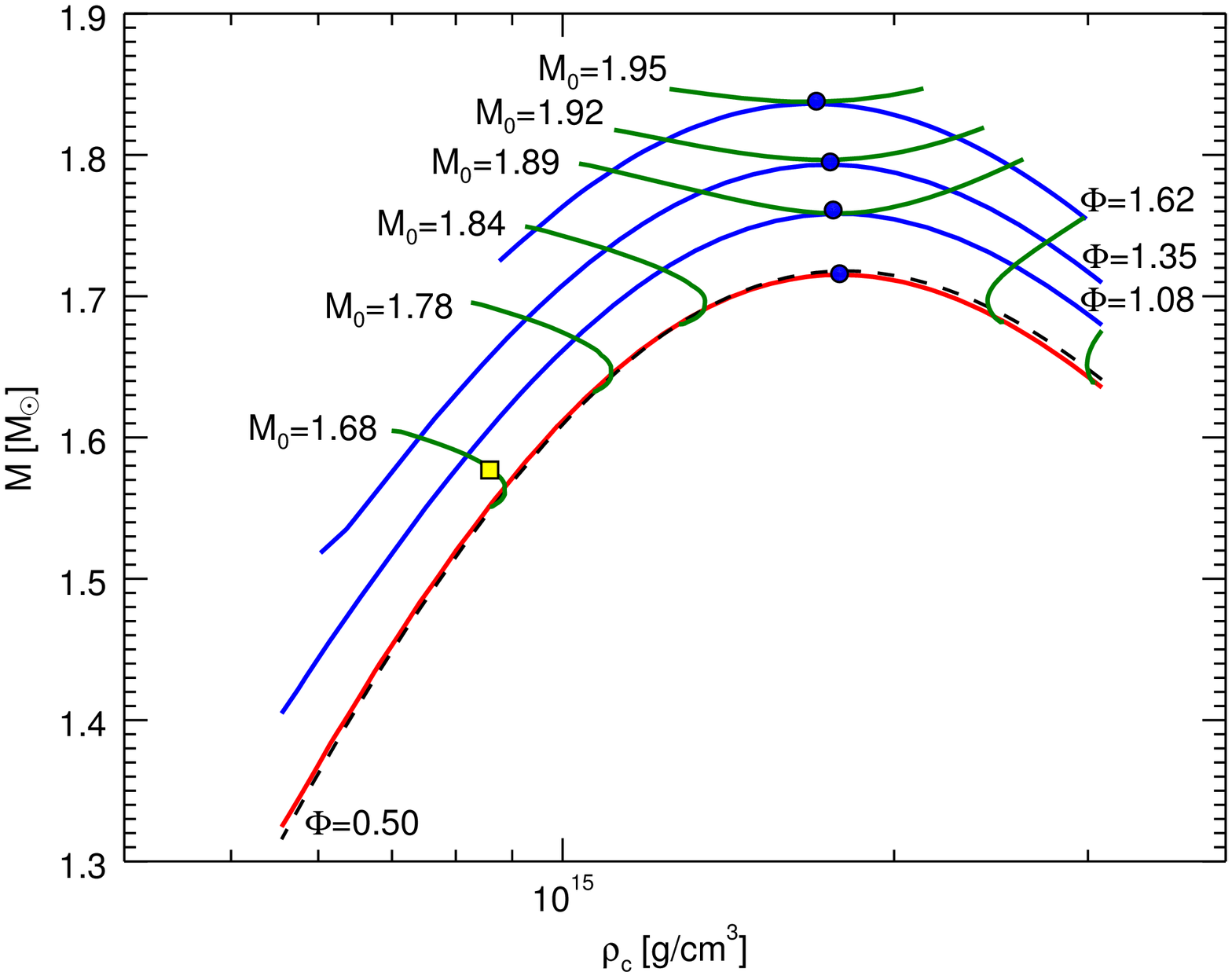}
    \includegraphics[width=.49\textwidth]{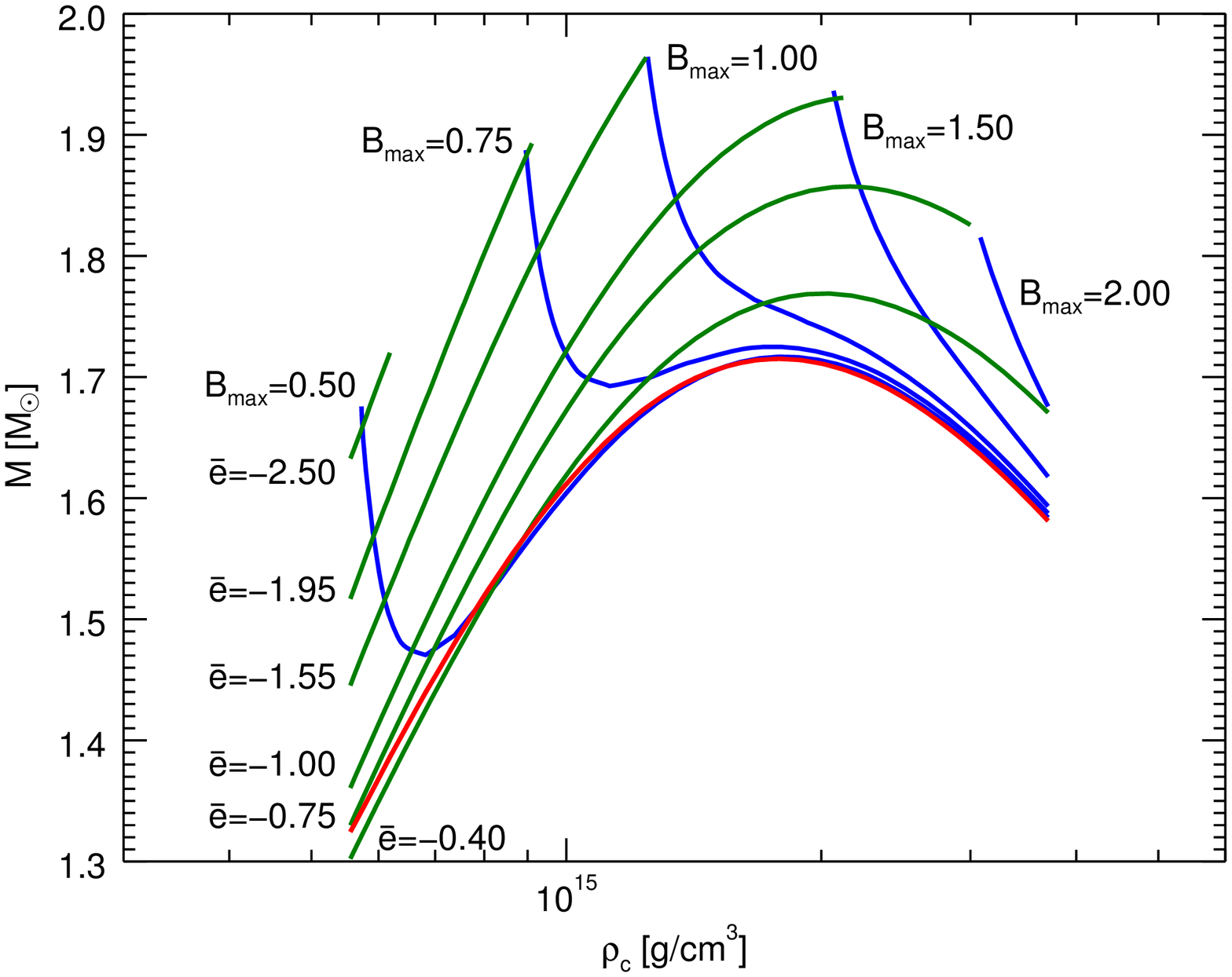}
    \includegraphics[width=.49\textwidth]{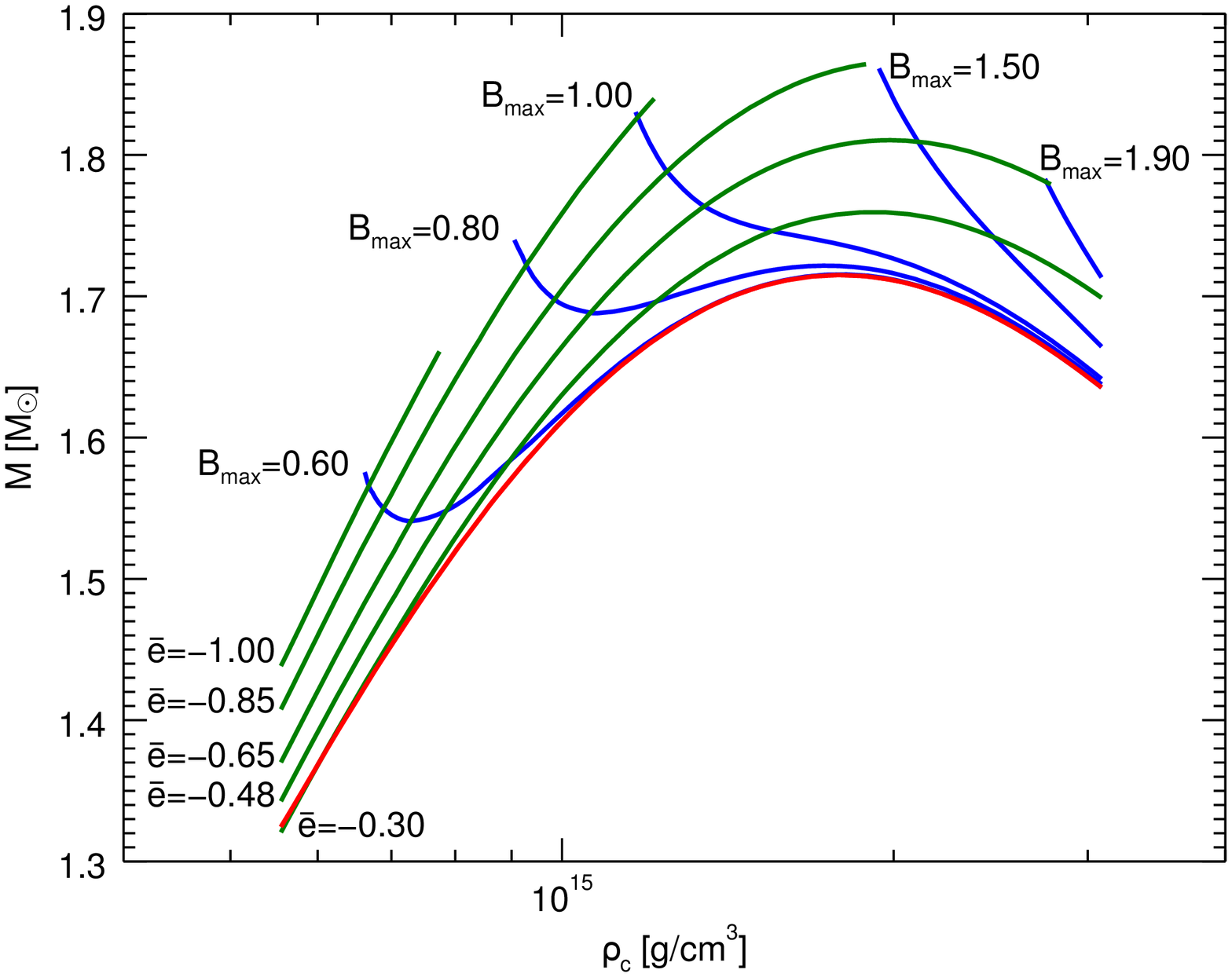}
    \caption{
   Sequences of equilibrium stellar models with purely toroidal field, for various fixed quantities. Top panel: with fixed baryonic mass $M_0$ and fixed magnetic
    flux $\Phi$. Bottom panel:  with fixed mean deformation rate $\bar e$ and fixed 
    maximum magnetic field strength $\Bmax$.
    Left panels show configurations with $m=1$ while right ones show configurations 
    with $m=2$.
    $M_0$ is expressed in unit of solar masses $M_{\sun}$, $\Phi$ in unity of 
    $10^{30}\mbox{G}\,\mbox{cm}^2$ and $\Bmax$ in unity of $10^{18} \mbox{G}$.
    The red line is the unmagnetized sequence while the black dotted lines represent
    equilibrium configurations with low magnetic flux $\Phi$.
    The filled circles locate the models with the maximum
    gravitational mass at fixed magnetic flux. Details of these models
    are listed in Table~\ref{tab:TPS}. The yellow squares represent
    the models shown in Fig.~\ref{fig:toroidal}.
    }
    \label{fig:TPS}
\end{figure*}

Following KY08 we have carried out a full sampling of the parameter
space. In Fig.~\ref{fig:TPS} we  plot the gravitational mass $M$ as a function of the 
central density $\rho_c$ both for sequences with a constant baryonic
mass $M_0$  and a constant 
magnetic flux $\Phi$. The first thing to notice is that  the maximum gravitational mass, at
fixed magnetic flux $\Phi$,
increases with  $\Phi$. Moreover for a given $\Phi$ the
model with the maximum  gravitational mass have also the maximum rest
mass. On the other hand the minimum gravitational mass, at
fixed rest mass $M_0$,
decreases with  $M_0$. Similarly, for a given $M_0$ the
model with the minimal  gravitational mass have also the minimum
magnetic flux. The filled circles locate the maximum gravitational mass models 
in the sequences of constant $\Phi$. The global quantities related to these
configurations are summarized in Table~\ref{tab:TPS}.

Interestingly, while for the vast majority of our magnetized models
the gravitational mass, for a given central density, is higher than in
the unmagnetized case,  for small values of $\Phi$ this is not true at
densities below $\sim 1.8\times 10^{15}$ g cm$^{-3}$ for $m=1$. This
is a manifestation of
the same effect discussed above in relation to the trend of the central
density in Fig.~\ref{fig:cfrKYFR}. This effect was already present to
a lesser extent in
KY08, but not discussed. 

Our set of models allows us also to construct sequences characterized
by a constant
magnetic field strength $\Bmax$ or a constant deformation rate
$\bar{e}$. It is evident that models with a higher central density,
which usually correspond to more compact stars, can harbour a higher
magnetic field with a smaller deformation. 


\begin{table*}
\caption{
Global quantities of the maximum mass models shown  in Fig.~\ref{fig:TPS}.
For the
definition of the various quantities see Appendix \ref{appendix}.}
\label{tab:TPS}
\begin{tabular}{l*{9}{c}}
\toprule
\toprule
Model &$\rho_c$& $M$ & $M_0$  & $R_{\rm circ}$& $r_p/r_e$ & $\mathscr{H}/\mathscr{W}$ & $ \bar e$ &$\Bmax$ & 
		$\Phi$\\
      & [$10^{14}\mbox{g}\,{\mbox{cm}}^{-3}$] & [$M_{\odot}$] & [$M_{\odot}$]  & [km] &  & [$10^{-1}$]  
                           & [$10^{-1}$] & [$10^{18}\,\mbox{G}$] & [$10^{30}\,\mbox{G}\,\mbox{cm}^2$] \\
\midrule
$m=0$ &  17.91  &  1.715  &  1.885  &  11.68 & 1.000  &  0.000  &  0.000  & 0.000  & 0.000 \\                        [.25cm]

 $m=1$      &  18.65  &  1.780  &  1.901  &  14.84 & 1.088  &  1.670  & -4.587  & 1.129   & 1.613 \\
      &  17.50  &  1.852  &  1.960  &  17.74 & 1.107  &  2.373  & -7.833  & 1.216   & 2.150  \\
      &  16.85  &  1.945  &  2.041  &  20.86 & 1.138  &  2.956  & -11.36  &  1.265  &2.690   \\[.25cm]
 $m=2$     &  17.69  &  1.761  &  1.890  &  13.22 & 1.067  & 1.330 & -3.041  & 1.133 & 1.080  \\
      &  17.78  &  1.795  &  1.916  &  13.98 & 1.094  & 1.747 & -4.311  & 1.262 & 1.350  \\
      &  17.00  &  1.838  &  1.950  &  15.07 & 1.115  & 2.158 & -5.944  & 1.291 & 1.620 \\
      
\bottomrule 
\end{tabular}
\end{table*}


\subsection{Purely Poloidal Field}

In this section we will discuss the properties of neutron star models with a purely
poloidal magnetic field. Models with a purely poloidal field have been presented in
the past by BB95
. However, a direct comparison can only be done
with one of their models. In fact they only present, with
full details,  two magnetized models with polytropic EoS. However one
of them has a very high magnetic field and strong deformation, and  
we could not reach those conditions in our code. The
polytropic index that they use is $\gamma_a=2$ while the polytropic constant is
$K_a=372$, different from the fiducial value we have adopted in this
study. For the model that we could reproduce, we found an agreement with the
BB95 results with deviations $\lesssim 1$\% for all quantities, except
the magnetic dipole moment, where the error is $\sim $ a few percents. We want however
to point out that our operative definition of magnetic dipole moment is
different than the one given by BB95, which is valid only in the
asymptotically flat limit, where magnetic field vanishes (see the
discussion in Appendix \ref{appendix}).
Given  that BB95 solve in the correct quasi-isotropic
metric,  the comparison is also a check on the accuracy of the CFC
approximation. It is evident that the CFC approximation gives results
that are in excellent agreement with what is found in the correct full
GR regime.

\begin{figure*}
	\centering
	\includegraphics[width=.35\textwidth,bb=8 0 415 440, clip]{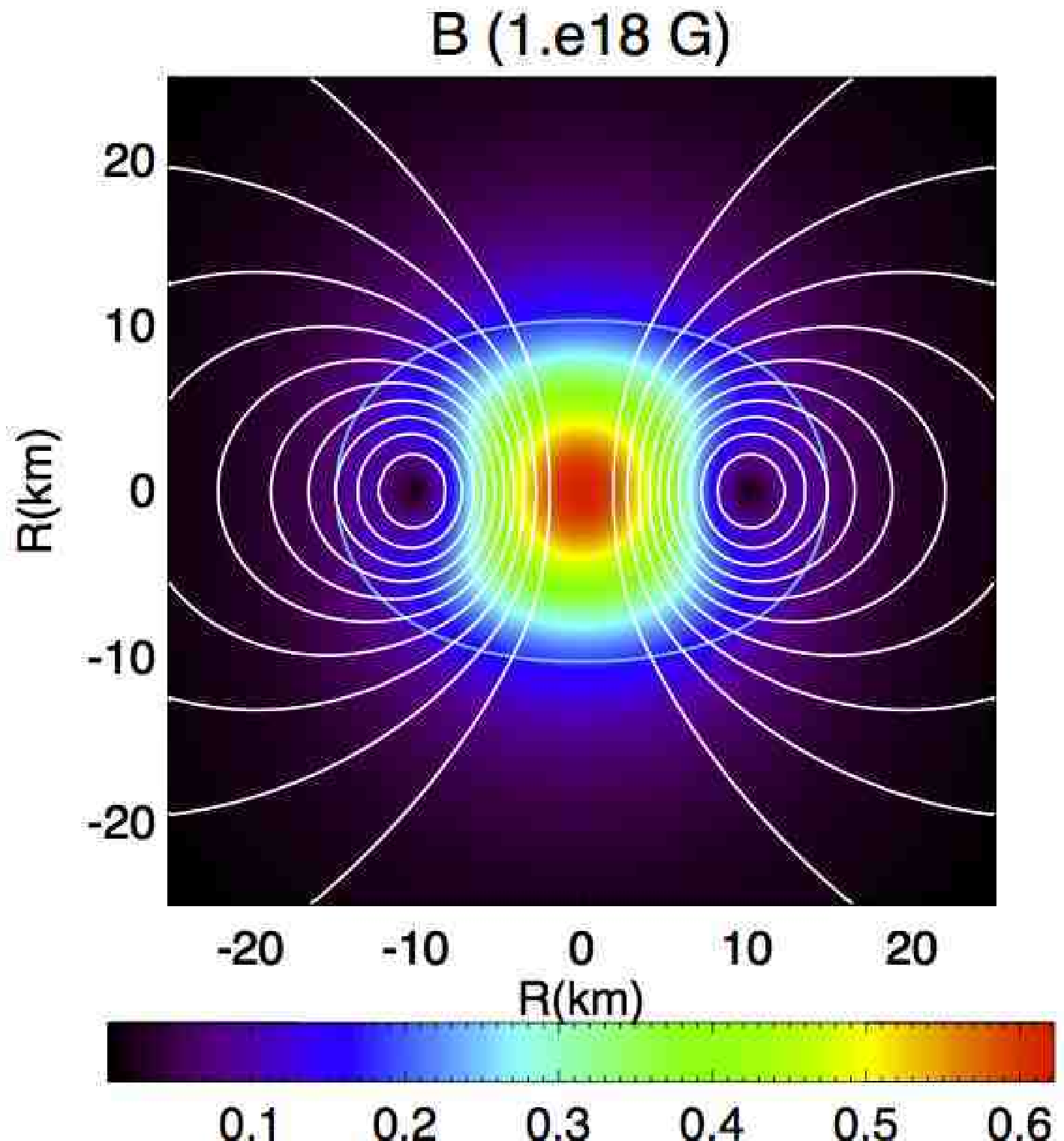} 
	\includegraphics[width=.35\textwidth,bb=8 0 415 440, clip]{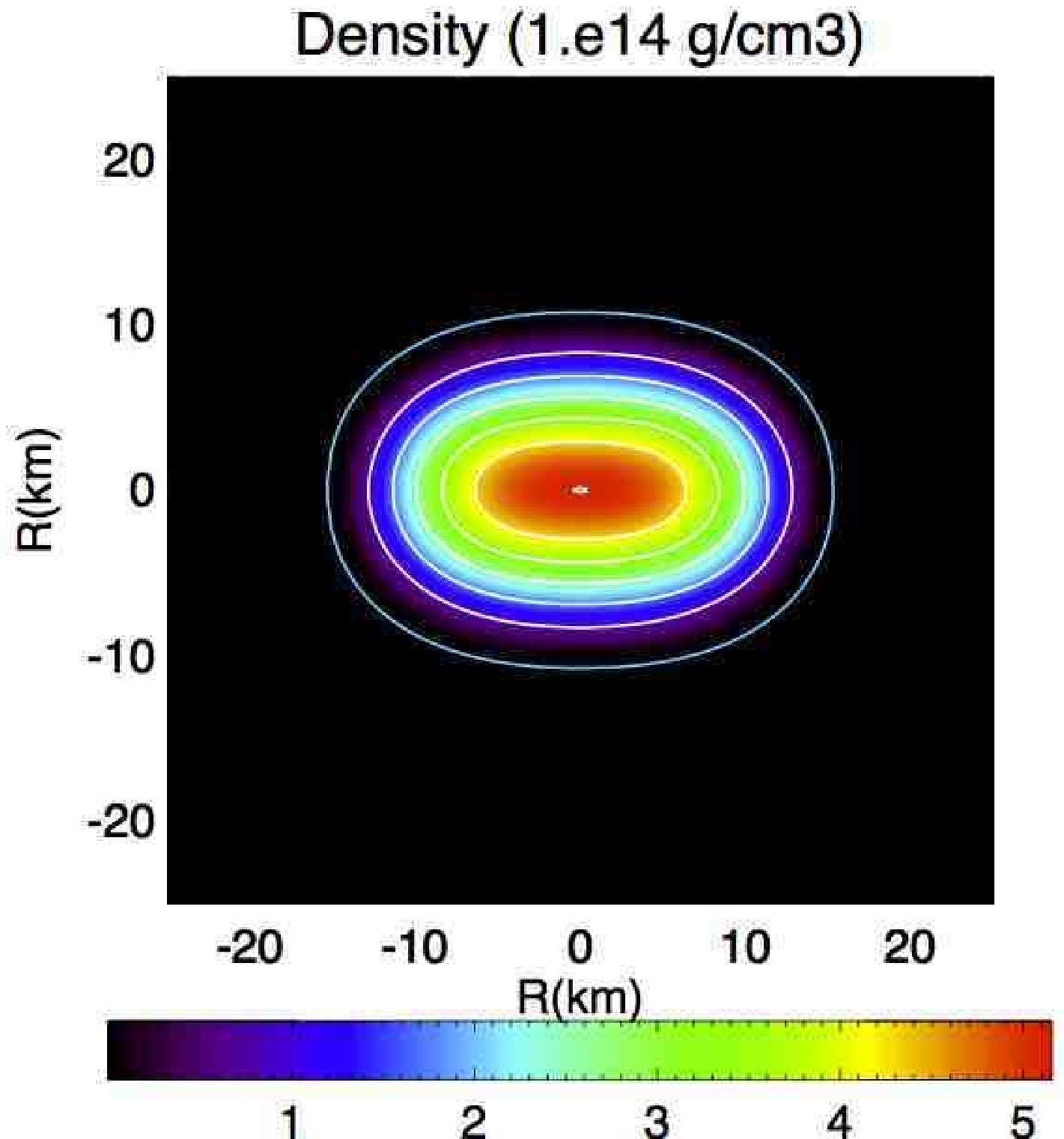}
	\caption{
	Left panel: \emph{magnetic field surfaces} (isocontours of $\tilde{A}_\phi$) and distribution of the magnetic strength 
	$B=\sqrt{B^r B_r+ B^\theta B_\theta}$. Right panel: baryonic density distribution.
	The blue curves represent the surface of the star.
	The model is characterized by $M_0=1.68 M_{\sun}$, $\Bmax=6.256 \times 10^{17} \, \mbox{G}$
	and magnetic dipole moment $\mu=2.188 \time 10^{35}\,\mbox{erg}\,\mbox{G}^{-1}$.
	}
	\label{fig:poloidal}
\end{figure*}

In Fig.~\ref{fig:poloidal} we present a model with a purely poloidal
field. The model  has been obtained in the simple case  $\xi=0$,
where only linear currents are present: $J^\phi=\rho h k_{\rm
  pol}$. The model has a rest mass  $M_0=1.680 M_{\sun}$, a maximum magnetic field $\Bmax=6.256\times
10^{17}$G, and a dipole moment $\mu=2.188\times 10^{35}$ erg G$^{-1}$.

\begin{figure}
 \centering
 \includegraphics[width=.36\textwidth]{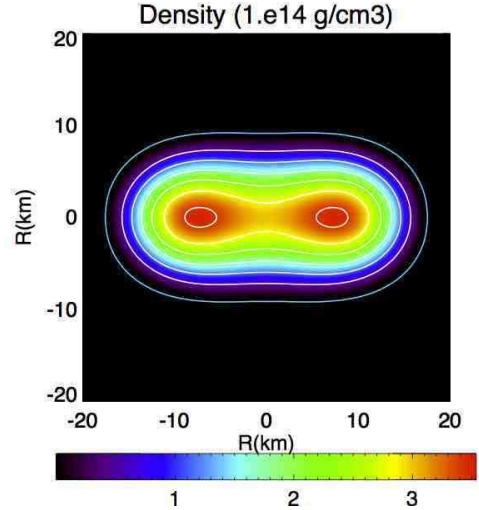}
 \caption{Baryonic density distribution for an extremely deformed configuration with a toroidal-like
  shape. This configuration is characterized by a baryonic rest mass $M_0=1.749 M_{\sun}$, a
  gravitational mass $M=1.661 M_{\sun}$, a maximum field strength
  $B_{\rm max}=5.815 \times 10^{17} \, \mbox{G}$, a magnetic dipole moment 
  $\mu=3.595 \times 10^{35}\,\mbox{erg}\,\mbox{G}^{-1}$, a circumferential radius 
  $R_{\rm circ}= 19.33 \mbox{km}$  and a mean deformation rate $\bar e = 0.386$.}
 \label{fig:toruslike}
\end{figure}

In contrast to the toroidal case, for a purely poloidal magnetic field
the NS acquires an oblate shape. The magnetic field threads the entire
star, and reaches its maximum at the very center. The pressure support
provided by the magnetic field, leads to a flattening of the density
profile in the equatorial plane. It is possible, for highly magnetized
cases, to build equilibrium models where the density has its maximum,
not at the center, but in a ring-like region in the equatorial
plane (see Fig~\ref{fig:toruslike}). Qualitatively, these effects are analogous to those produced by
rotation. Rotation leads to oblate configurations, and for a very fast
rotator, to doughnut-like density distribution. The main difference
however, is that rotation acts preferentially in the outer stellar
layers, leaving the central core unaffected in all but the most extreme
cases. A poloidal magnetic field instead acts preferentially in the core, where
it peaks. 

Another difference with respect to cases with a purely toroidal field, is the
fact that the magnetic field extends smoothly outside the NS
surface. Surface currents are needed to confine it entirely within the
star. As a consequence, from an astrophysical point of view, the
dipole moment $\mu$ is a far more important parameter than the
magnetic flux $\Phi$, because it is in principle an observable (it is
easily measured from spin-down).

\begin{figure*}
	\centering
    \includegraphics[width=.4\textwidth]{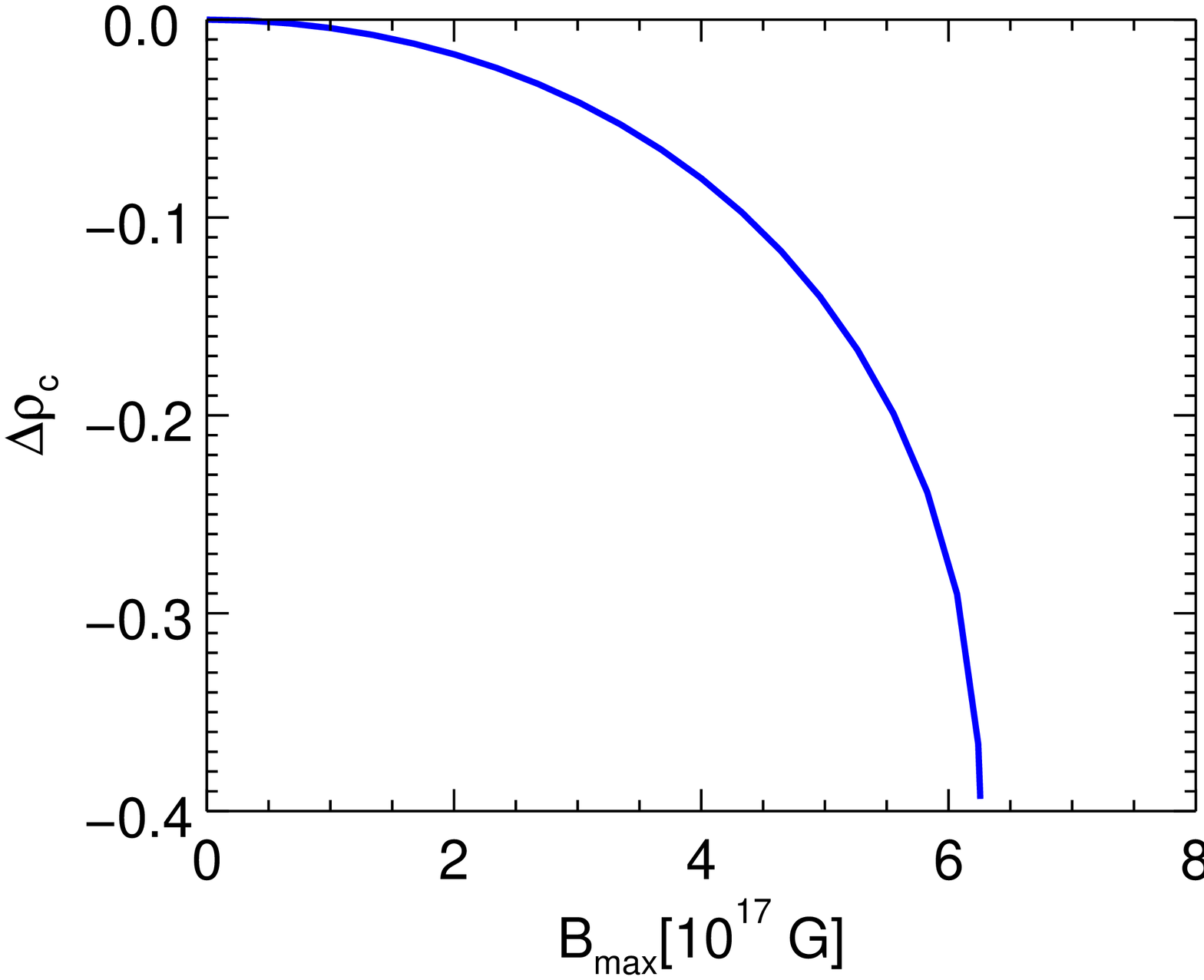}
	\includegraphics[width=.4\textwidth]{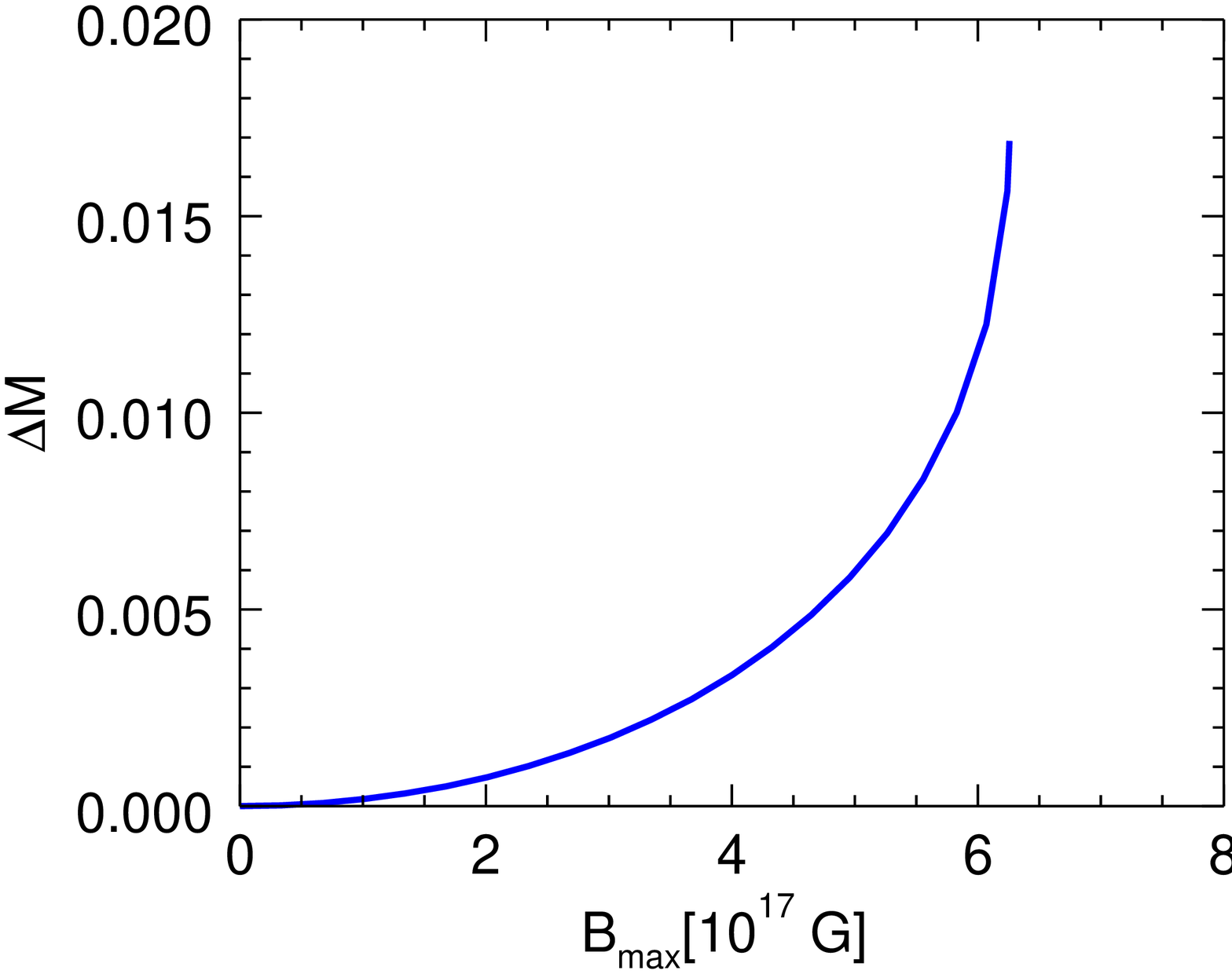} \\
	\includegraphics[width=.4\textwidth]{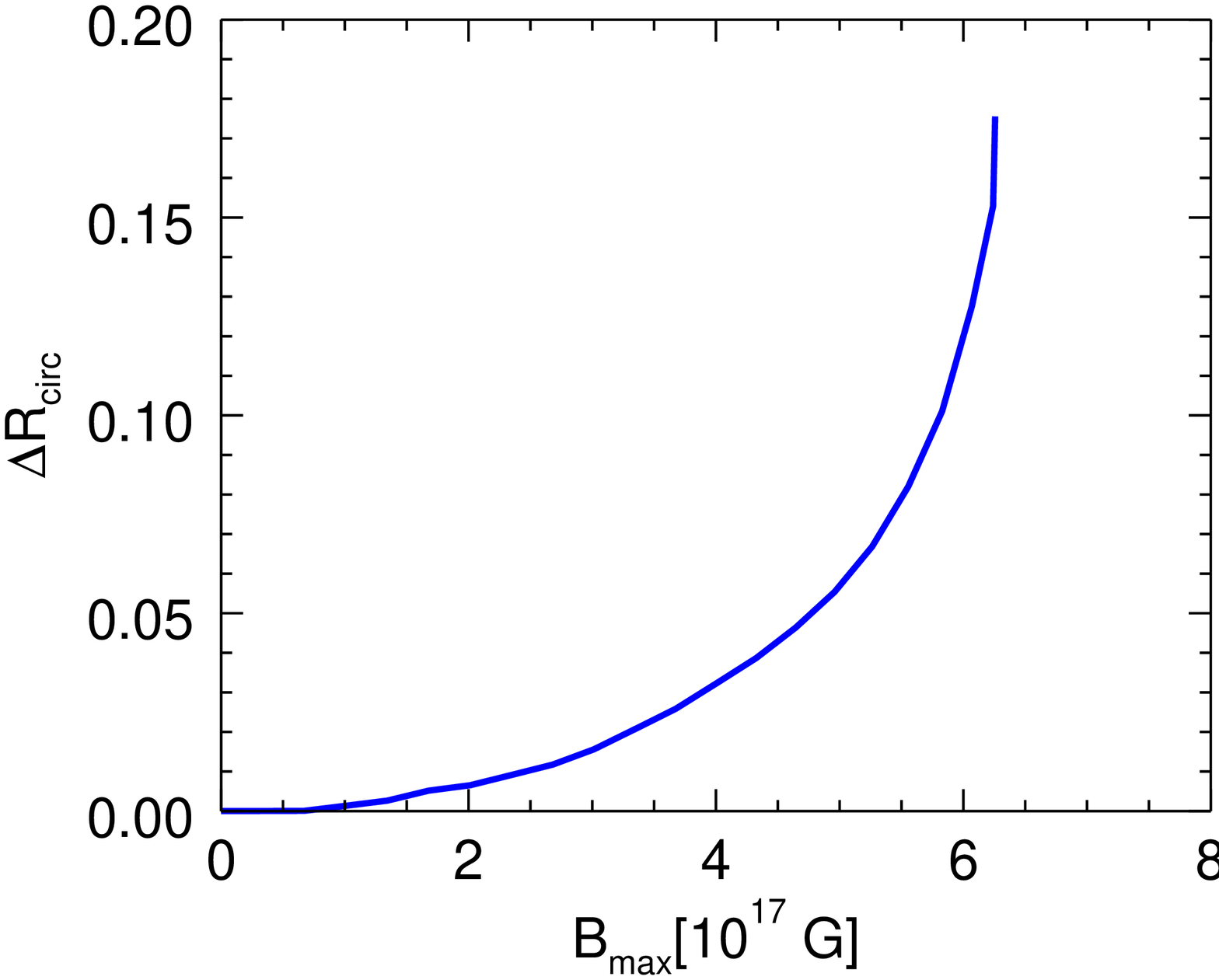}
	\includegraphics[width=.4\textwidth]{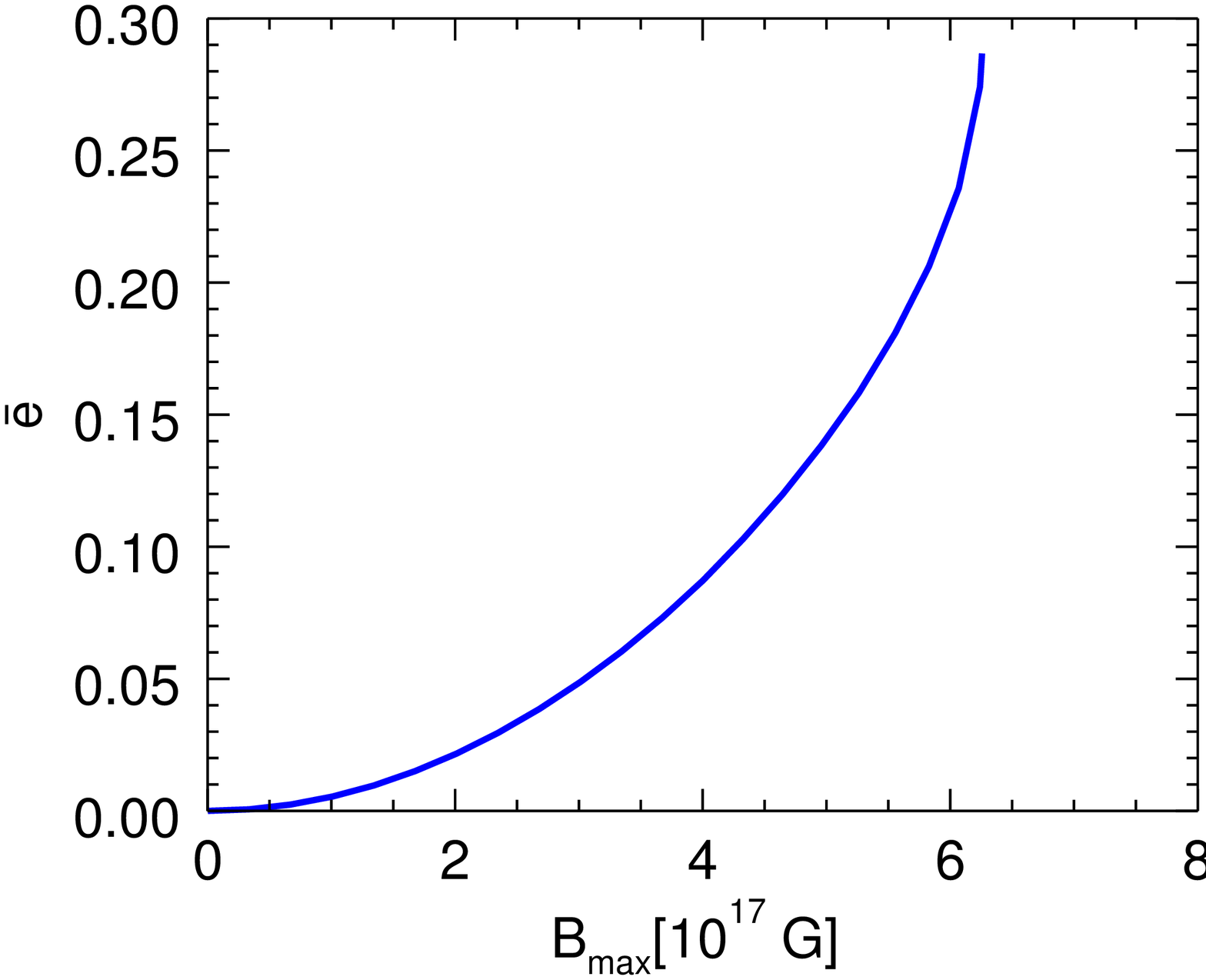}
	\caption{
	Variations of global quantities with respect to a
        non-magnetized configuration along an
	equilibrium sequence at fixed baryon mass $M_0=1.680
        M_{\sun}$, for  models with a  purely poloidal
        magnetic field.
	Notation is the same as in Fig.~\ref{fig:cfrKYFR}. }
	\label{fig:seqpol1}
    \end{figure*}

\begin{figure*}
	\centering
	\includegraphics[width=.4\textwidth]{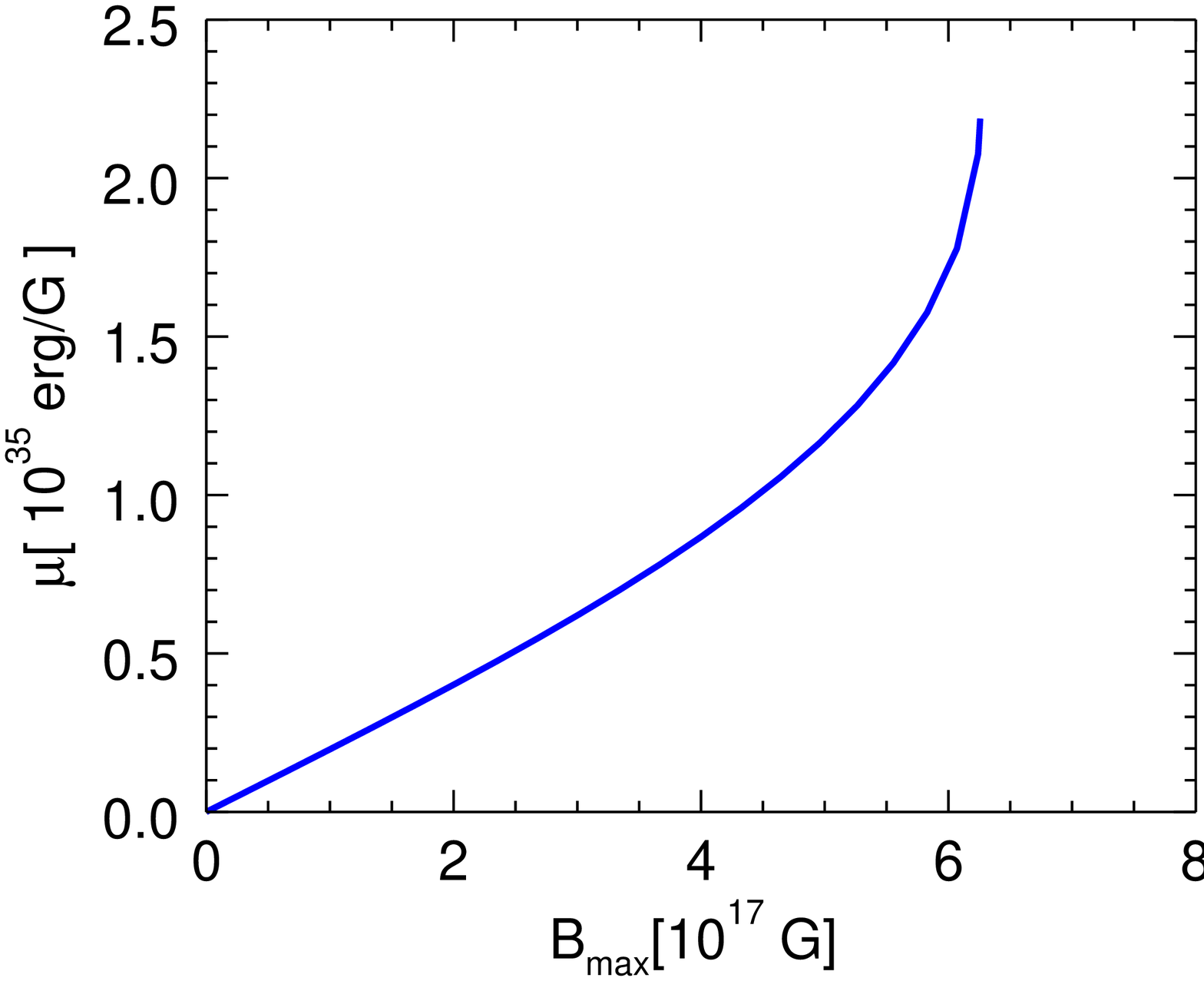}
	\caption{ Magnetic dipole moment $\mu$ as a
	 function of the maximum field strength inside the star
         $\Bmax$  for an equilibrium sequence with the
	purely poloidal magnetic  field and fixed baryon mass $M_0=1.680 M_{\sun}$.
	}
	\label{fig:mdm}
    \end{figure*}

Similarly to what was done in the case of a purely toroidal magnetic
field,  we have built an equilibrium sequence, in the simplest case $\xi=0$, at fixed baryonic mass
$M_0=1.680 M_{\sun}$ (Fig~\ref{fig:seqpol1}). Changes in the various global quantities are shown as a
function of the maximum magnetic field inside the  star $B_{\rm max}$.
The results in Fig~\ref{fig:seqpol1} show that the central density $\rho_c$ 
decreases with $B_{\rm max}$ while the gravitational mass $M$, the circumferential
radius $R_{\rm circ}$ and the mean deformation rate $\bar e$, which is
now 
positive (oblateness), grow. As in the toroidal case, for this
sequence, there appears to
be a maximum value of magnetic field $B_{\rm max} \approx  6.25\times
10^{17}$ G. However, we have not been able to
build models with higher magnetization, and so we cannot say if such value is
reached asymptotically, or, as in the toroidal case, increasing
further the magnetization, leads to a reduction of the maximum
field strength. The other main qualitative difference with respect to the
toroidal case is the trend of the central density, which is now a
monotonic function of the maximum magnetic field. From a quantitative
point of view
we notice that the central density is more affected by the magnetic
field. In
Fig.~\ref{fig:mdm} we also  display the variation of the magnetic dipole moment $\mu$ 
along the same sequence as a function of the maximum field strength
$B_{\rm max}$. The trend is linear for weak magnetic fields, and then
seems to increase rapidly once the field approaches its maximum.

Our choice for the magnetic function $\cal{M}$, allows us to
investigate the effects of non-linear currents terms  $J^\phi=\rho h
k_{\rm pol} \xi A_\phi$. Unfortunately we cannot treat configurations
with just non-linear currents,  because in this situation the
Grad-Shafanov equation has always a trivial solution $A_\phi=0$, and
our numerical algorithm always converges to it. It is not
clear if non-trivial solutions of the Grad-Shafranov equation exist in any case, and it is just the
numerical algorithm that fails to find them, or if they only exist for
specific values of the background quantities  $(\rho,\phi,\alpha)$,
and in this case it well could be that no self-consistent model can be
build. So to model cases with $\xi\neq 0$, is it necessary to add a
stabilizing linear current. This can be done either by adding a
distributed current term 
$J^\phi=\rho h k_{\rm pol} $, or by introducing singular currents, for
example surface currents. We will not consider here this latter
possibility and we will investigate configurations with
distributed currents alone. As anticipated, the non-linear current terms
can in principle produce multipolar magnetic configurations. However,
the symmetry of the magnetic field geometry is dictated by the
stabilizing linear currents. Given that a current $J^\phi=\rho h
k_{\rm pol}$, always gives dipolar dominated magnetic fields, this
geometry will be preserved also by including non-linear terms. To
obtain prevalent
quadrupolar magnetic fields, one needs, for example, to
introduce singular 
currents that are antisymmetric with respect to the equator.
Depending on the sign of $\xi$ the non-linear current terms can be
either additive or subtractive.

\begin{figure*}
	\centering
	\includegraphics[width=.35\textwidth]{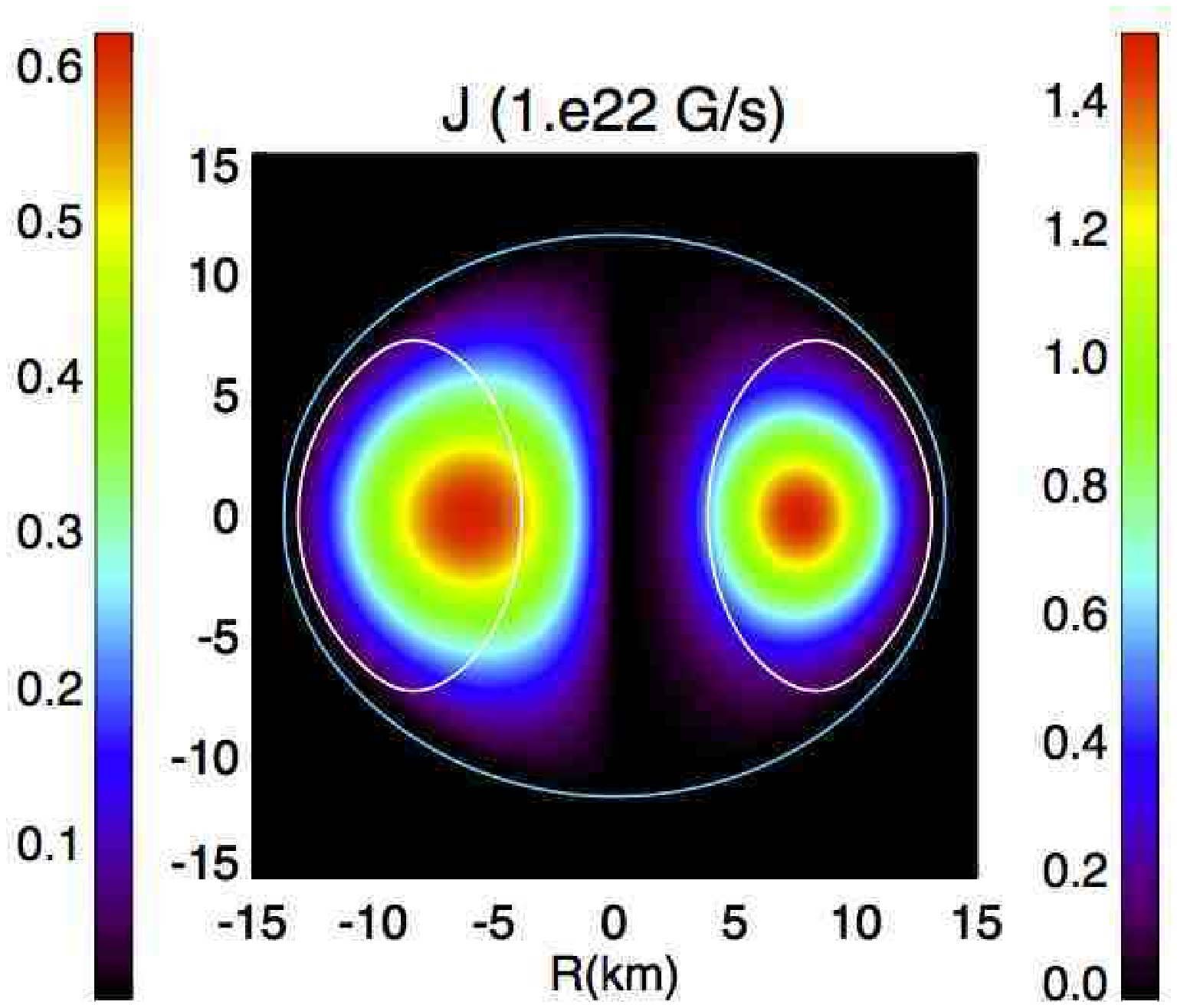}
	\includegraphics[width=.35\textwidth]{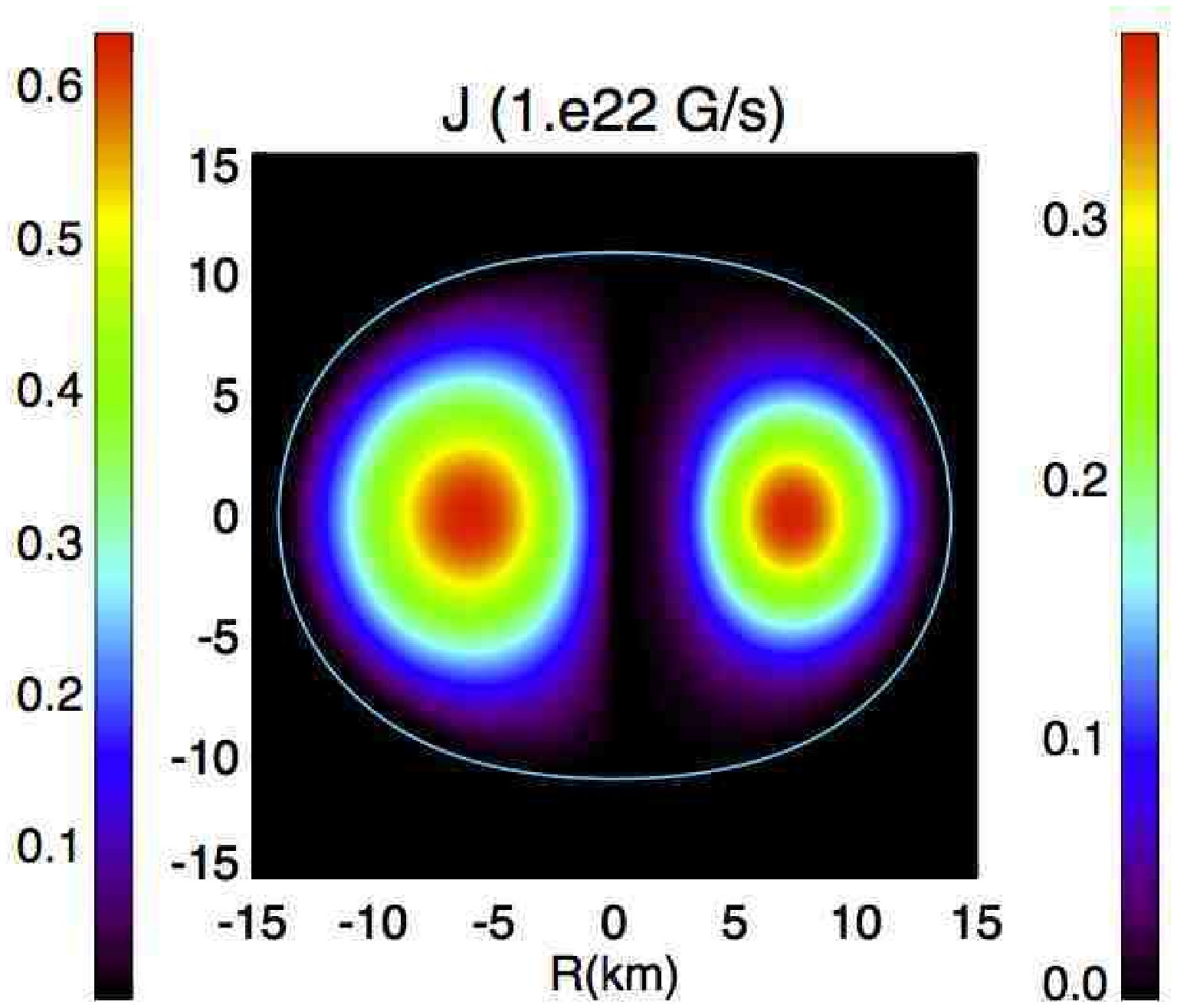}
	\caption{
	Comparison among models with different current distributions.
	Left panel:  modulus of the zero-current term $J^\phi_0=\rho h
        k_{\rm pol}$ (left-half) and first-order 
        one $J^\phi_1=\rho h k_{\rm pol}\xi A_{\phi}$ (right-half) for an equilibrium configuration with $\xi=20$,
	$M=1.551 M_{\sun}$ and $\mu=1.477\times 10^{35}\, \mbox{erg}\,\mbox{G}^{-1}$.
	Right panel: same as the left panel but for a model with
        $\xi=-5$, same mass $M=1.550 M_{\sun}$  and magnetic dipole moment
	$\mu=1.510\times 10^{35}\,\mbox{erg}\,\mbox{G}^{-1}$ .
	The white line locates the points where $ |J^\phi_1| / |J^\phi_0|\sim 1$.
	The blue line represents the stellar surface.}
	\label{fig:current}
\end{figure*}

In Fig.~\ref{fig:current} we show the distribution of the linear and
non-linear currents inside the star, both in the additive and
subtractive cases. Non-linear currents are more concentrated and they
peak at larger radii. In the additive case, we succeeded in building
model where non-linear currents are dominant in the outer stellar
layers. On the contrary, for subtractive currents,  we could not reach
configurations with current inversions, and the level of the non-linear currents are at most half of the linear term.

\begin{figure*}
	\centering
	\includegraphics[width=.4\textwidth]{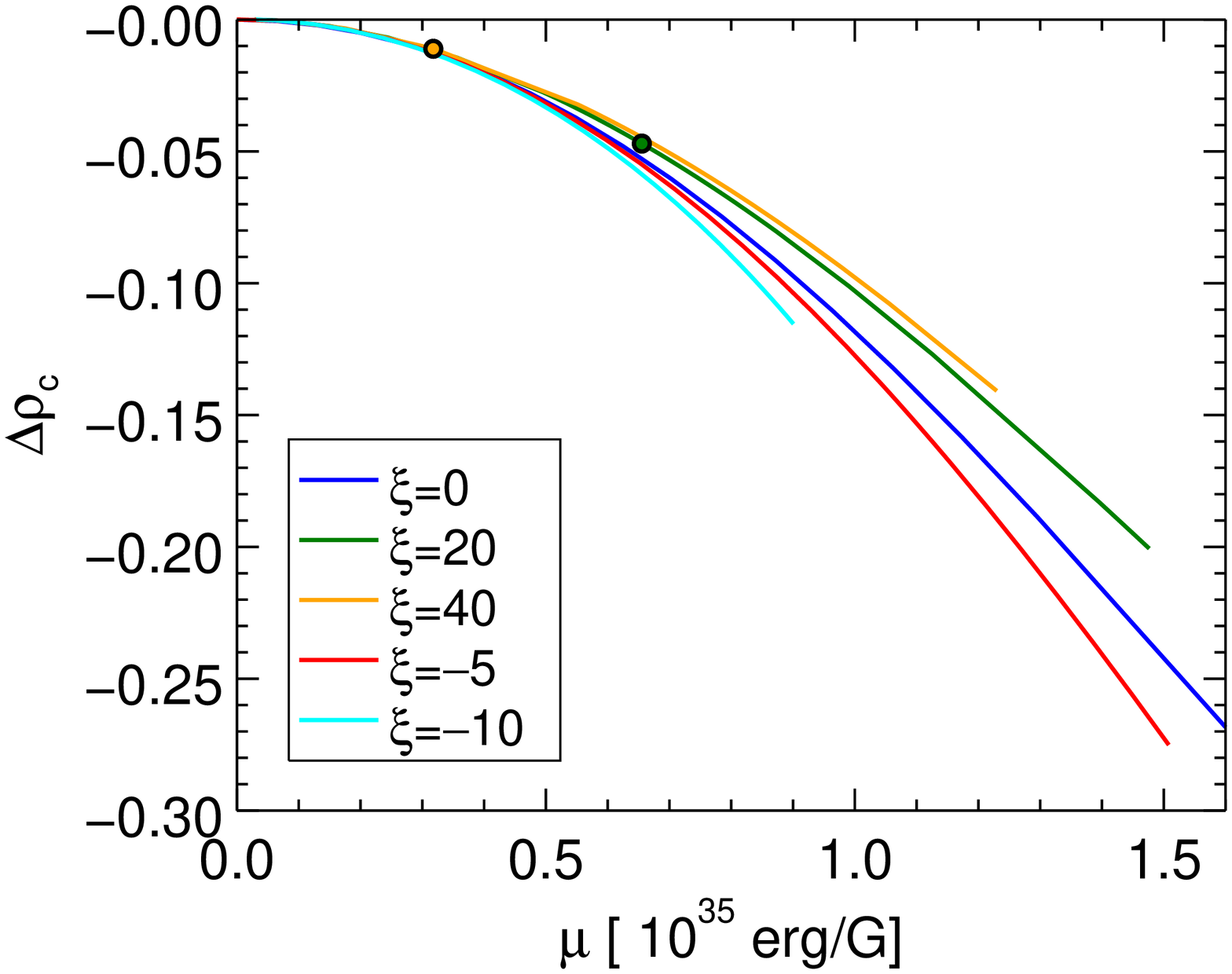} 
	\includegraphics[width=.4\textwidth]{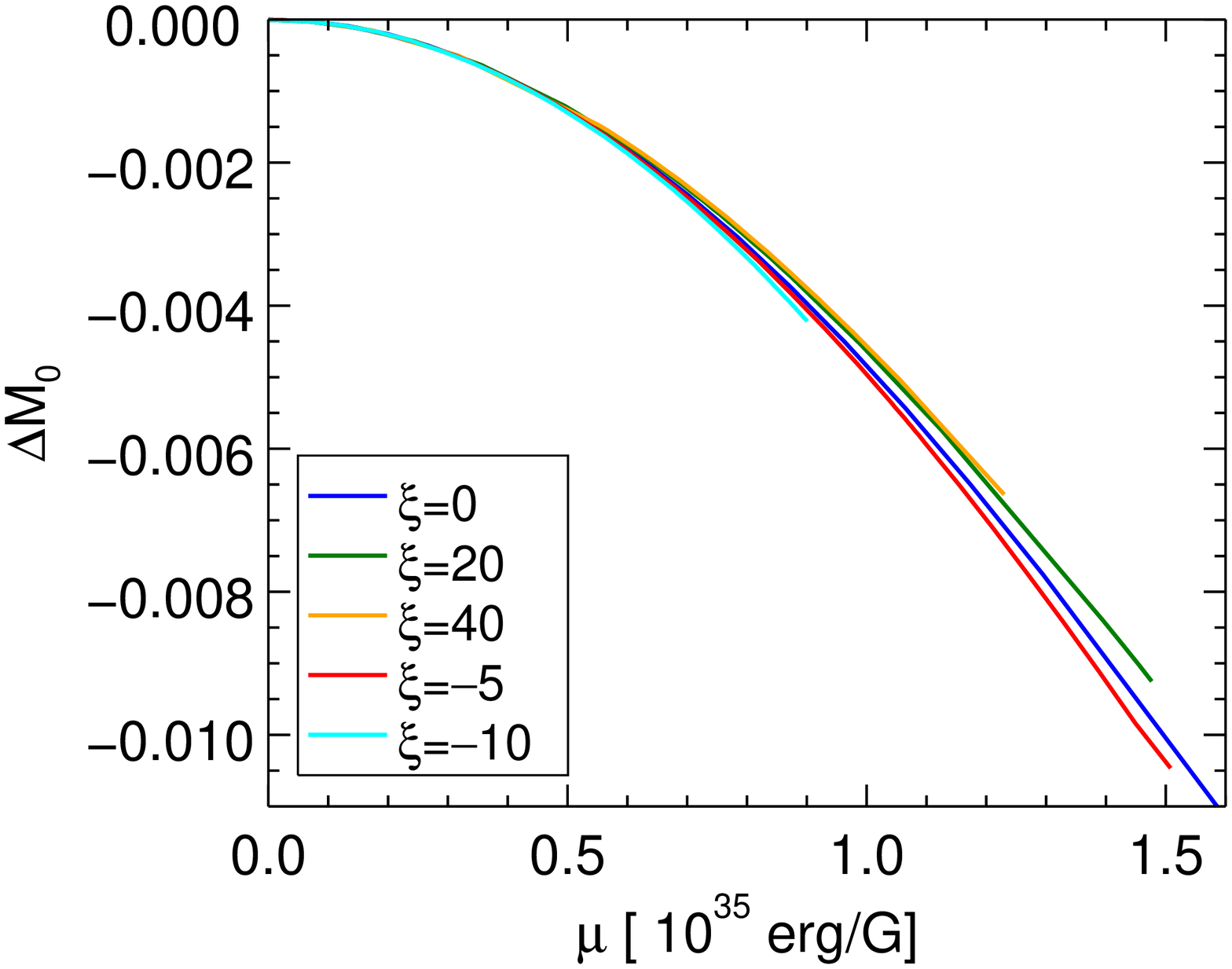} \\
    \includegraphics[width=.4\textwidth]{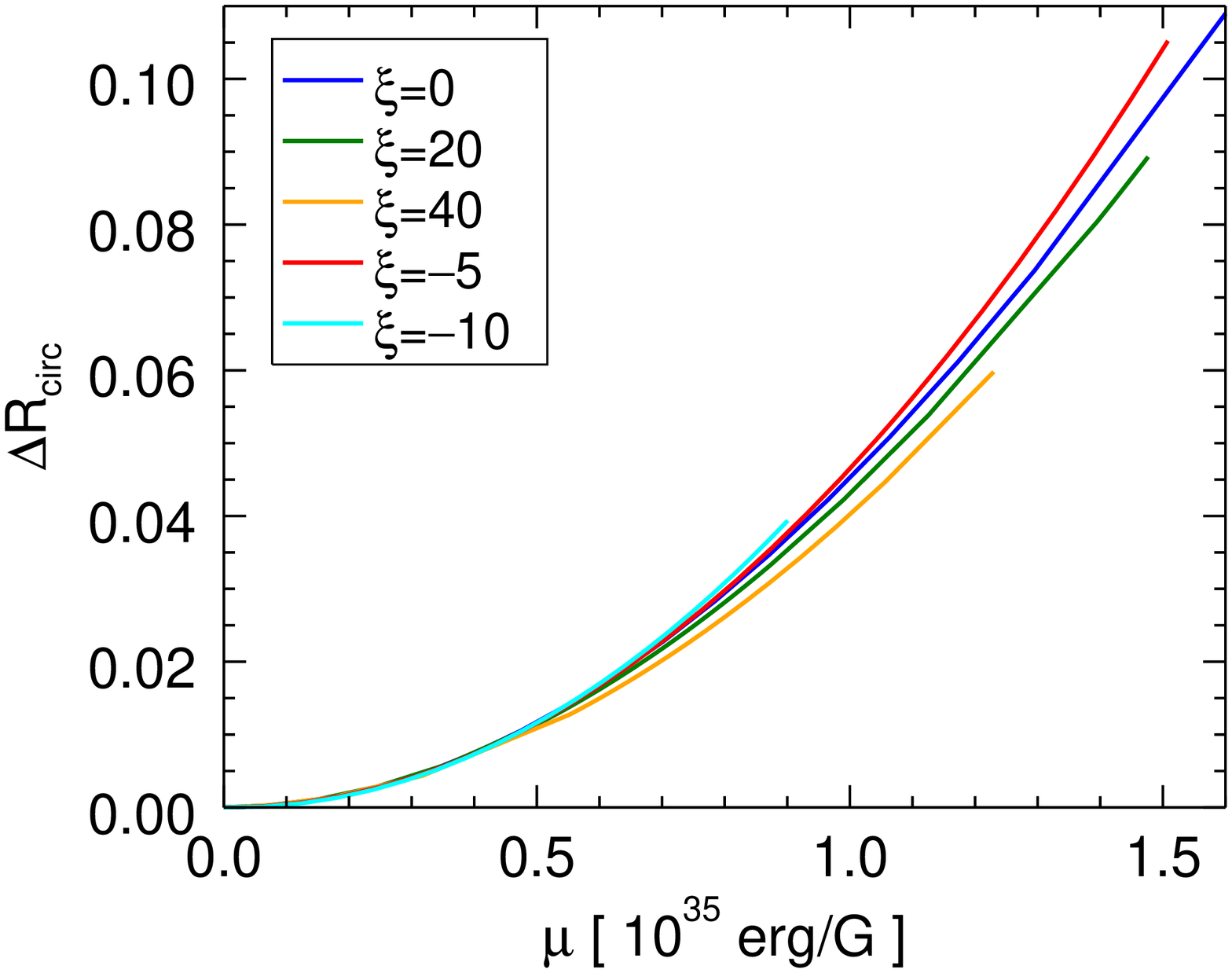}
	\includegraphics[width=.4\textwidth]{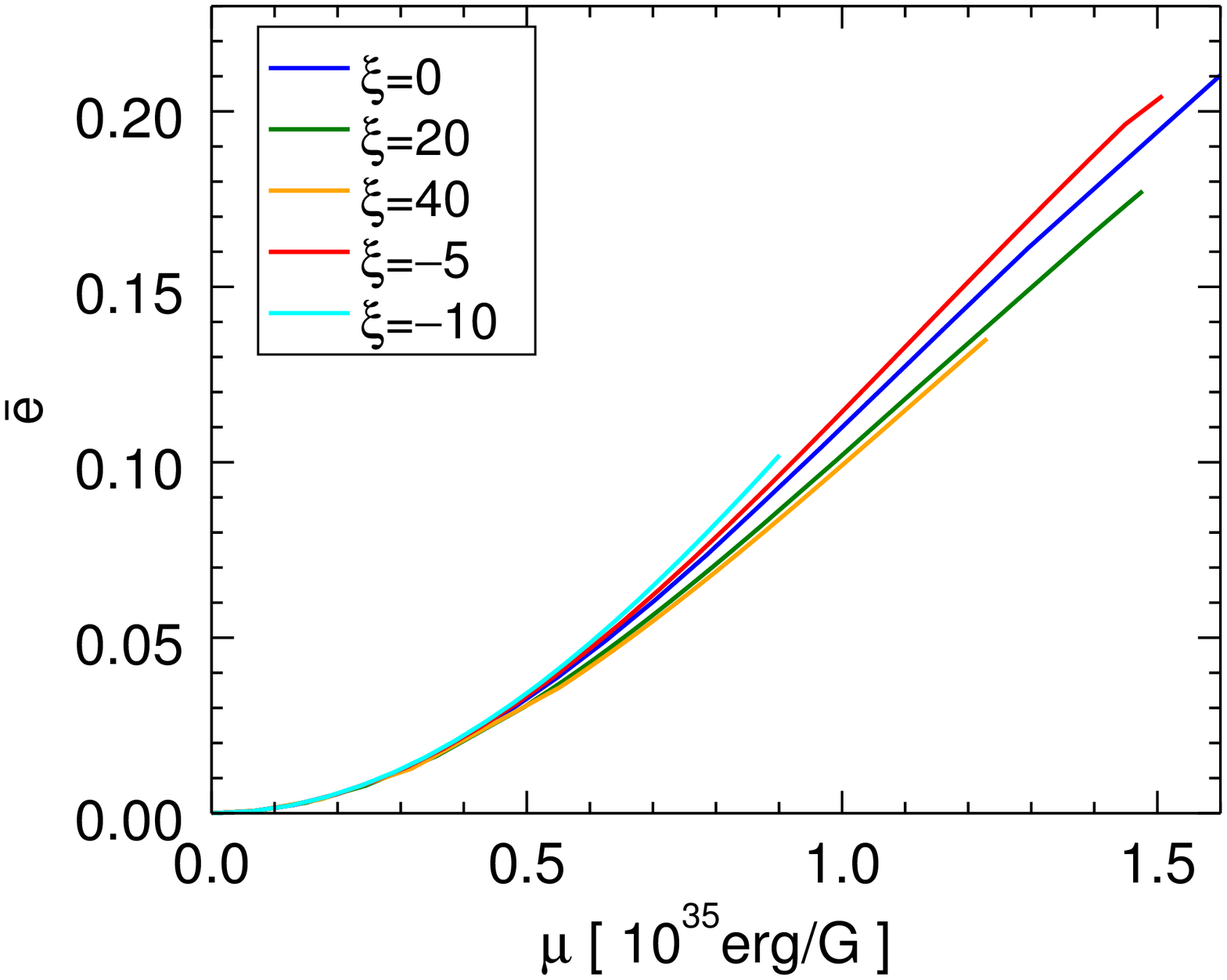} 
	\caption{
     Variations of global quantities with respect to the
     non-magnetized configuration, as a function
     of the magnetic dipole moment, along an
	 equilibrium sequence with fixed
	 gravitational mass $M=1.551 M_{\sun}$, and purely poloidal field.
	 Notation is the same as in Fig.~\ref{fig:cfrKYFR}.	
	 Filled dots locate the points where the maximum strength of
         zeroth-order term $J^\phi_0=\rho h k_{\rm pol}$
	 is equal to the maximum strength of first-order term $J^\phi_1=\rho h k_{\rm pol}\xi A^{\phi}$.
	 Details concerning these configurations and those which show the higher value of $\mu$
	 for each sequence are listed in Table~\ref{tab:xi}.
	}
	\label{fig:seqpol2}
    \end{figure*}

In  Fig.~\ref{fig:seqpol2} we compare how various global quantities
change, as a function of the magnetic dipole moment $\mu$ for  NSs
with fixed gravitational mass $M=1.551 M_\odot$, and for various values
of the parameter $\xi \in
\lbrace-10,-5,0,20,40\rbrace$. We opted for a parametrization in terms of
$\mu$ and $M$ instead of $\Bmax$ and $M_0$, because the former are in
principle observable quantities, and as such of greater astrophysical relevance, while the latter are not.

Here we can note that, for a fixed dipole moment $\mu$, the addition of negative current terms ($\xi<0$) leads to less compact and
more deformed configurations,	conversely the presence of a positive current term ($\xi>0$) 
makes the equilibrium configurations more compact and less oblate. This
might appear as contradictory: increasing currents should make
deformation more pronounced. However this comparison is carried out
at fixed dipole
moment $\mu$. This means than any current added to
the outer layers, must be compensated by a reduction of the current in
the deeper ones (to keep $\mu$ constant). Giving that deformations are
dominated by the core region, this explains why the star is less
oblate. The opposite argument applies for subtractive currents.

\begin{figure*}
	\centering
   \includegraphics[width=.49\textwidth]{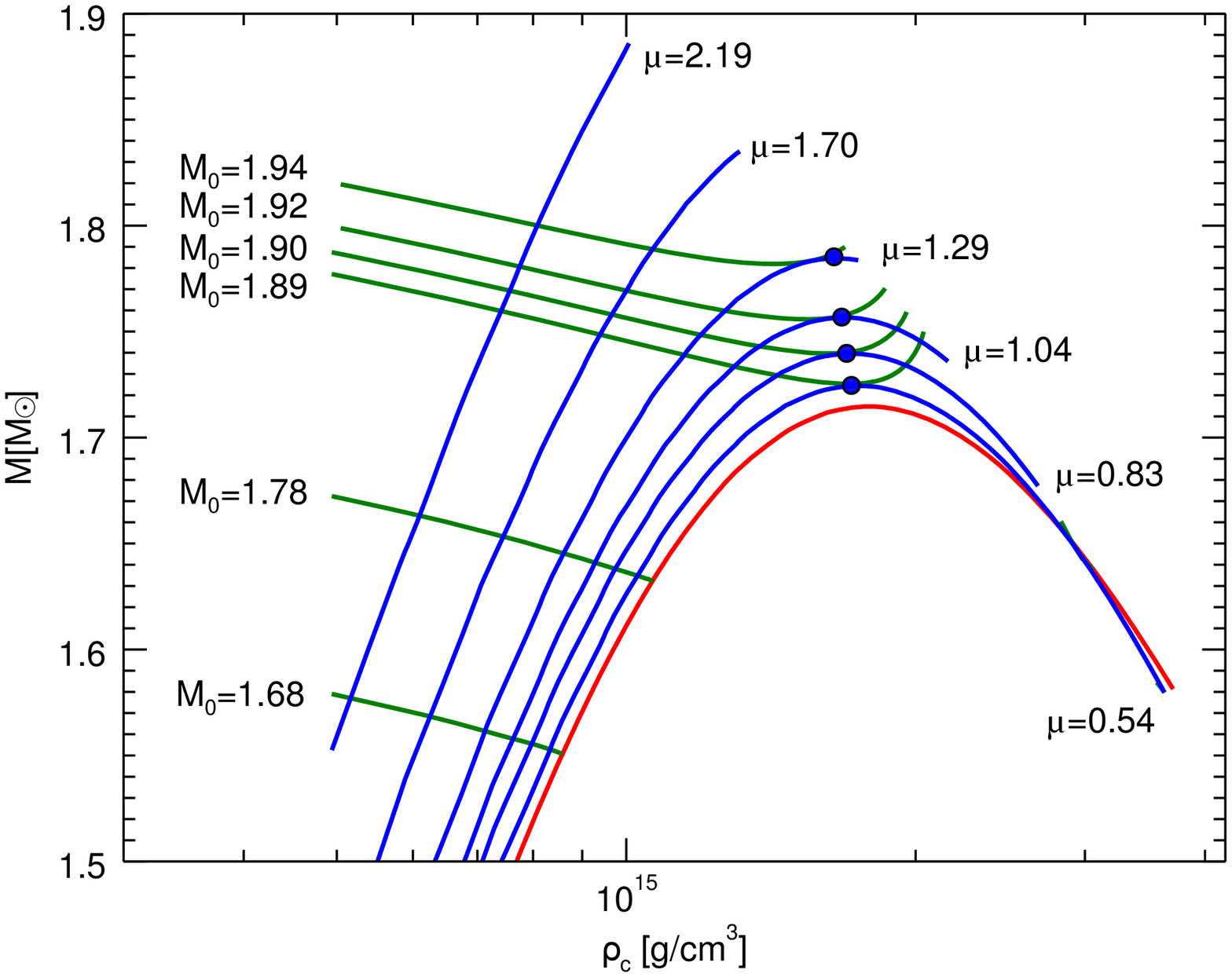}
   \includegraphics[width=.49\textwidth]{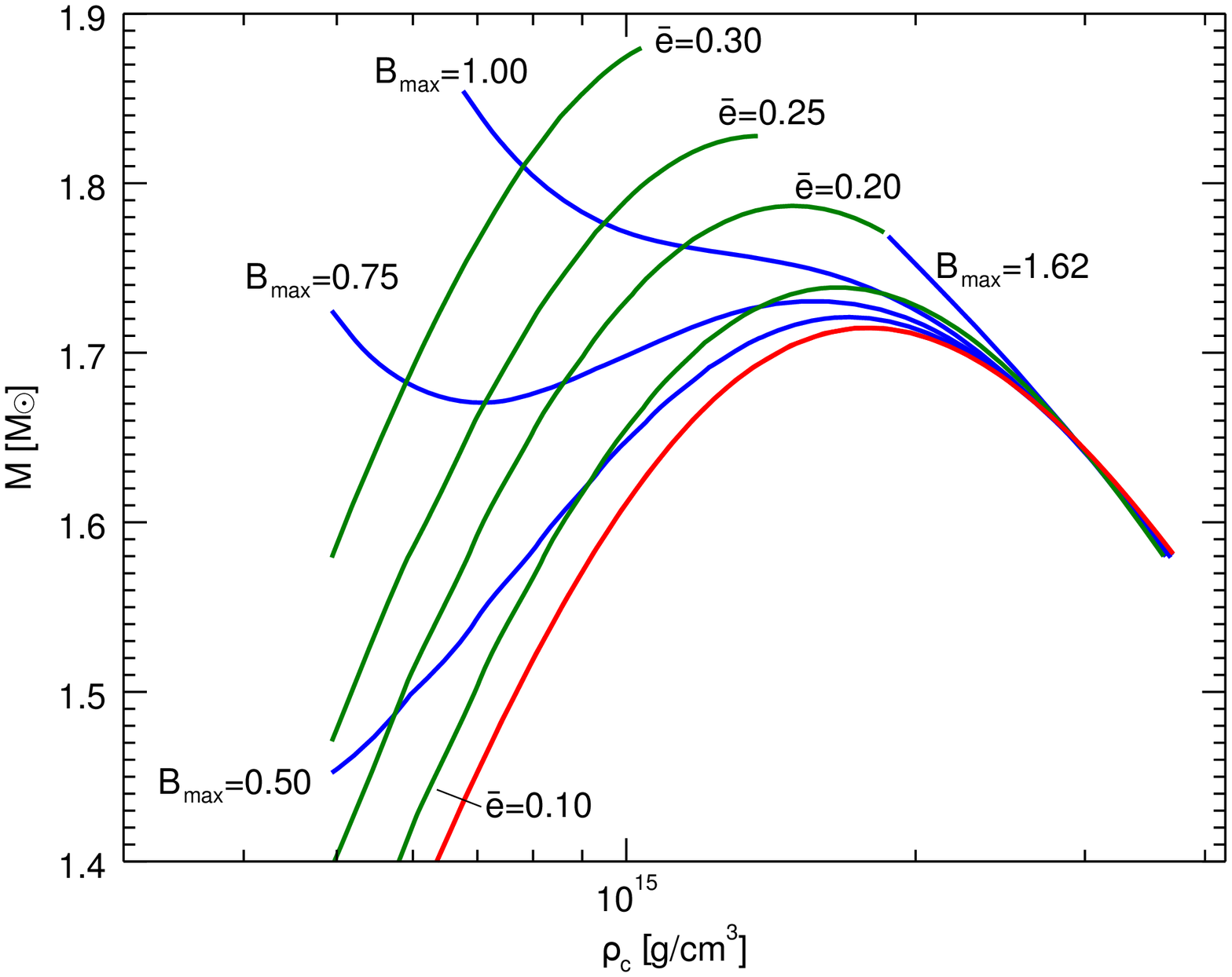}
    \caption{  \label{seqPOL}
    Left panel: equilibrium sequences with fixed magnetic field moment $\mu$ and fixed baryonic mass $M_0$. 
    Right panel: equilibrium sequences with fixed deformation rate $\bar e$ and maximum field strength $\Bmax$.
    The baryonic mass is expressed in units of $M_{\sun}$, the magnetic dipole moment in units of
    $10^{35}\,\mbox{erg}\,\mbox{G}^{-1}$ and the maximum field strength in units of $10^{18} \mbox{G}$.
   The red line shows the unmagnetized sequence
    while the filled dots locate the configurations with maximum mass
    for a given dipole moment $\mu$. Parameters for these
    configurations  are listed in Table~(\ref{tab:pol2}).}
    \label{fig:pspol}
\end{figure*}

Finally we have repeated a detailed parameter study, in analogy to
what has been presented in the previous section, to explore the space
$(\rho_c,k_{\rm pol})$.  In Fig.~\ref{fig:pspol}  we show various
sequences characterized by either a  constant baryonic mass $M_0$, or a
constant magnetic dipole moment $\mu$, or a constant
maximum field strength $\Bmax$, or a constant deformation rate $\bar
e$. We have limited our study to models with $\xi=0$, because the
addition of other currents leads in general to minor effects.
Again, it is found that systems with lower central densities are in general
 characterized by larger deformation, for a given magnetic field
and/or magnetic moment. There is, however, no inversion trend analogous to the
one found for purely toroidal configurations.

\begin{table*}
\caption{
Global quantities from selected configurations belonging to the
equilibrium sequences shown in Fig~\ref{fig:seqpol2}, at $M=1.551 M_\odot$. For each value
of $\xi$ we show the details for the configuration with the maximal
magnetic dipole moment. For cases with $\xi=20,40$ we also present
those configurations where ratio $ |J_1| / |J_0| \simeq 1$. For the
definition of the various quantities see Appendix~\ref{appendix}.
}
\begin{tabular}{l*{9}{c}}
\toprule
\toprule
Model &$\rho_c$& $M_0$ & $R_{\rm circ}$ &
 $r_p/r_e$ & $\bar e$ &  $\mathscr{H}/\mathscr{W}$ &
$\Bmax$ & $\mu$ & $|J_1|/|J_0|$  \\

&[$10^{14}\mbox{g}\,{\mbox{cm}}^{-3}$] & [$M_{\sun}$] & [km] &
 [$10^{-1}$] & [$10^{-1}$] & [$10^{-2}$]&
[$10^{17}$G] & [$10^{35}$erg G$^{-1}$] & \\

\midrule

$\xi=20$ &8.149 & 1.678 & 14.48 & 9.656 & 0.468 & 1.443 & 2.692 & 0.629 & 0.989\\ 
         &6.810 & 1.665 & 15.54 & 8.420 & 1.773 & 6.656 & 4.595 & 1.477 & 2.421\\[.25cm]
$\xi=40$ &8.426 & 1.680 & 14.35 & 9.827 & 0.127 & 0.979 & 1.417 & 0.118 & 0.990\\
		 &7.320 & 1.670 & 15.11 & 8.857 & 1.352 & 4.837 & 3.964 & 1.230 & 4.023\\[.25cm]
$\xi=-5$ &6.176 & 1.663 & 15.75 & 7.774 & 2.067 & 7.482 & 6.243 & 1.510 & 0.585\\[.25cm]
$\xi=-10$&7.543 & 1.674 & 14.79 & 8.996 & 1.014 & 3.099 & 4.782 & 0.911 & 0.691 \\
\bottomrule 
\end{tabular}
\label{tab:xi}
\end{table*}  
    
\begin{table*}
\caption{
Global quantities from the poloidal models with maximum gravitational mass in sequences with
fixed magnetic dipole moment $\mu$, shown in Fig.~\ref{fig:pspol}.  For the
definition of the various quantities see Appendix~\ref{appendix}.}
\begin{tabular}{c*{9}{c}}
\toprule
\toprule
$\rho_c$ & $M$ & $M_0$ & $R_{\rm circ}$ & $\mathscr{H}/\mathscr{W}$  &$B_{\mbox{\tiny{max}}}$&
$\bar e$ & $r_p/r_e$ & $\mu$\\

[$10^{14}\mbox{g}\,{\mbox{cm}}^{-3}$] & [$M_{\odot}$] & [$M_{\odot}$]  & [km] &   [$10^{-2}$]  & 
[$10^{17}\,\mbox{G}$]& [$10^{-1}$]  & [$10^{-1}$] & [$10^{35}\,\mbox{erg}\,\mbox{G}^{-1}$]\\

\midrule
                                                                                          
17.29 &  1.725  &  1.892  &  11.96  &  1.821  & 6.162  & 0.481 & 9.551 & 0.543 \\
17.19 &  1.740  &  1.903  &  11.89  &  4.275  & 9.406  & 1.036 & 8.961 & 0.833 \\
16.76 &  1.757  &  1.916  &  11.93  &  6.647  & 11.70  & 1.481 & 8.442 & 1.041 \\
16.45 &  1.785  &  1.938  &  12.00  &  10.17  & 14.45  & 2.012 & 7.922
& 1.290 \\

\bottomrule 
\end{tabular}
\label{tab:pol2}
\end{table*}

\subsection{Mixed  Field}

Finally, in this subsection we will illustrate in detail the properties of TT configurations. For all the
cases we present, we have adopted a functional form for
$\mathcal{M}$ identical to the one used in the  purely poloidal
case [see Eq.~(\ref{eq:mbern})] but only assuming linear terms for the
toroidal currents, $\xi=0$. Note, however, that the presence of a
toroidal field is equivalent to the existence of an effective non-linear current term. The toroidal magnetic field is instead
generated by a current term $\mathcal{I}$, given by Eq.~(\ref{eq:fbern}).
Again we have selected the simplest case: $a\neq 0$ and 
$\zeta=0$. We focus here on fully non-linear solutions in the strong
magnetic field limit. A study of the low magnetic field limit is presented in
Appendix~\ref{appendixB}.

In Fig.~\ref{fig:TT} we present a typical TT model, and in particular
this configuration corresponds to the one with the highest toroidal
magnetic field among all our models. As anticipated, the structure of
the poloidal magnetic field closely resembles what was found in the
previous section, on purely poloidal models: it threads the entire
star, reaches its maximum value at the center, vanishing only in
ring-like region in the equatorial plane, and crosses smoothly the
stellar surface. The magnetic field outside the star is dominated by
its dipole component.
On the other hand, the toroidal magnetic field has now a rather
different structure, with respect to purely toroidal cases. It does
not fill completely the interior of the star, but it is confined in a torus
tangent to the stellar surface at the equator. It reaches its maximum
exactly in the ring-like region where the poloidal component
vanishes. Of course this behaviour is related to our choice of the
poloidal current distribution, and to our requirement that they should
be confined within the star. In principle it is possible to build
models where the toroidal magnetic field fills the entire star, but
this can only be achieved if one allows  the presence of
magnetospheric currents, extending beyond the stellar surface.

In the same Fig.~\ref{fig:TT} we also show the distribution of the baryonic density.
As we pointed out in Sec.~\ref{subsec:ptf} a magnetic field that extends prevalently
into the outer layers of the star has minor effects on  the stellar properties with respect
to one that penetrates also in the core region. Therefore, it is the
poloidal component of the magnetic field, which is also dominant, that is mostly responsible for the deformation of 
the star in the TT configuration: the baryonic
density distribution in fact resembles closely what we obtained in the purely poloidal configuration and the
stellar shape is oblate and the external layers have a {\it lenticular} aspect.

\begin{figure*}
	\centering
	\includegraphics[width=.75\textwidth]{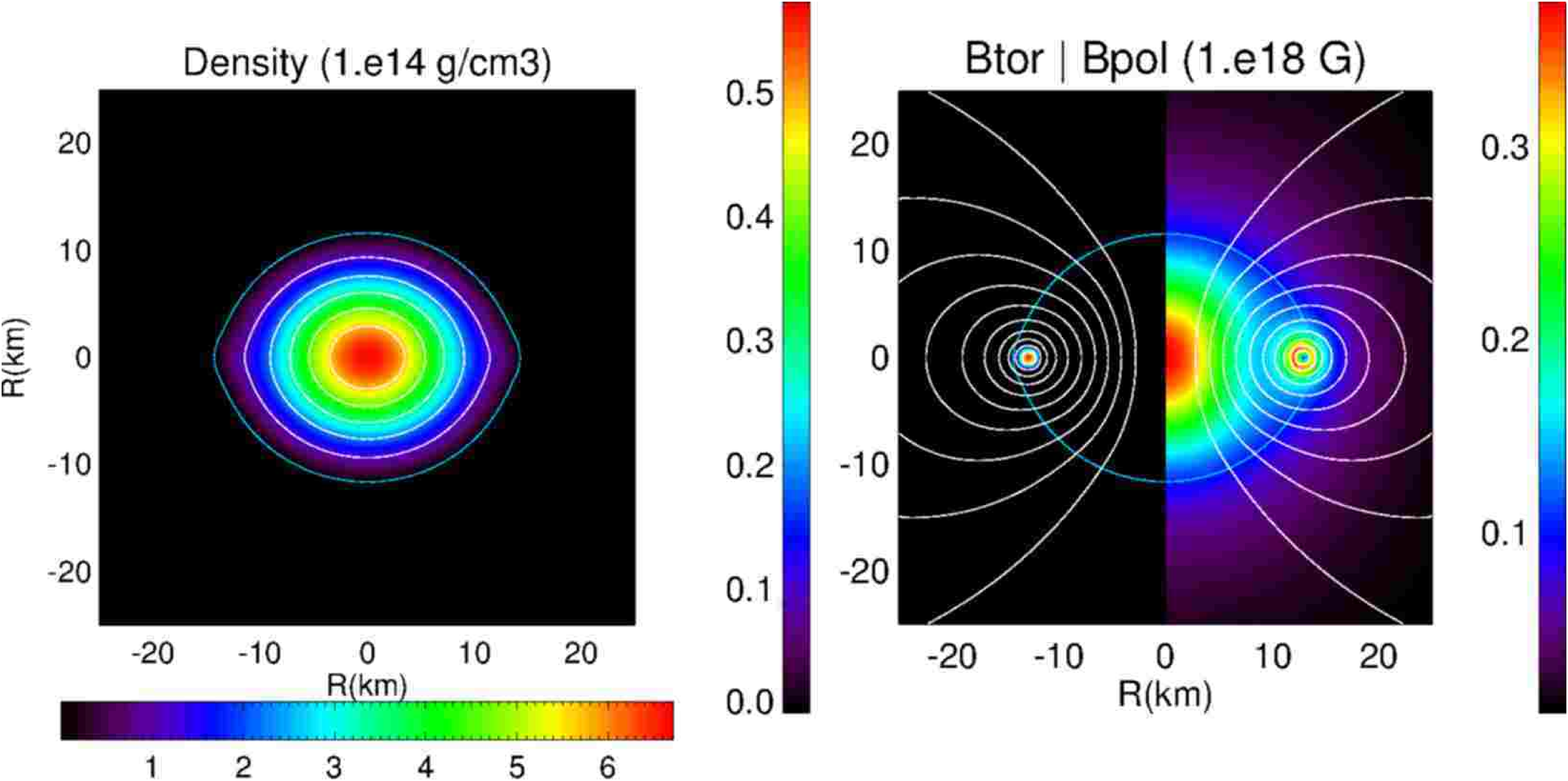} 
	\caption{TT configuration with a gravitational mass $M=1.551 M_{\sun}$, a
	baryonic mass $M_0=1.660$, a maximum field strength $\Bmax =5.857\times10^{17}\,\mbox{G}$.
	Left panel: baryonic density distribution. Right panel: strength of the toroidal (left half) and poloidal
        (right half) magnetic field 
	components, superimposed to \emph{magnetic field surfaces}
        (isocontours of $\tilde{A}_\phi$.
	The blue curve locates the stellar surface.
	The other global physical quantities of this configuration are 
	listed in the last line of Table~\ref{tab:TT}. }
	\label{fig:TT}
\end{figure*}

\begin{figure*}
	\centering
	{\includegraphics[width=.3\textwidth]{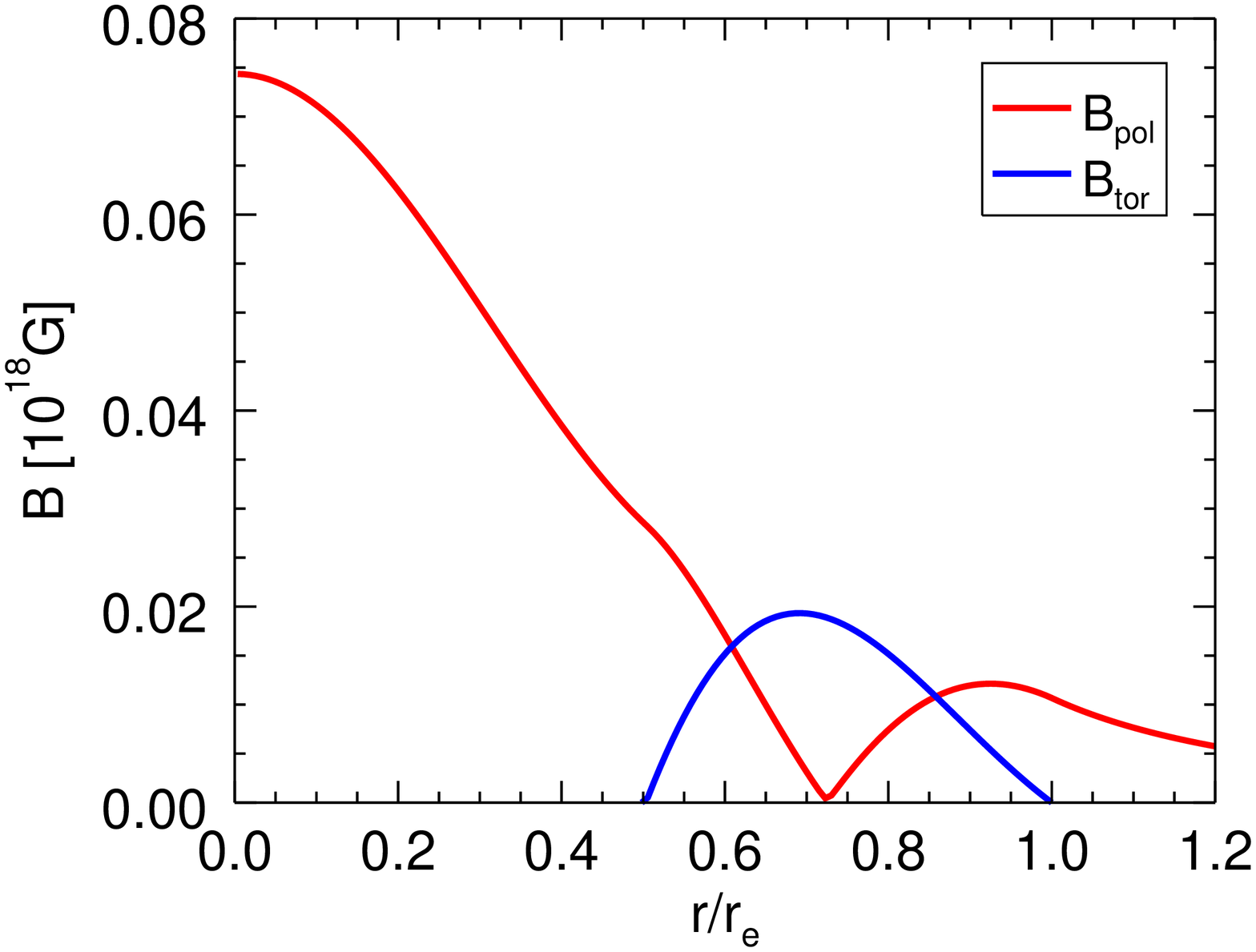}} 
	{\includegraphics[width=.3\textwidth]{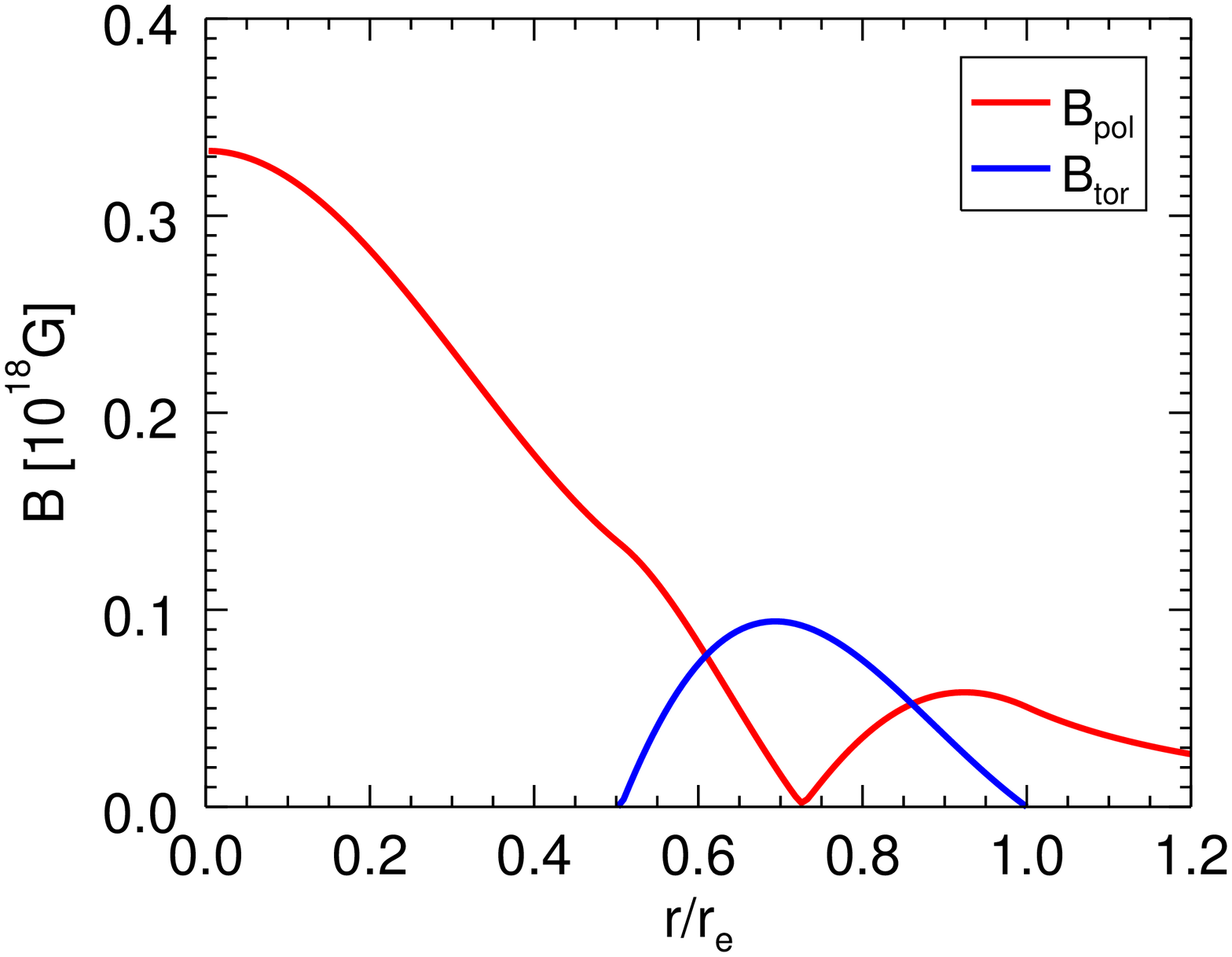}} 
	{\includegraphics[width=.3\textwidth]{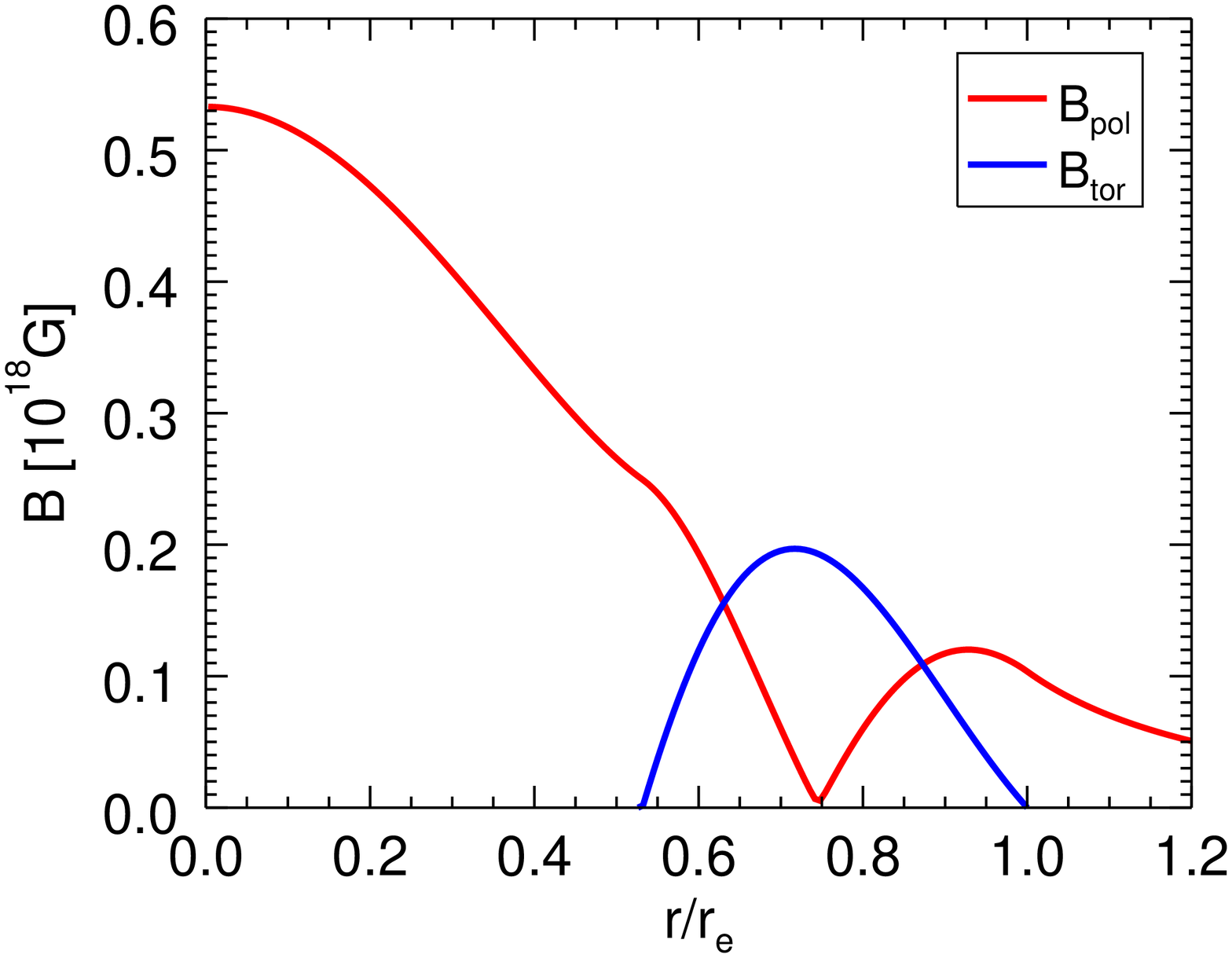}}\\ 
	{\includegraphics[width=.3\textwidth]{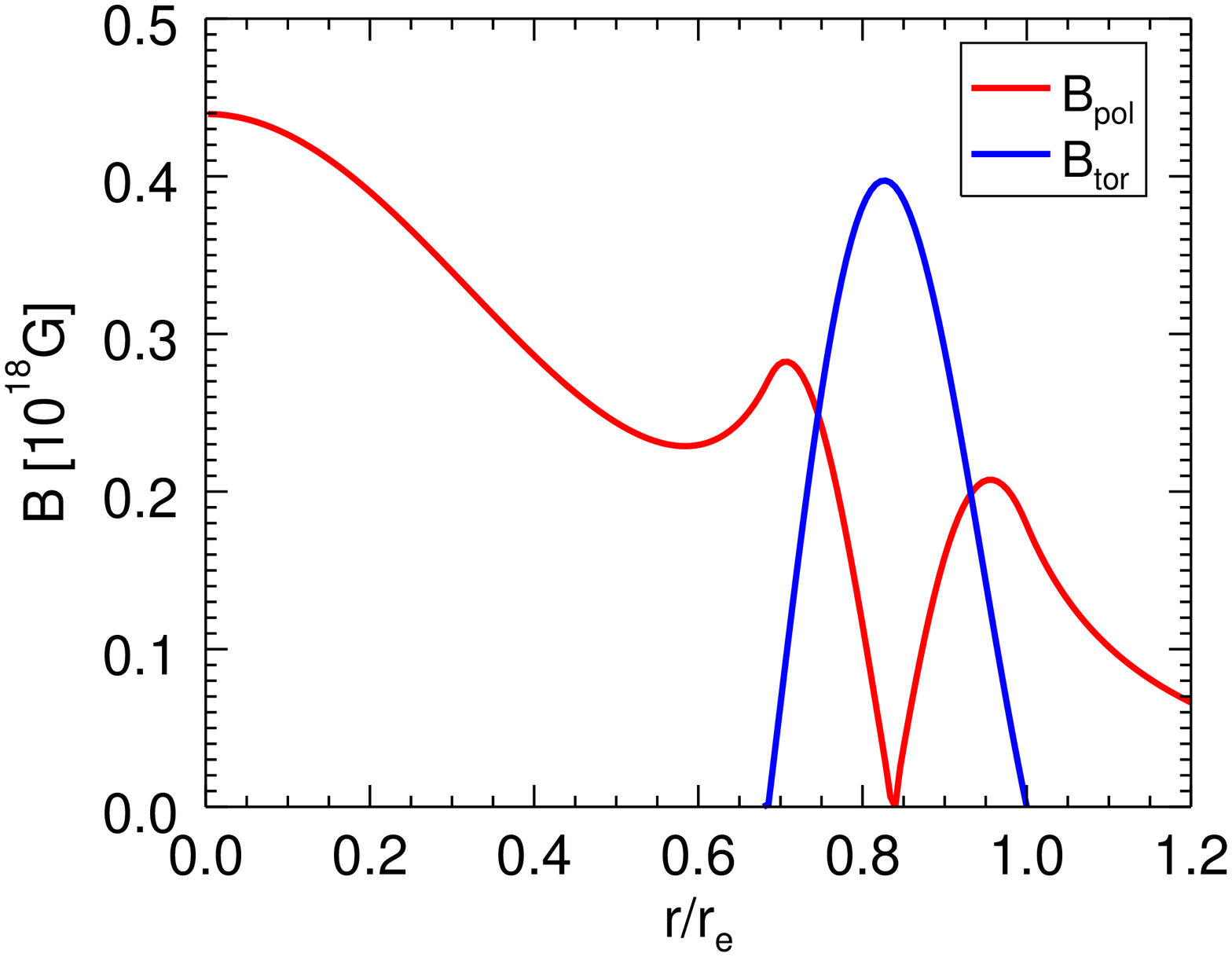}}
	{\includegraphics[width=.3\textwidth]{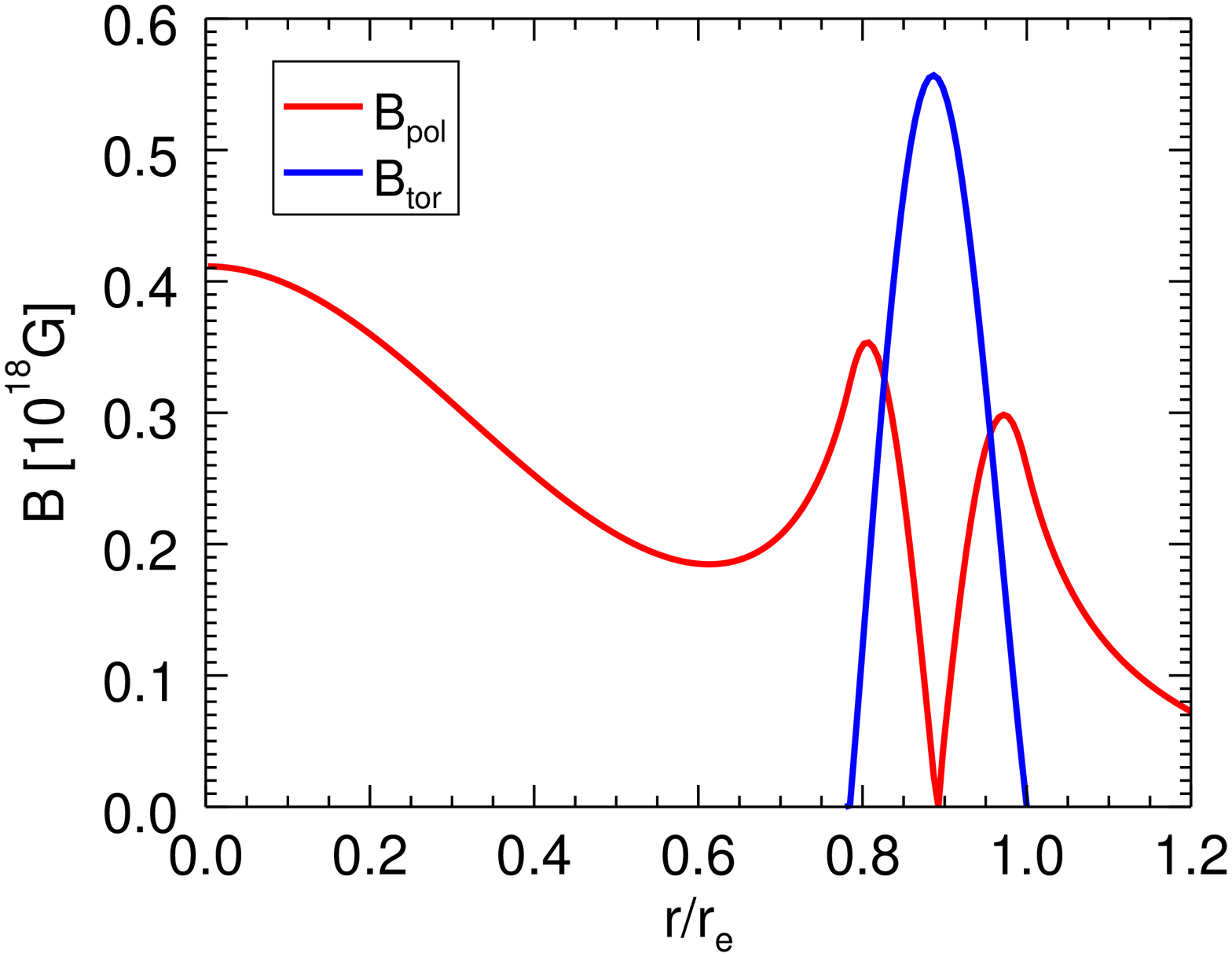}}
	{\includegraphics[width=.3\textwidth]{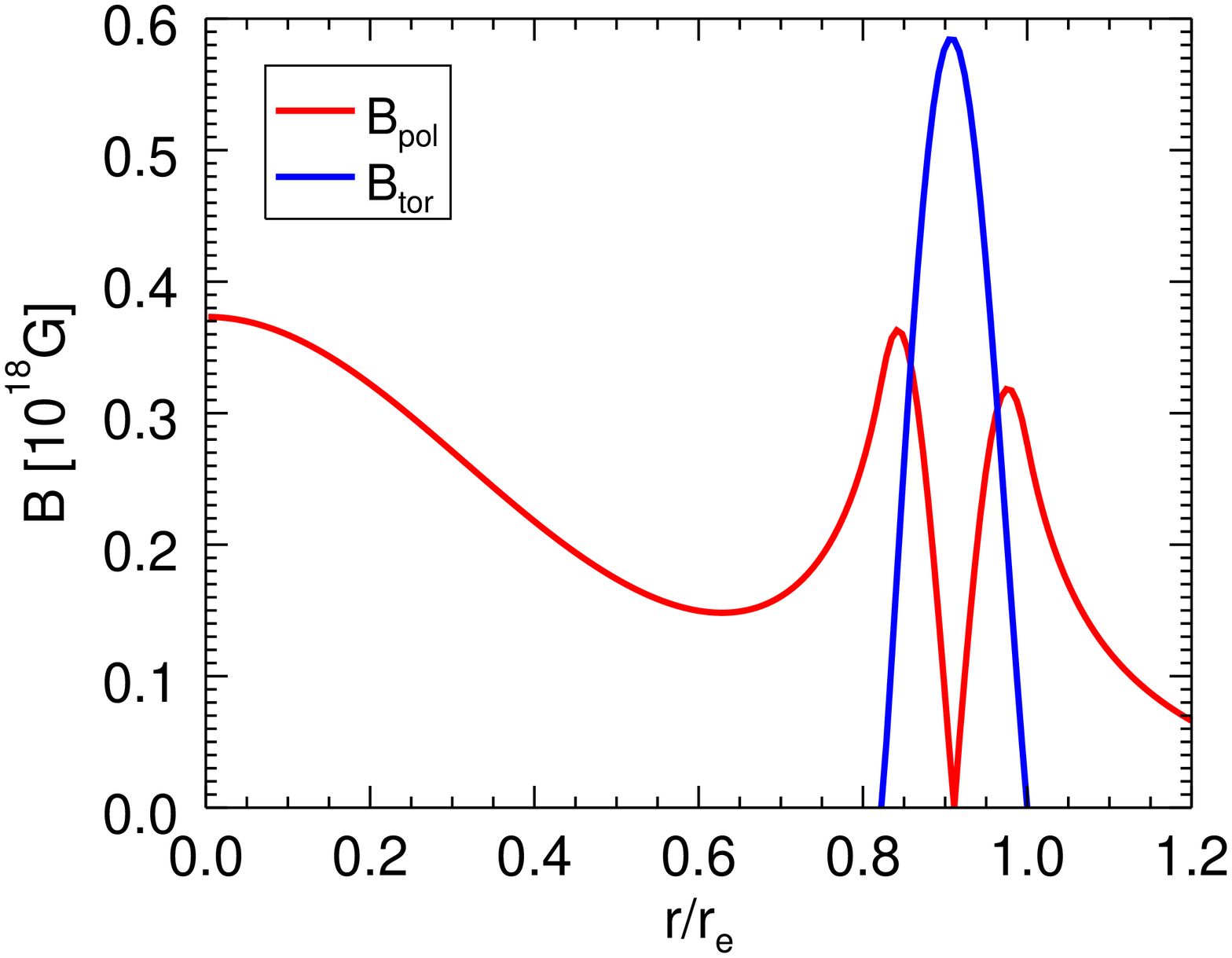}}
	\caption{
	Profiles of the strength of the poloidal and toroidal
        components of the magnetic field, along the
        equator. $r_e$ is the equatorial radius.
	All models have the same gravitational mass $M=1.551M_{\sun}$.
        Top panels show three models with
	$a=0.5$ and $k_{\rm pol}=0.04$ (left), $k_{\rm pol}=0.18$ (center) or $k_{\rm pol}=0.31$ (right). The left bottom
        panel shows a model with $a=1.0$ and $k_{\rm pol}=0.23$, the central bottom panel with $a=1.5$ and $k_{\rm pol}=0.22$, and the right
        bottom panel with $a=2.0$ and $k_{\rm pol}=0.19$.
	The global physical quantities of these configuration are listed in Table~\ref{tab:TT}.	}
    \label{fig:TTprof}
    \end{figure*}

Fig.~\ref{fig:TTprof} shows a comparison between the strength of the
toroidal and poloidal magnetic field at the equator, for various
models characterized by the same  gravitational mass 
$M=1.551 M_{\sun}$, but different values of the magnetization 
constants $k_{\rm pol}$ and $a$.
We found that, at a fixed value of $a$, the strength of both the toroidal
and poloidal field grows with $k_{\rm pol}$, while 
if one keeps fixed the maximum strength of the  poloidal field, then
the region occupied by the torus shrinks as $a$ grows. 

\begin{figure*}
	\centering
	{\includegraphics[width=.4\textwidth]{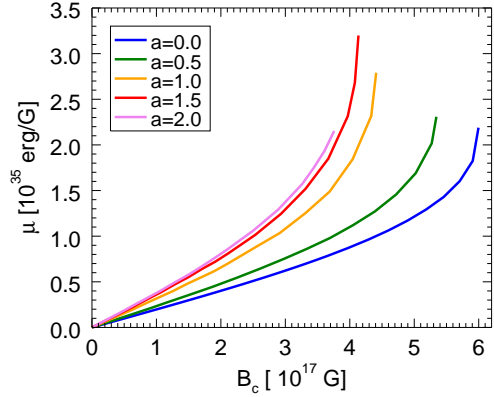}} 
	\caption{Magnetic dipole
    moment $\mu$  as a function of $B_{\rm c}$ for various values of
    the parameter $a$. 
	All models have the same gravitational mass $M=1.551M_{\sun}$.	}
    \label{fig:BcenMuseqTT}
    \end{figure*}

In Fig.\ref{fig:BcenMuseqTT} we show the relation between the magnetic
dipole moment and the value of the magnetic field strength in the
centre$B_{\rm c}$ along equilibrium
sequences where the
gravitational mass has been kept fixed, $M=1.551M_{\sun}$, for various
values of $a=\lbrace 0.0,\, 0.5,\, 1.0 ,\,1.5, \,2.0 \rbrace$. It is
evident that at fixed magnetic dipole moment, the field strength
decreases with $a$. This can be understood if one recalls that at
higher values of $a$ there is an increasing contribution to the
magnetic dipole moment from currents associated to the toroidal
field (the same value of $\mu$ corresponds to a lower value of $k_{\rm pol}$).
As a result the value of the magnetic field at the center,
which is mostly determined by the current term  $\rho h k_{\rm pol}$,
drops. Moreover it is also evident that there appears to be a maximum
asymptotic value that the central field can reach, as we discussed in
the previous section on purely poloidal configurations, and that such
value is smaller for higher values of $a$.

\begin{figure*}
	\centering
	\includegraphics[width=.35\textwidth]{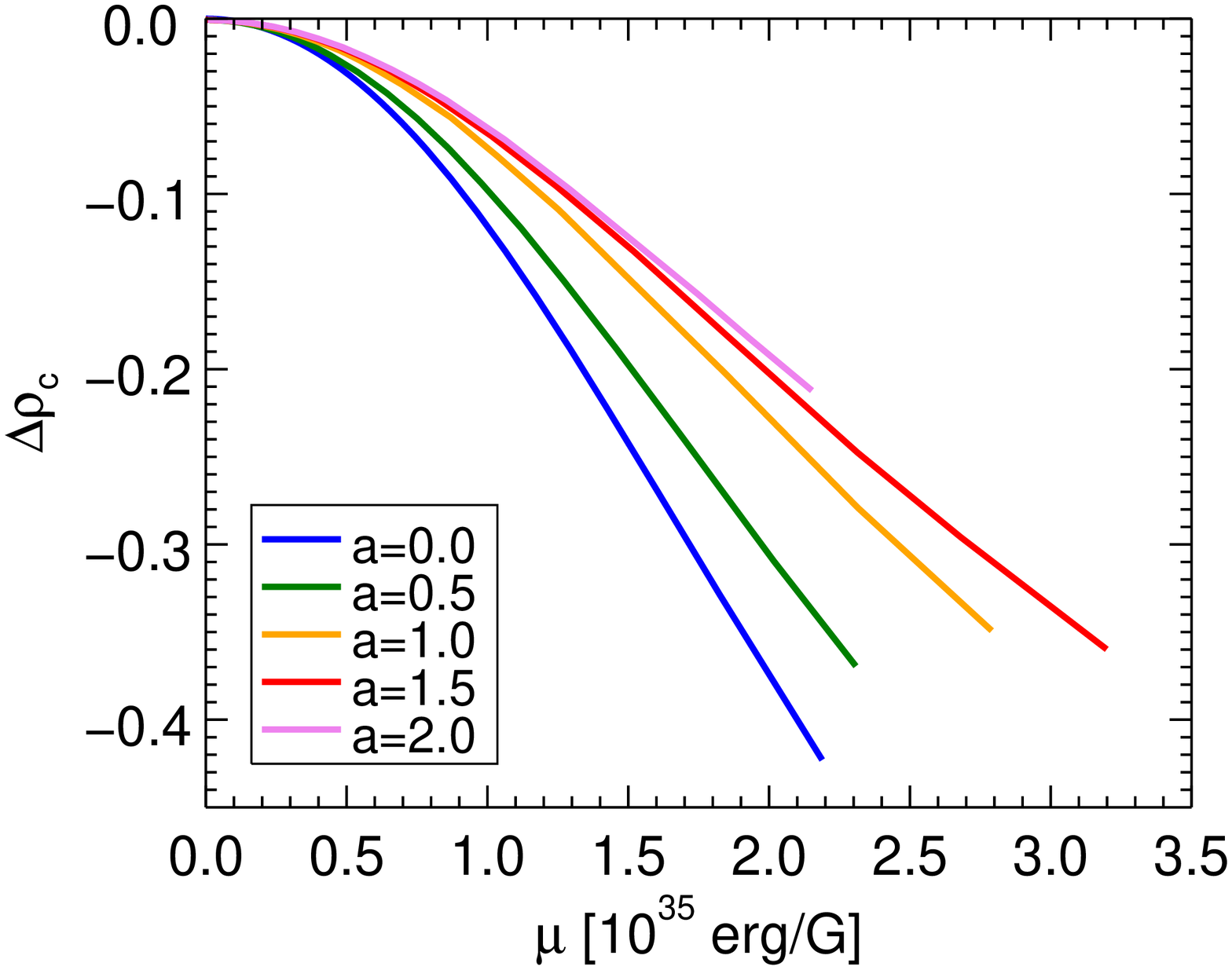}
	\includegraphics[width=.35\textwidth]{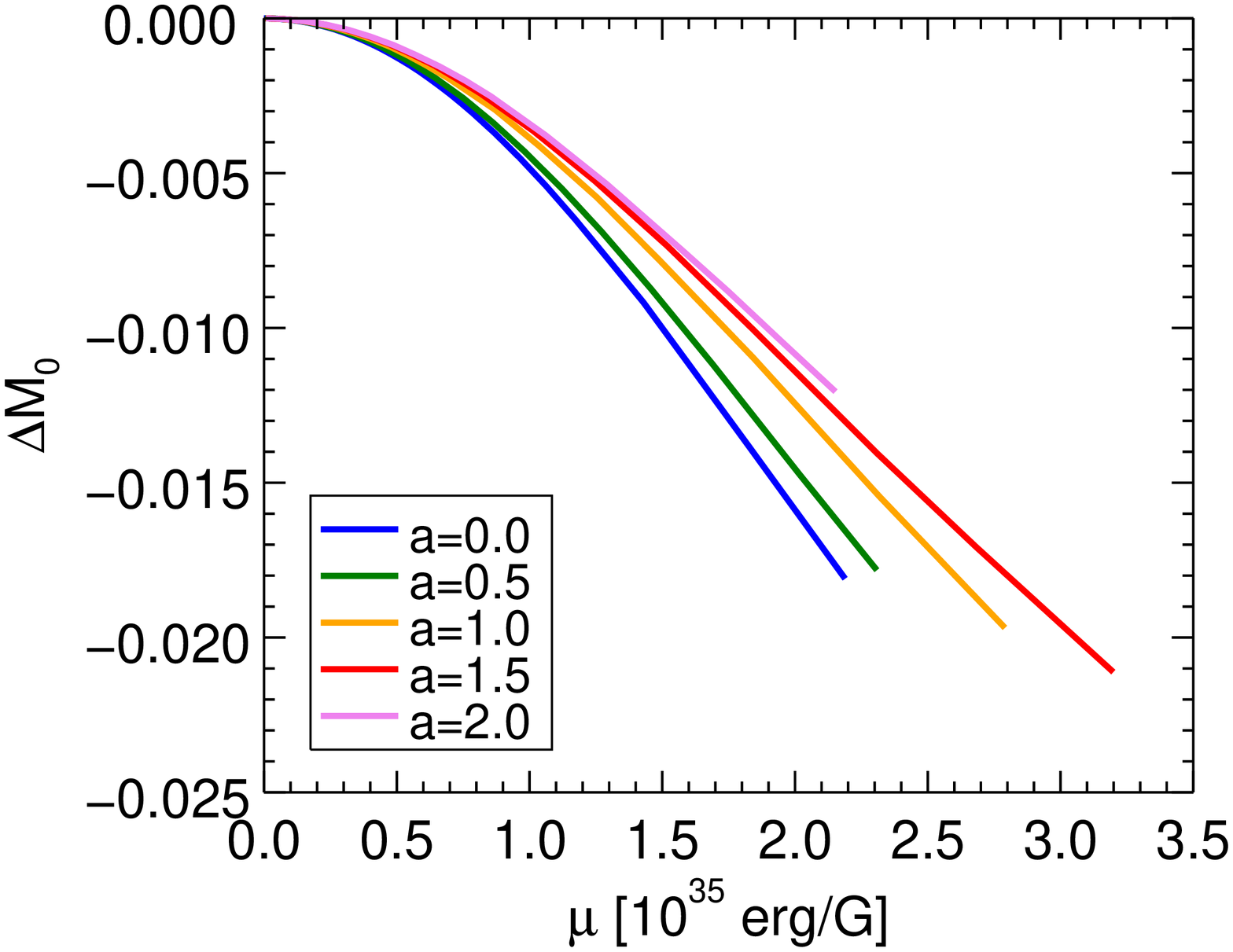} \\
	\includegraphics[width=.35\textwidth]{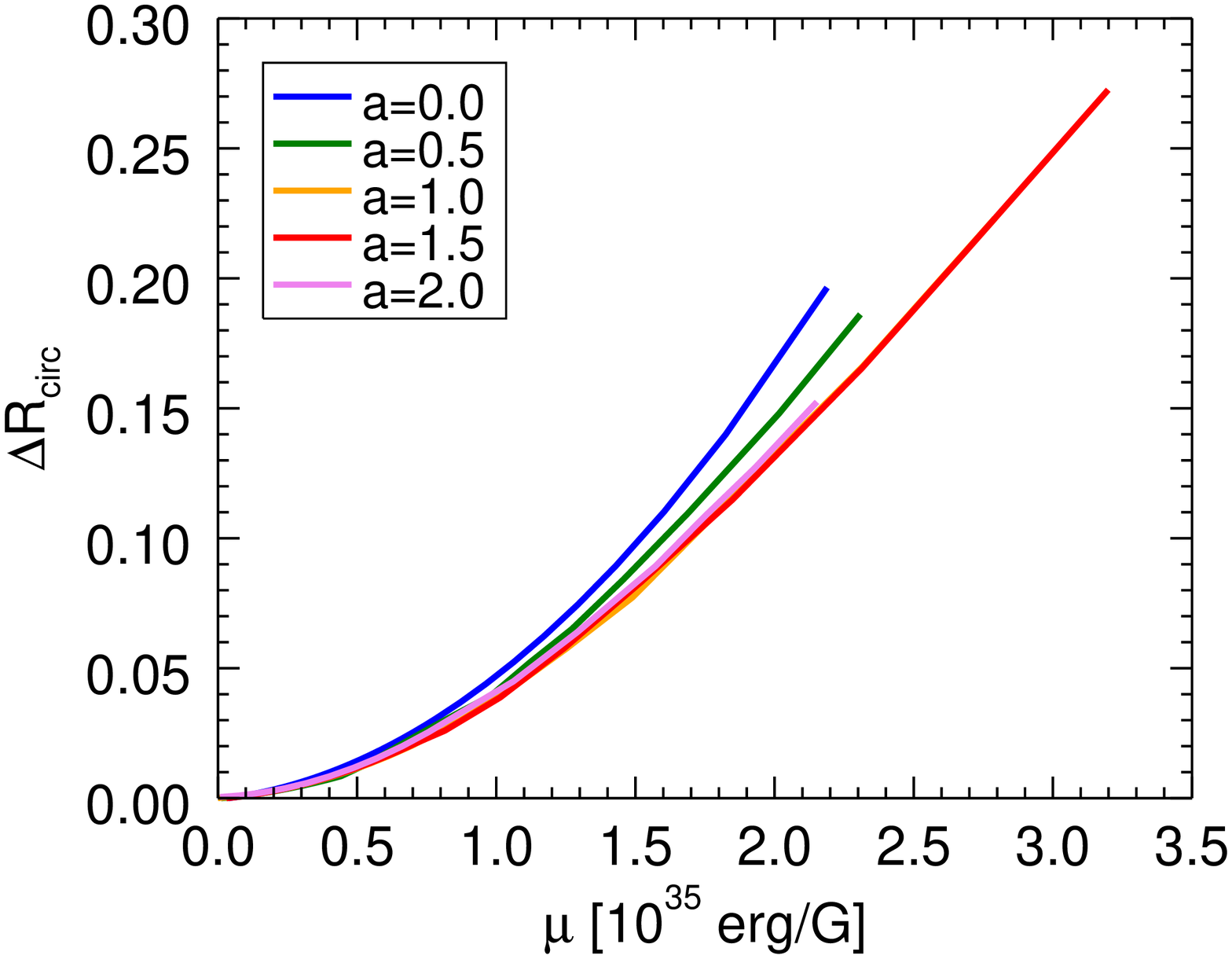} 
	\includegraphics[width=.35\textwidth]{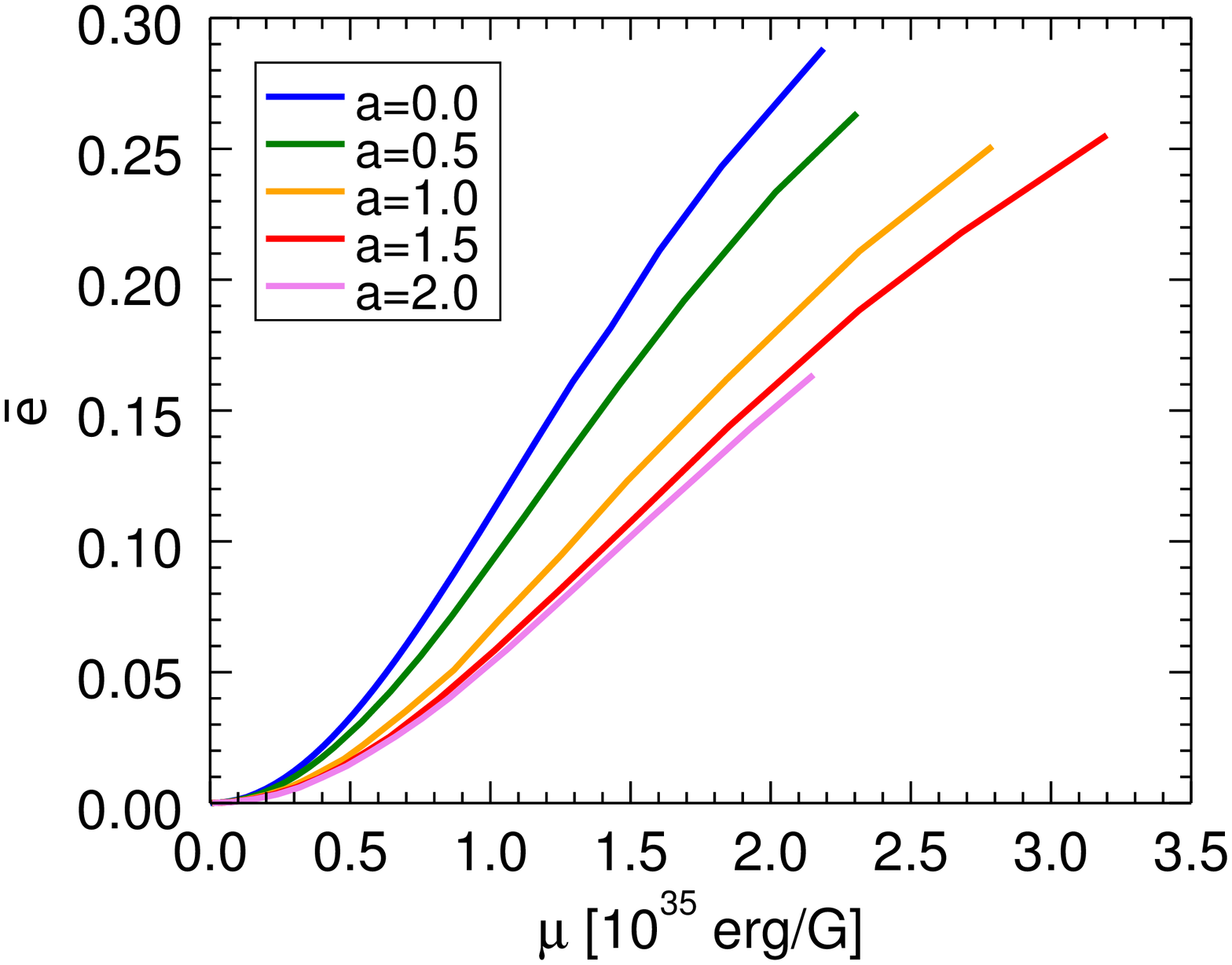}
	\caption{ 
	Behaviour of the baryonic central density $\rho_c$, of the baryonic mass $M_0$, of 
	the circumferential radius $R_{\rm circ}$ and the mean deformation rate for
        TT equilibrium sequences with a fixed gravitational mass $M=1.551M_{\sun}$.
 All quantities are shown as a function of the magnetic dipole moment $\mu$.  The models
 corresponding to the extreme cases for each sequence are presented in
 details in the last four lines in Table~\ref{tab:TT}.}
	 \label{fig:seqTT}
    \end{figure*}

In Fig.~\ref{fig:seqTT}
we display, along the same sequences, how  some global quantities change as a function of the
magnetic dipole moment $\mu$. We stress here that this parametrization
is not equivalent to the one in terms of the strength of the magnetic
field at the center. We can notice that for a fixed $\mu$ the deviation
from the unmagnetized case is progressively less pronounced at increasing values of
$a$. This happens for the same reasons discussed above for $B_{\rm
  c}$. Peripheral currents, that contribute to the magnetic dipole
moment, have minor effects on the magnetic field at the center. On the
other hand it is the poloidal field that penetrates the core and
dominates the enrgetics which
 is mostly responsible for these deviations.
Moving to higher values of $\mu$ along the TT sequences in Fig.~\ref{fig:seqTT}
the mean deformation rate $\bar e$ and the circumferential radius $R_{\rm circ}$ increase 
whereas  the central density $\rho_c$  diminishes, just as in the purely poloidal
configuration.

It is also interesting to look at the same quantities as parametrized
in terms of the strength of the magnetic field, either the toroidal or
the poloidal component. In our models, for $a<1$,
the maximum magnetic field inside the star is associated
with the poloidal component, and it is coincident with the central
value $B_{\rm c}$, while for $a>1$ the maximum strength of the magnetic field
is associated to the toroidal component. This does not seem to depend
on the overall strength of the magnetic field. For the highest values
 the strength of the poloidal component of the magnetic field migth reach 
its maximum in the torus region (see the trend in Fig.~\ref{fig:TTprof}). 

\begin{figure*}
	\centering
	\includegraphics[width=.35\textwidth]{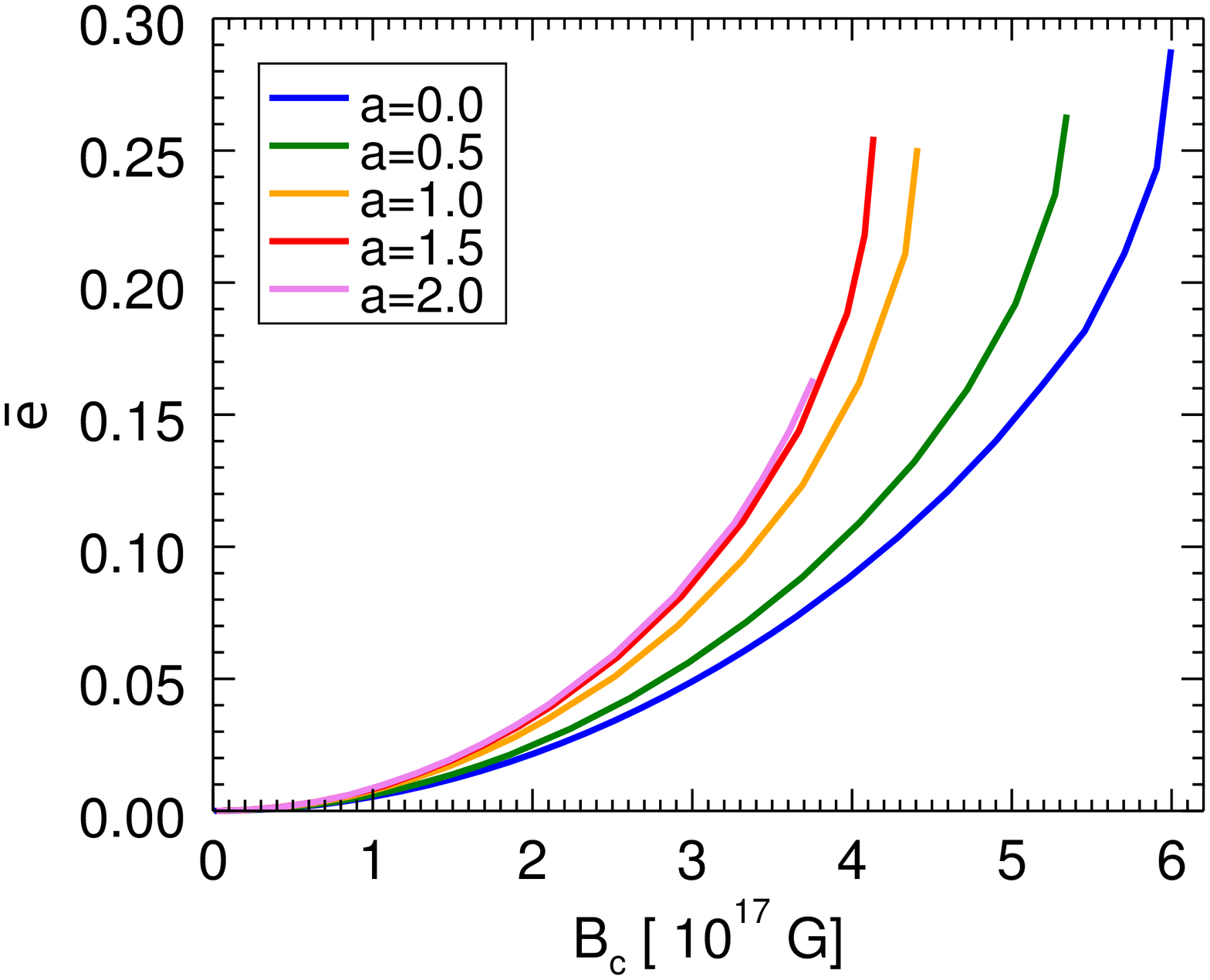}
	\includegraphics[width=.35\textwidth]{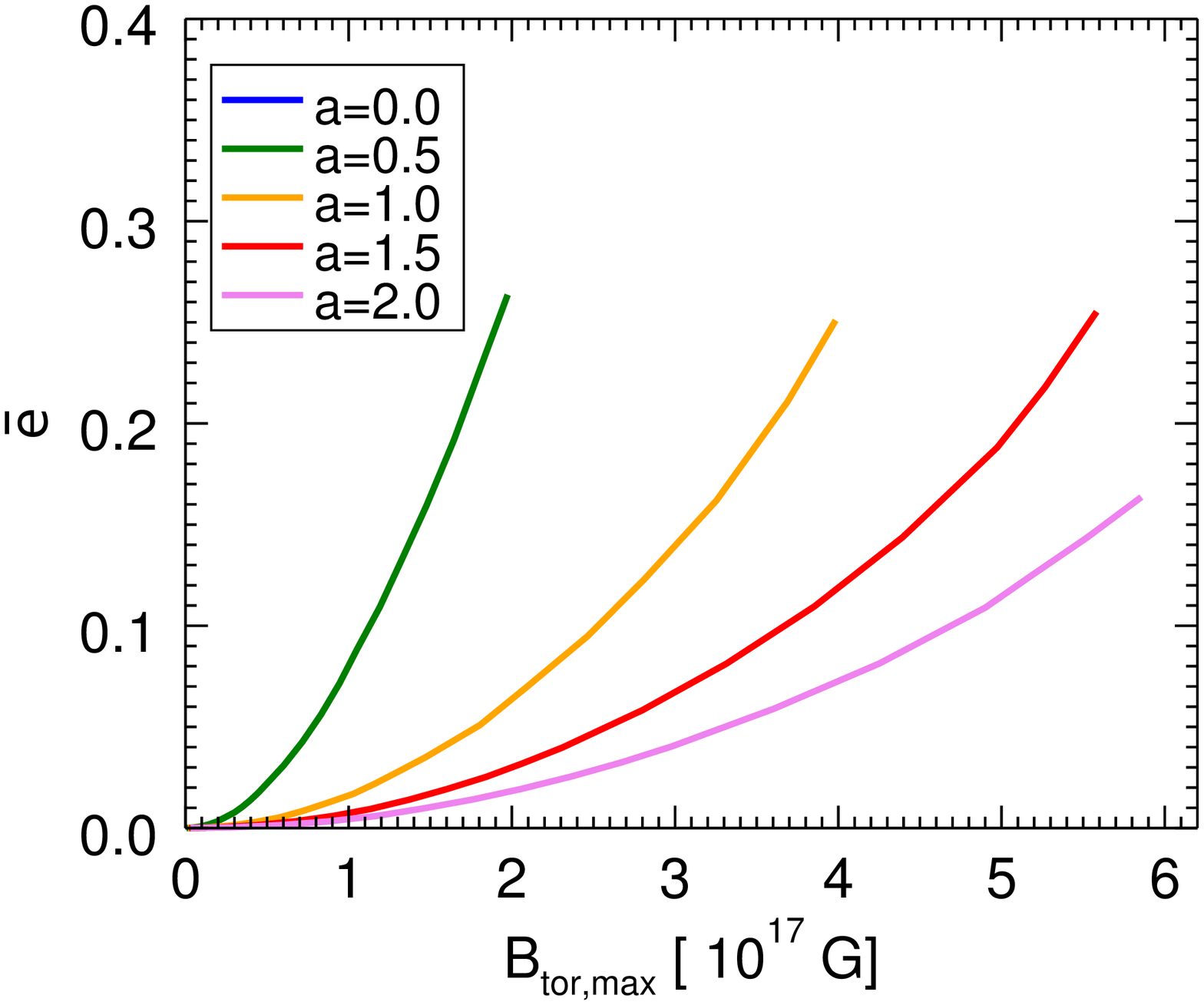}\\
	\includegraphics[width=.35\textwidth]{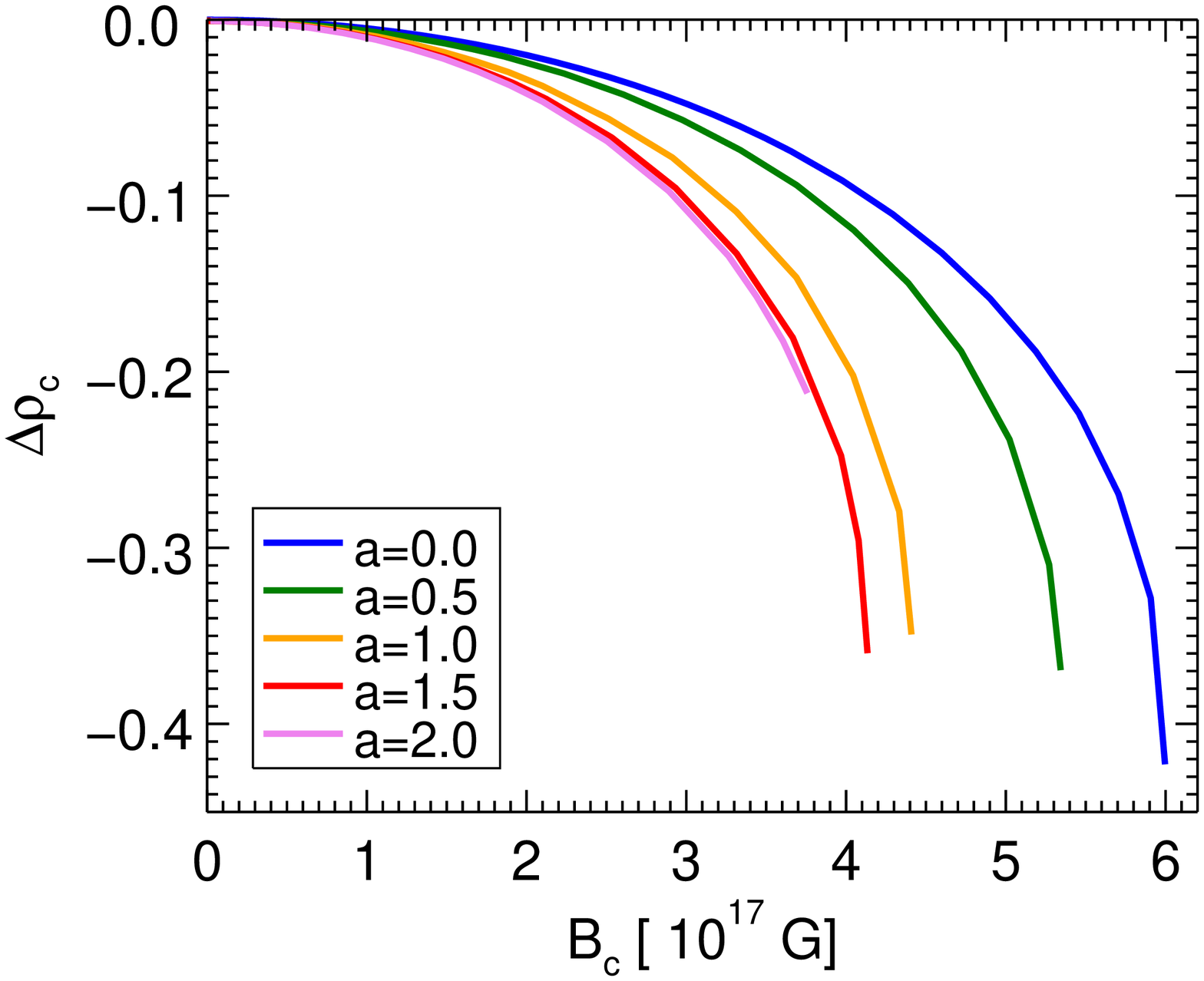}
	\includegraphics[width=.35\textwidth]{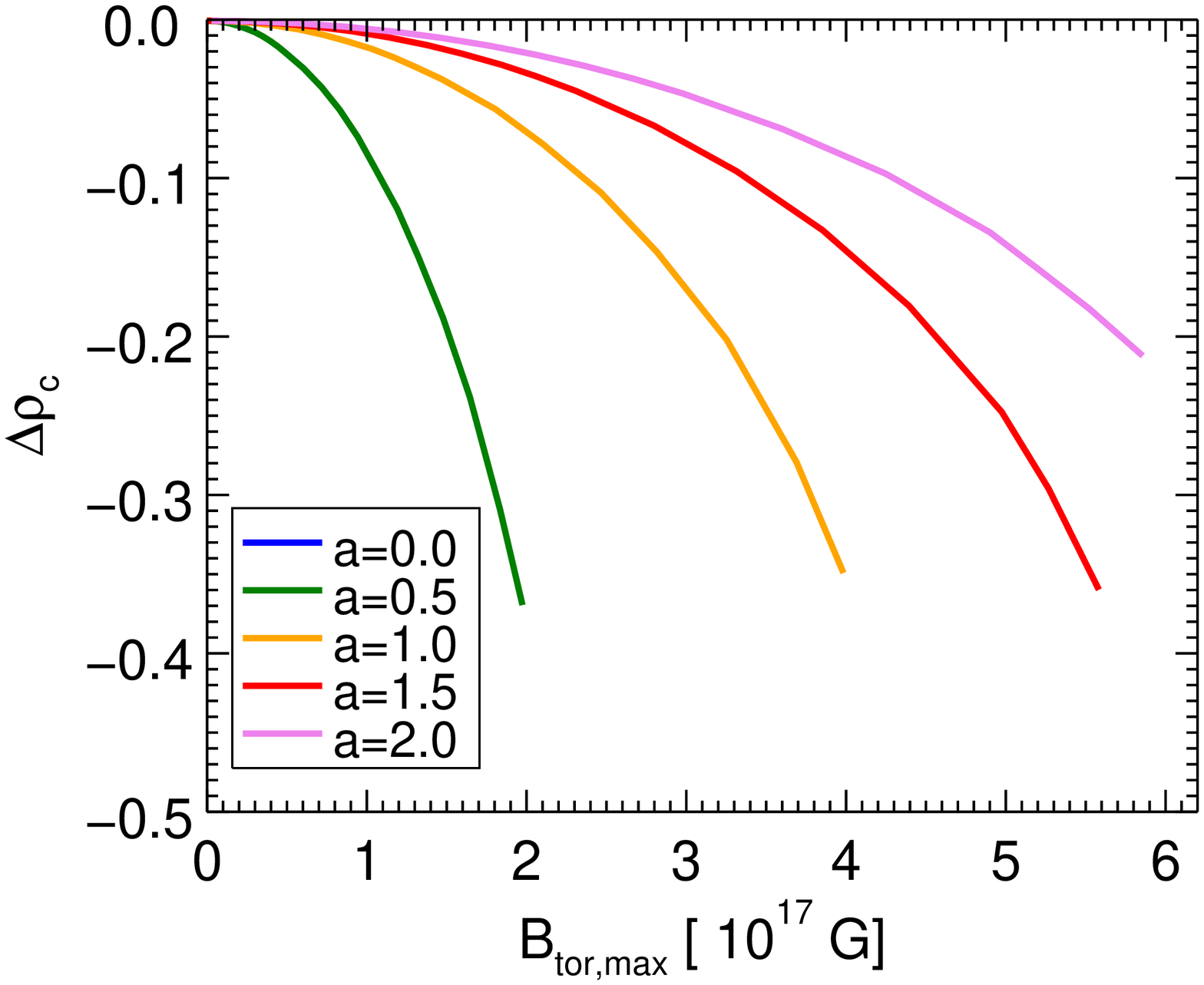}
	\caption{
	Mean deformation rate $\bar e$ (top) and $\Delta \rho_c$
        (bottom) displayed as
	a function of $B_{\rm c}$ and $B_{\rm tor,max}$ along the same sequences shown 
	in Fig.~\ref{fig:seqTT}.
    }
	\label{fig:BcTT}
\end{figure*}

In Fig.~\ref{fig:BcTT} we show $\Delta \rho_c$ and $\bar e$ as a function
of $B_{\rm c}$ and $B_{\rm tor,max}$. We can notice that for a fixed
$B_{\rm c}$ the trend with $a$ is
exactly the opposite than the one shown previously for fixed
$\mu$. This might seem counter-intuitive, given that both quantities are
parametrizations of the strength of the poloidal field. However,
models with higher $a$, at fixed $\mu$, have weaker central fields, and
smaller deviations, while models with higher $a$, at fixed $B_{\rm c}$
have higher total magnetic energy, and as such higher deviations. The effects due to the tension of
the toroidal field (that would lead to a less deformed star), are
dominated by the drop in the central density due to the increase of
magnetic energy. For the same reason, when shown as a function of
the maximum strength of the toroidal magnetic field, models show that
higher values of $a$ imply smaller deviations from the unmagnetized
case.

\begin{table*}
\caption{ \label{tab:TT} Global quantities for various TT
  models 
with the same gravitational mass $M=1.551M_{\sun}$ but different values of both
$\Bmax$ and $a$. In the last three lines we present the models with
the highest maximum magnetic field that we could build, for each value
of $a$. For the definition of the various quantities see Appendix~\ref{appendix}.}
\begin{tabular}{lccccccccccc}
\toprule
\toprule
 $a$&$\rho_c$&$M_0$&$R_{\rm circ}$& $r_e/r_p$ & $\bar e$ 
 &$\mathscr{H}/\mathscr{W}$&$B_{\mbox{\tiny{c}}}$& $B_{\mbox{\tiny{tor,max}}}$ & $\mu$ &
 $H_{\rm m}$& $\mathscr{H}_{\mbox{\tiny{tor}}}/\mathscr{H}$\\
 &[$10^{14}\mbox{g}\,{\mbox{cm}}^{-3}$]&[$M_{\odot}$]&[km]& & [$10^{-1}$]
 &[$10^{-1}$] & [$10^{17}$G] & [$10^{17}$G] & [$10^{35}\,\mbox{erg}\,\mbox{G}^{-1}$]
 &[$10^{42}\,\mbox{G}^2\,\mbox{cm}^4$] & [$10^{-2}$] \\
\midrule
0.5 & 8.488 & 1.680 & 14.24 & 1.000 & 0.033 & 0.011 & 0.745 & 0.194 & 0.173 & 0.031 & 2.893 \\
0.5 & 7.890 & 1.675 & 14.70 & 0.935 & 0.715 & 0.251 & 3.338 & 0.944 & 0.862 & 0.791 & 3.228 \\
0.5 & 5.373 & 1.650 & 16.88 & 0.723 & 2.636 & 1.285 & 5.344 & 1.974 & 2.308 & 5.512 & 4.082 \\
1.0 & 5.545 & 1.647 & 17.41 & 0.733 & 2.510 & 1.455 & 4.409 & 3.983 & 2.790 & 8.079 & 7.262 \\
1.5 & 5.454 & 1.645 & 18.11 & 0.711 & 2.552 & 1.566 & 4.134 & 5.582 & 3.199 & 7.752 & 7.282 \\
2.0 & 6.713 & 1.660 & 16.40 & 0.816 & 1.636 & 0.880 & 3.758 & 5.857 & 2.152 & 3.234 & 6.696 \\
\bottomrule 
\end{tabular}
\end{table*}   

\begin{figure*}
    \centering
	\includegraphics[width=.35\textwidth]{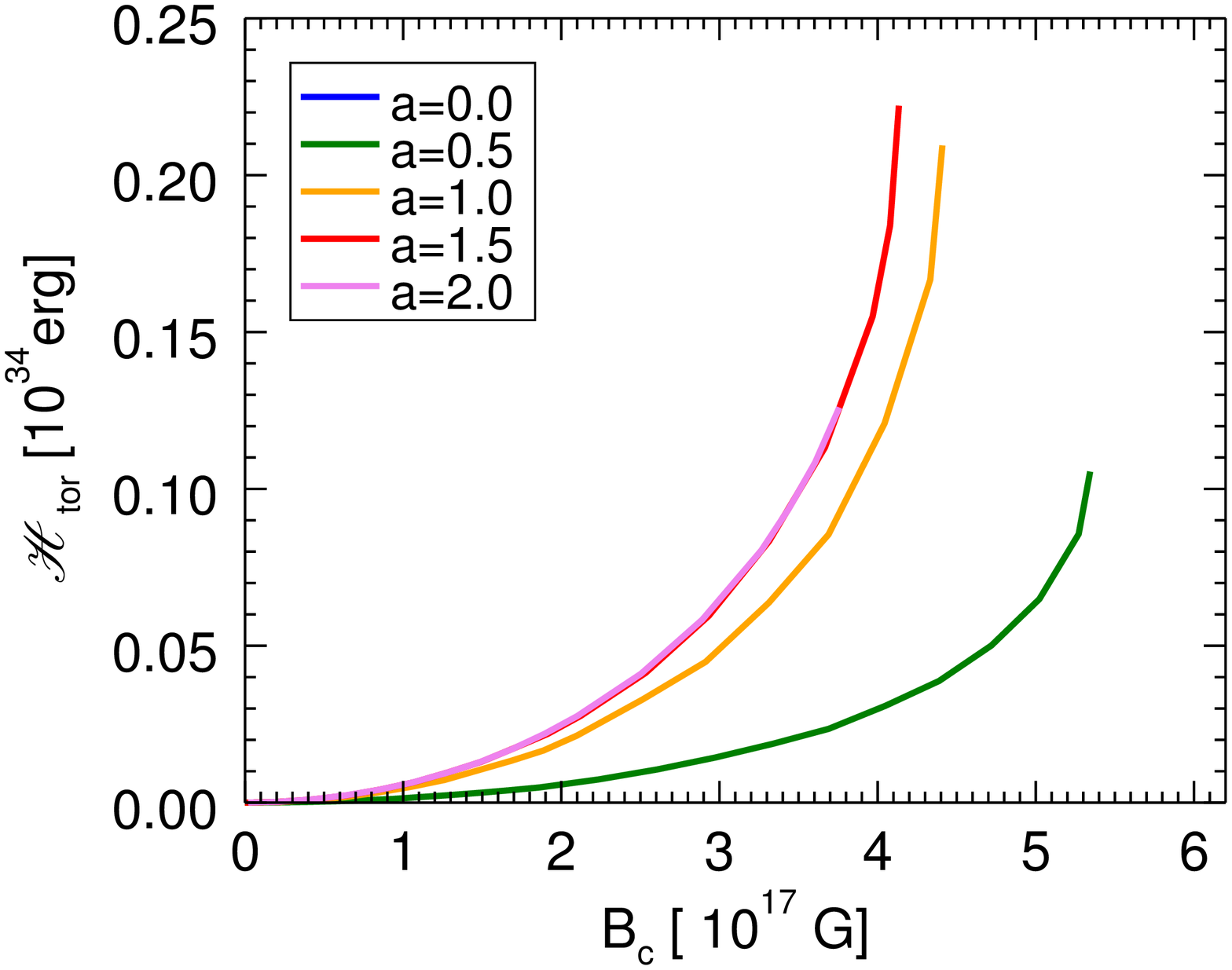} 
	\includegraphics[width=.35\textwidth]{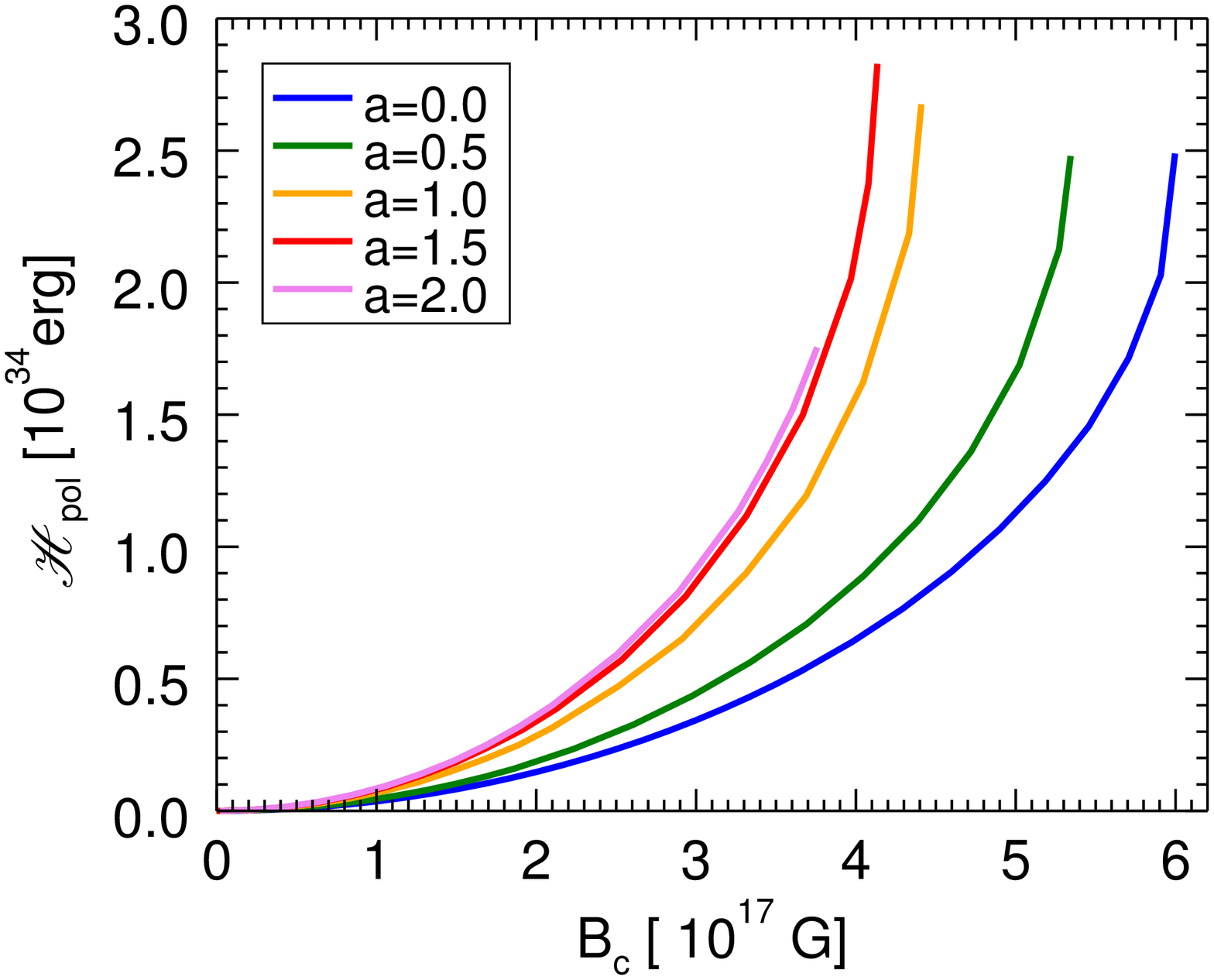} \\
	\includegraphics[width=.35\textwidth]{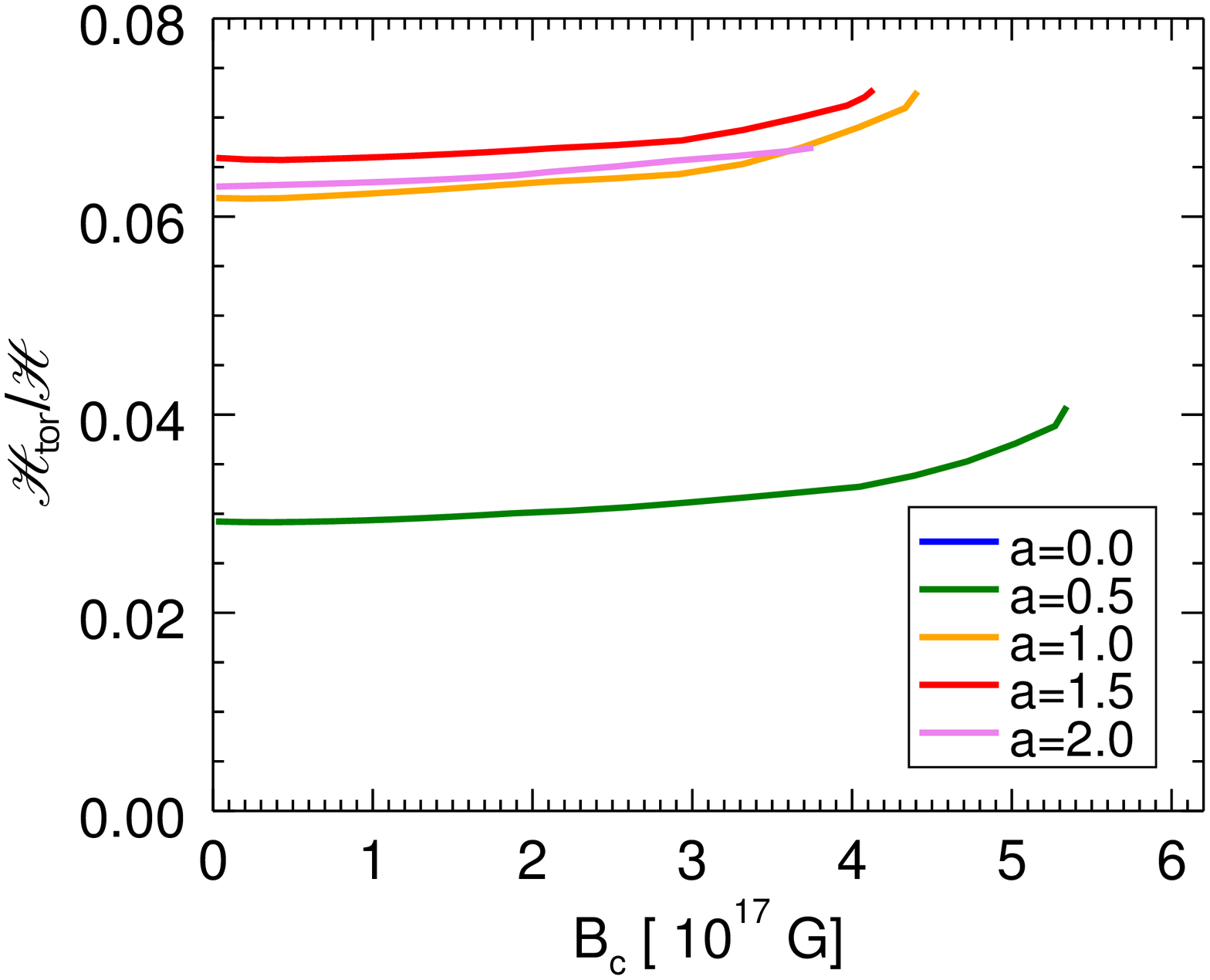}
	\includegraphics[width=.35\textwidth]{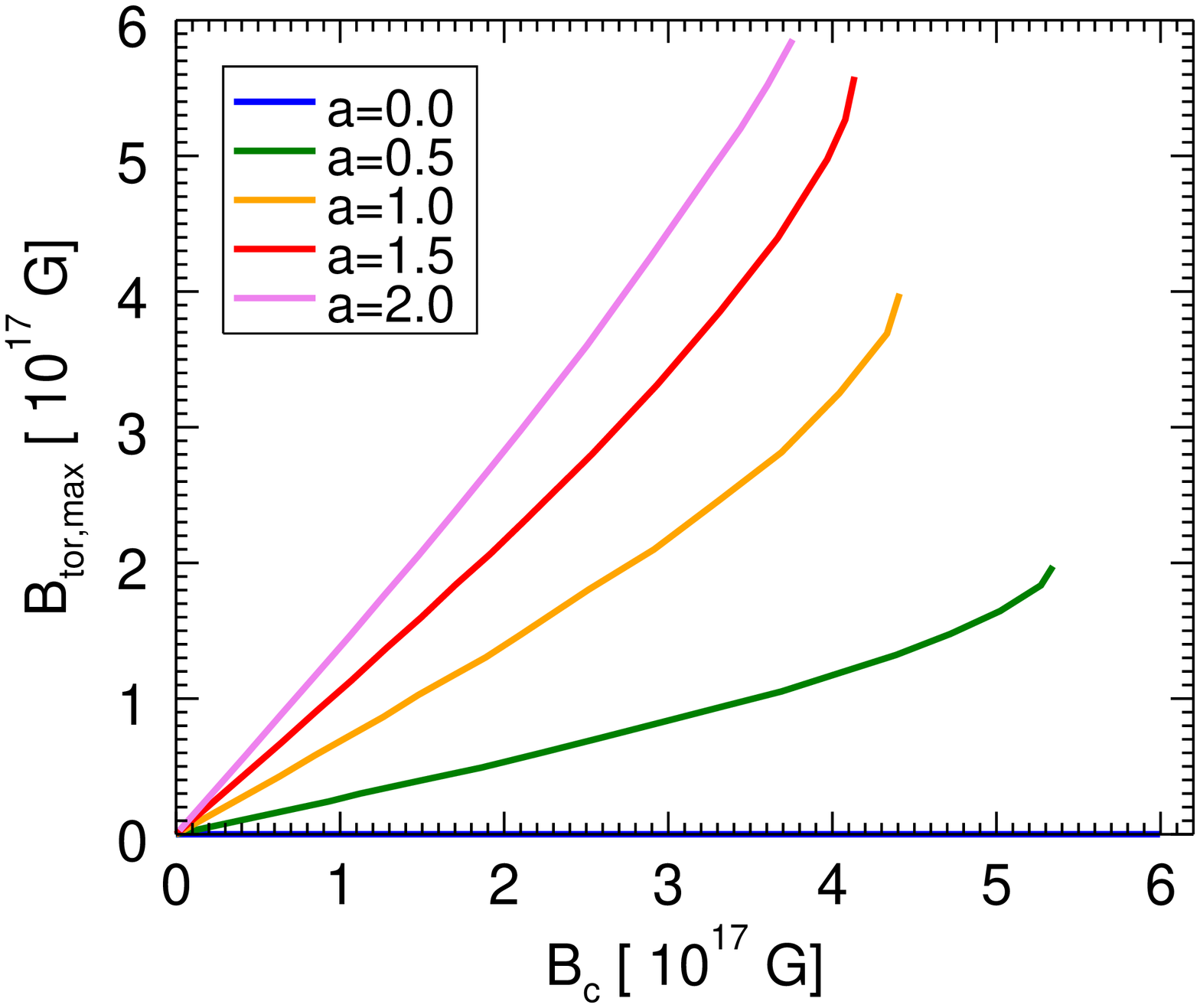}
	\caption{
	Top left panel: toroidal magnetic energy $\mathscr{H}_{\rm
          tor}$. Top right panel: 
	poloidal magnetic energy $\mathscr{H}_{\rm pol}$. Bottom left panel: 
     ratio of the toroidal magnetic energy $\mathscr{H}_{\rm tor}$ to 
    the total magnetic energy $\mathscr{H}$. Right bottom panel: maximum value of
    the toroidal magnetic field strength $B_{\rm tor,max}$.
    All quantities are plotted  as a function of the central magnetic field strength $B_{\rm c}$ 
    along the same sequences shown in Fig.~\ref{fig:seqTT}. 
	}
	 \label{fig:TTenergy}
\end{figure*}  

\begin{figure}
    \centering
	\includegraphics[width=.45\textwidth]{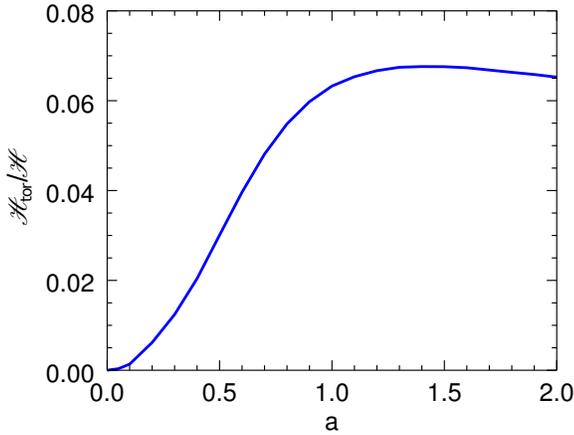}
	\caption{Ratio of the toroidal magnetic energy $\mathscr{H}_{\rm tor}$ to 
    the total magnetic energy $\mathscr{H}$ as a function of the parameter $a$ 
    along a sequence with fixed gravitational mass $M=1.551 M_\odot$ and central 
    magnetic field strength $B_{\rm c} = 2 \times 10^{17}$ G.
	}
	 \label{fig:HoH-a}
\end{figure}

In Fig.~\ref{fig:TTenergy} these same sequences are shown in terms of
their energy content. We note that, at fixed $B_{\rm c}$, the 
equilibrium configurations with higher $a$ are characterized 
by a higher value of both  the total toroidal magnetic field energy
$\mathscr{H}_{\rm tor}$, and the poloidal magnetic field energy
$\mathscr{H}_{\rm tor}$, as expected. It is also evident that the parameter $a$
regulates the ratio  of energy in the toroidal and poloidal components of
the magnetic field, $\mathscr{H}_{\rm tor}/\mathscr{H}$. We
see that the ratio $\mathscr{H}_{\rm tor}/\mathscr{H}$ tends to a
constant in the limit of a negligible magnetic field. 
In the last panel in Fig.~\ref{fig:TTenergy} we also show the
relation between  $B_{\rm c}$ and the maximum strength of the toroidal
magnetic field $B_{\rm tor,max}$. The ratio  $\mathscr{H}_{\rm
  tor}/\mathscr{H}$ shows a clear maximum at $\sim 0.07$  for $a\simeq
1.5$. For smaller values of $a$ this ratio increases because the
strength of the toroidal field increases, however, for $a\gtrsim 1$,
the volume taken by the torus, where the toroidal field is confined,
begins to drop substantially, and this leads to a smaller total energy
of the toroidal component. The net effect of the torus shrinkage over
$\mathscr{H}_{\rm tor}/\mathscr{H}$ is also evident
from Fig.~\ref{fig:HoH-a} where the magnetic energy ratio is shown as a function of 
the parameter $a$ along a sequence with fixed $B_{\rm c} = 2 \times 10^{17}$~G.

\begin{figure*}
	\centering
	{\includegraphics[width=.35\textwidth]{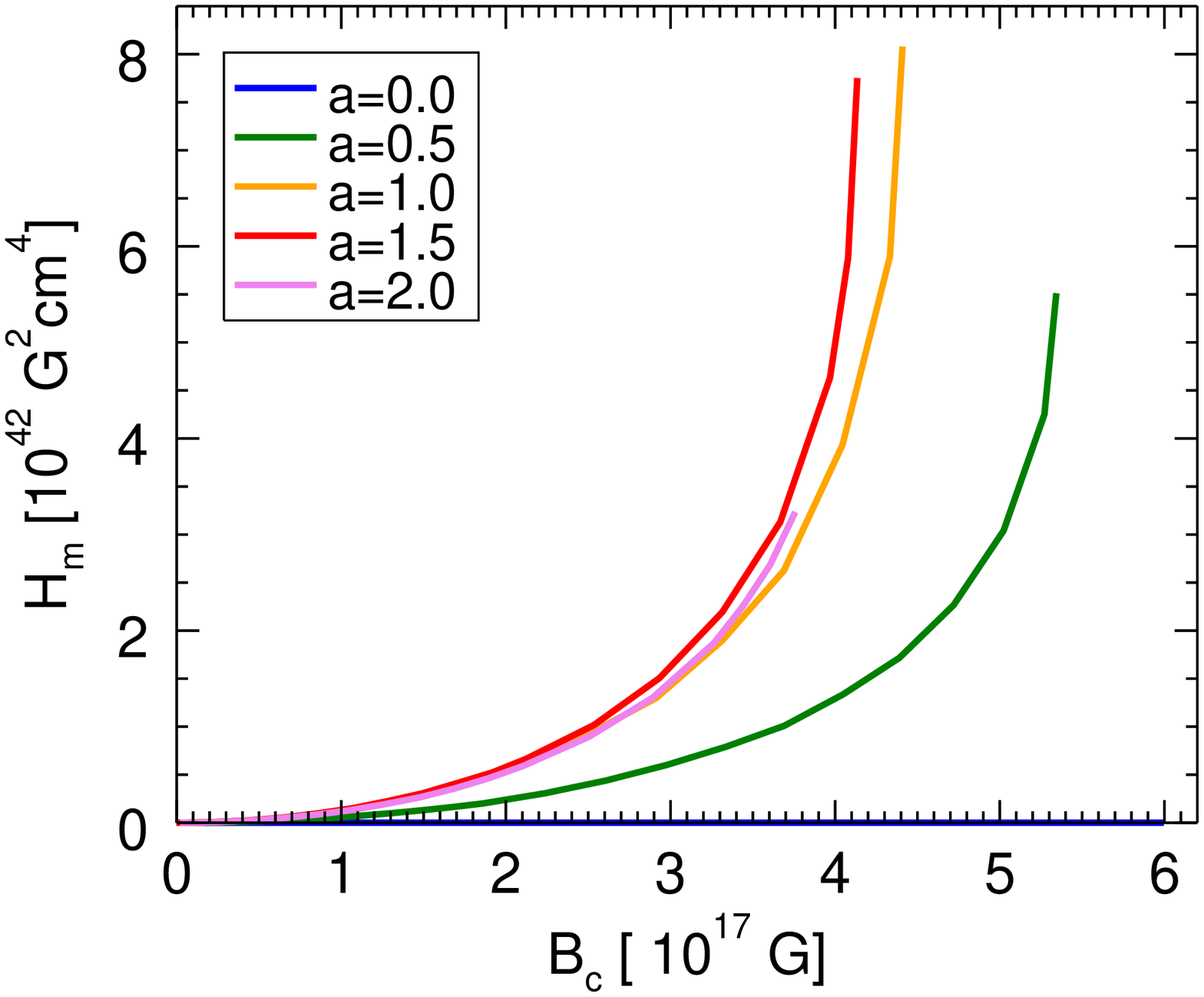}} 
	{\includegraphics[width=.35\textwidth]{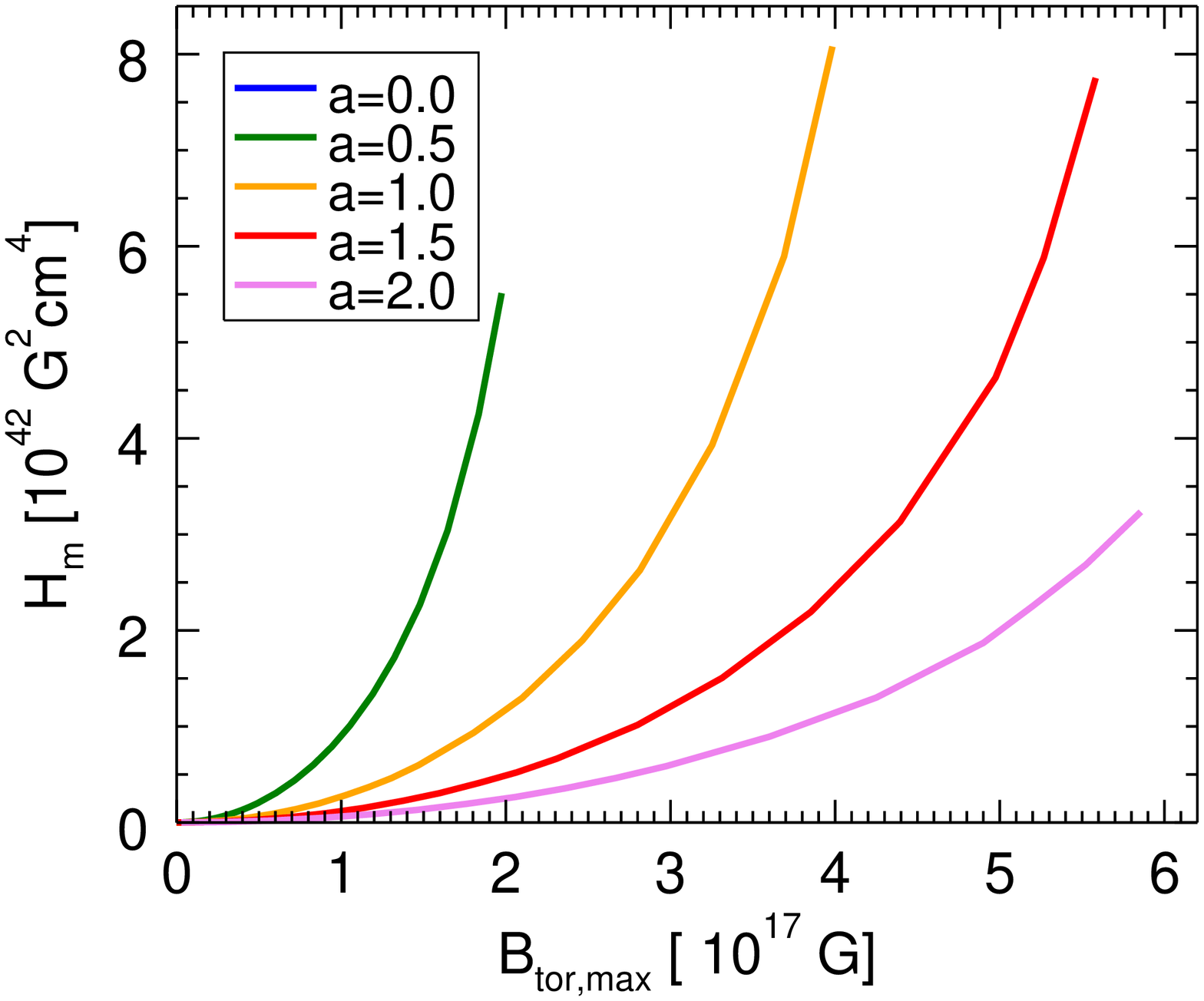}} 
	\caption{Magnetic helicity $H_m$ as a function of the central magnetic field
	strength $B{\rm c}$ (left) and as a function of the maximum toroidal 
	magnetic field strength $B_{\rm tor,max}$ (left)  along
	the same equilibrium sequences shown 
	in Fig.~\ref{fig:seqTT}.
	}
	\label{fig:helicity}
\end{figure*}

Finally in Fig~\ref{fig:helicity} we show  the magnetic helicity $H_m$ as a function
of either the field strength at the centre$B_{\rm c}$ or the maximum
strength of the toroidal
magnetic field $B_{\rm tor,max}$. The magnetic helicity is an important quantity in MHD
because it is conserved in the limit of infinite conductivity, and it
can be shown that it is dissipated on a much longer timescale than the
magnetic energy in the resistive case \citep{Candelaresi_Brandenburg11a}. It is
generally expected that MHD will rapidly relax to configurations that
minimize magnetic energy, keeping fixed the magnetic helicity.  
At a fixed $B_{\rm c}$, $H_m$ increases up to $a\simeq 1.5$, then
drops, for the same reason discussed above for the energetics. Instead,
at a fixed $B_{\rm tor,max}$,  $H_m$ decreases with $a$ since, in this
case, the same toroidal magnetic strength, corresponds to a weaker
poloidal field. 

In general we found a qualitative agreement with previous results
\citep{Lander_Jones09a,Ciolfi_Ferrari+09a,Ciolfi_Ferrari+10a,Ciolfi_Rezzolla13a}, concerning the shape, deformation, and expected
distribution of the poloidal and toroidal components of the magnetic field. In all of our models, the poloidal component is
dominant and the ratio $\mathscr{H}_{\rm tor}/\mathscr{H}<0.07$. This
agrees with previous results where it was shown  that only poloidally dominated models could be built for simple electric current distributions, although recently a more complicated prescription for the currents allowed to build toroidally dominated models \citep{Ciolfi_Rezzolla13a}. For strong fields,
inducing an appreciable deformation, a direct comparison is possible
only with
previous results by \citet{Lander_Jones09a}. They adopt a
different value of $\zeta=0.1$ instead of $0$, their values of $a$ are not directly comparable with ours due to the
different choice of units, and their reference unmagnetized model is
different. Notwithstanding these differences, our results agree with
theirs, on many aspects.
A direct quantitative, comparison with results by \citet{Ciolfi_Ferrari+09a,Ciolfi_Ferrari+10a,Ciolfi_Rezzolla13a} is also not
straightforward, because their choice for the functional form of
the current associated with the toroidal field, Eq.~\ref{eq:fbern}, is
different from our (they assume that the current is a cubic function
of the vector potential while we assume it to be linear). Their
perturbative approach in principle corresponds to a low magnetic
field limit. A more detailed discussion in this limit
is presented in
Appendix~\ref{appendixB}. 

In the fully non linear regime, given that we do not impose any constrain on
the shape of the stellar surface, and allow for oblate
configurations, our field may adjust to this change in shape. Indeed,
as shown in Fig.~\ref{fig:seqTT} we found that, for strong fields,
inducing an appreciable deformation, the ratio  $\mathscr{H}_{\rm
  tor}/\mathscr{H}$ is higher that for the weak field limit by about $10-15\%$.

We want to stress here that the Grad-Shafranov equation Eq.~(\ref{eq:gs}), in
cases where the currents are non linear in the vector potential
$A_\phi$, becomes a non linear Poisson-like equation, that in
principle might admit multiple solutions (\emph{local uniqueness} is
not guaranteed). This is a known problem \citep{Ilgisonis_Pozdnyakov03}, so that we
cannot safely say that these are the only possible equilibria. 

\section{Conclusions}
\label{sec:conclusions}

Magnetic fields are a key element in the physics and phenomenology of
NSs. Virtually nothing of their observed properties can be understood
without considering their effects. In particular, the geometry of the
magnetic field plays an important role, and even small differences
can lead to changes in the physical processes that might be important
for NS phenomenology \citep{Harding_Muslimov11a}. Here we have investigated the role
that a very strong magnetic field has in altering the structure, by
inducing deformations. For the first time we have derived equilibrium
configurations, containing magnetic field of different geometries,
assuming the metric to be Conformally Flat. This is a further
improvement on previous works, which where either done in a Newtonian
or perturbative regime, and allow us to handle very strong fields, and
to take into account the typical non-linearity of Einstein equations.

We have presented a general formalism to model magnetic field of
different geometry, and illustrated our numerical technique. The comparison
with previous results (when available)  has shown that the assumption of a conformally flat
metric leads to results that are indistinguishable, within the
accuracy of the numerical scheme, from those obtained in the correct regime. 
The simplifications in our approach do not compromise the accuracy of the
results, while greatly simplifying their computation.

For the first time we have carried out a
detailed parameter study, where the role of current distributions was
analyzed, for various geometries of the magnetic field. We briefly
summarize here the key results:
\begin{itemize}
\item the characteristic deformation induced by a purely toroidal field, fully
  confined below the stellar surface, is prolate: the magnetic field acts
  by compressing the internal layers of the star around its symmetry
  axis, causing, on the other hand, an expansion of the outer layers;
\item given the same strength, magnetic fields concentrated in the
  outer part of the star, lead to smaller deformations, with respect
  to magnetic fields concentrated in the internal regions;
\item a purely poloidal field, that in our case extends also outside
  the star, leads to oblate equilibrium configurations: the magnetic
  stresses act preferentially in the central regions, where the field
  peaks, leading to a flatter density profile perpendicularly to the
  axis itself. We can also obtain doughnut-like configurations where the density
  maximum is not at the center;
\item the presence of additional currents located in the outer layers
  of the stars, leads only to marginal changes in its structure, and on
  the shape of the magnetic field lines outside the stellar surface;
\item for the same maximum magnetic field inside the star, purely
  poloidal configuration, can be characterized by smaller deformations,
  than purely toroidal ones (about a factor one half in the $m=1$
  case). However for higher values of $m$ this trend might be
  reversed;
\item we have computed \emph{Twisted-Torus}
configurations in the non-perturbative regime. We confirm previous
results, in either the Newtonian or the perturbative regime, that only
models where the poloidal component is
energetically dominant can be built for simple electric current
distributions [this limitation could be avoided using more complex
prescriptions for the currents as shown by \citet{Ciolfi_Rezzolla13a}]. These show oblate deformations that are almost completely
due to the poloidal field, acting on the interior;
\item for a fixed central density, a higher magnetic field gives a
  higher eccentricity, a higher radius and a higher gravitational mass;
\item the more compact configurations, having a higher central
  density, can support stronger magnetic fields, and show much smaller deformations.
\end{itemize}

Our results are clearly indicative, that the magnetic energy, or the
maximum strength of the magnetic field, are in general not good
indicators of the possible deformation of the NS. The current
distribution is a key parameter: magnetic field concentrated in the
outer layers of the stars are less important than similar fields
located in the deeper interior. Given that the magnetic field
geometry, might strongly depend on the details of the NS formation
(the stratification of differential rotation, the location of the
convective region, etc...), one should be careful to make general
statements based only on energetic arguments. 

We plan to further extend this work, by investigating also rotating
configurations and/or NS models with magnetospheric currents, that we
have not touched upon here, and trying to provide some more quantitative
estimates on the possible GW emission from this objects and its
dependence on the strength and structure of the magnetic field. 

The updated XNS code for building
magnetized neutron star equilibria is publicly available for the community
at \texttt{www.arcetri.astro.it/science/ahead/XNS/}.

\section*{Acknowledgments}
We thank the referee for his/her useful comments and suggestions. 
This work has been done thanks to a EU FP7-CIG grant issued to the
NSMAG project (P.I. NB). 

\bibliography{my}{}
\bibliographystyle{mn2e}

\appendix
\section{Limit of Weak Magnetic Fields}
\label{appendixB}

In the limit of a weak magnetic
field (i.e. for $\mathscr{H} \ll M$), one can safely assume that the
metric terms $\alpha$ and $\psi$ are the same as in the unmagnetized
case (up to corrections of the order of  $\mathscr{H}/M$). For our
models, which are also static, these are only function of the radial
coordinate $r$. In this limit, for our choice of magnetic
current distributions, $\xi=0$ and $\zeta=0$, both the currents
associated to the toroidal field and the
magnetic field itself become linear functions of the vector potential
$A_\phi$. For a given value of the twisted torus magnetization constant $a$, the Grad-Shafranov
equation, Eq.~(\ref{eq:gs}), contains only terms linear  in $A_\phi$
($A_\phi$ is now a linear function of the poloidal magnetization constant
$k_{\rm pol}$). This implies that in the weak magnetic field limit,
the magnetic field structure and the geometry of the magnetic field
lines are independent of the strength of the magnetic field. It is
thus meaningful to talk about a \emph{low magnetization limit}, without
reference to the exact value of the magnetic field. This is quite
different from previous results, published in literature. For example
the works by \citet{Ciolfi_Ferrari+09a} and by \citet{Glampedakis_Andersson+12a}, following the choice initially suggested by
\citet{Tomimura_Eriguchi05a}, all
assume  that the function $\mathcal{I}$ is quadratic in
$A_\phi$ [qualitative analogous to taking $\zeta=1$ in our formalism,
even if their functional form for $\mathcal{I}$ is different from our
generic form Eq.~(\ref{eq:fbern})]. This implies that the currents associated to the toroidal
field are cubic in $A_\phi$, and the Grad-Shafranov
equation now contains terms that are  nonlinear in $A_\phi$. The same
holds for the choice presented by \cite{Lander_Jones09a} which is equivalent to
take $\zeta=0.1$. In these cases the
magnetic field structure is also a function of the magnetic field
strength, and one cannot talk of a generic low field limit, but the
exact value of the field strength must be specified.

Let us briefly describe here the properties of our solution in the
limit of a small field. We will consider a fiducial model, with a
central density $\rho_{\rm c}=8.515 \times 10^{14}$ g cm$^{-3}$, corresponding to a gravitational mass
$M=1.551 M_\odot$, and a radius $R_{\rm circ}=14.24$ km. For convenience, all our results are shown in the case of a
magnetic field with a typical strength  $\approx
10^{12}$G (they can however be rescaled to higher/lower values because of
the linearity implied by our choice for the distribution of the currents).

\begin{figure}
	\centering
	{\includegraphics[width=.45\textwidth]{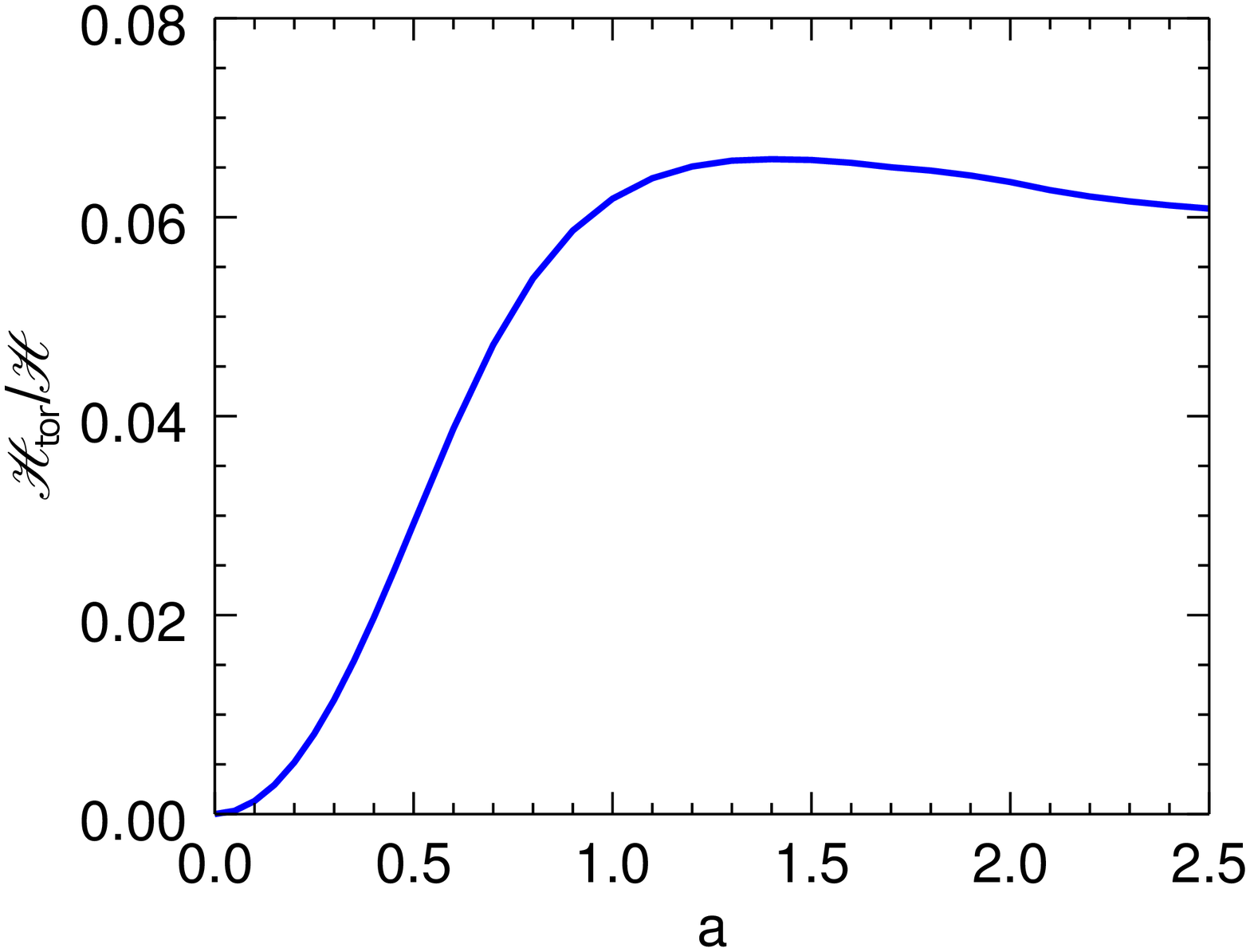}}
	\caption{Ratio of the toroidal magnetic energy $\mathscr{H}_{\rm tor}$ to 
    the total magnetic energy $\mathscr{H}$ in the weak field limit,
    as function of the parameter $a$,
    for our fiducial NS model with $M=1.551 M_\odot$.
	}
	\label{fig:magenlowB}
\end{figure}

\begin{figure*}
	\centering
	{\includegraphics[width=.35\textwidth]{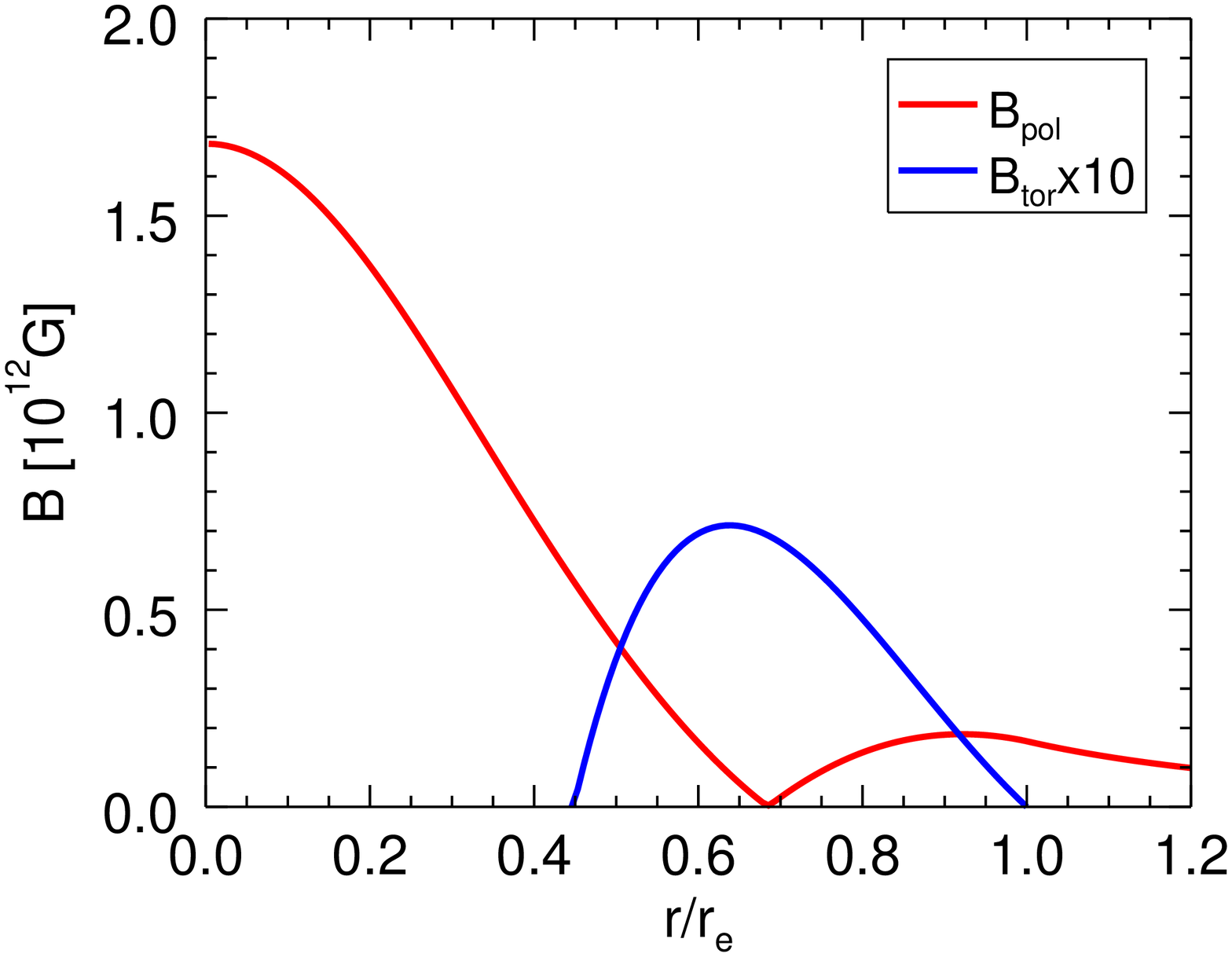}} 
	{\includegraphics[width=.35\textwidth]{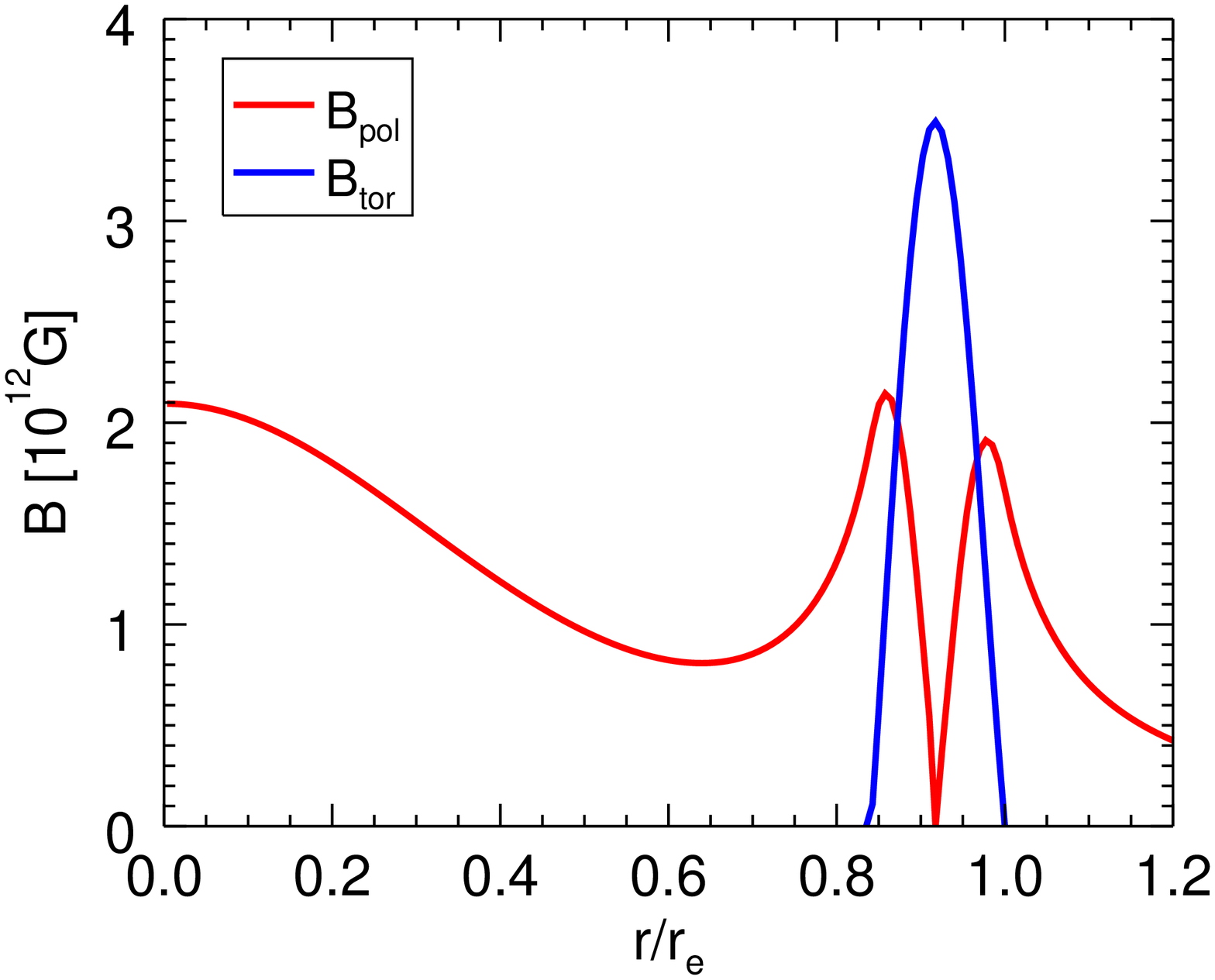}}\\
	{\includegraphics[width=.35\textwidth]{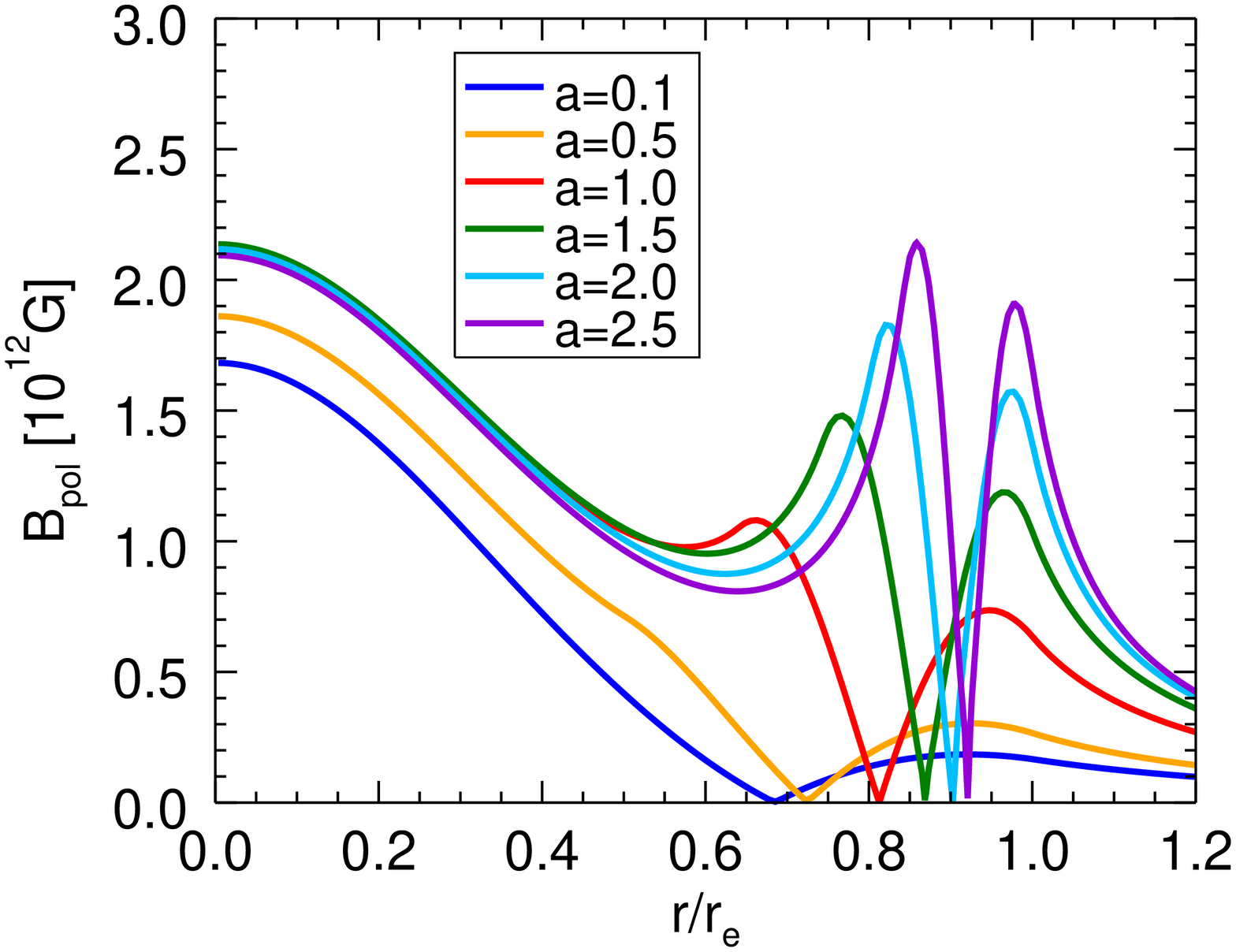}} 
	{\includegraphics[width=.35\textwidth]{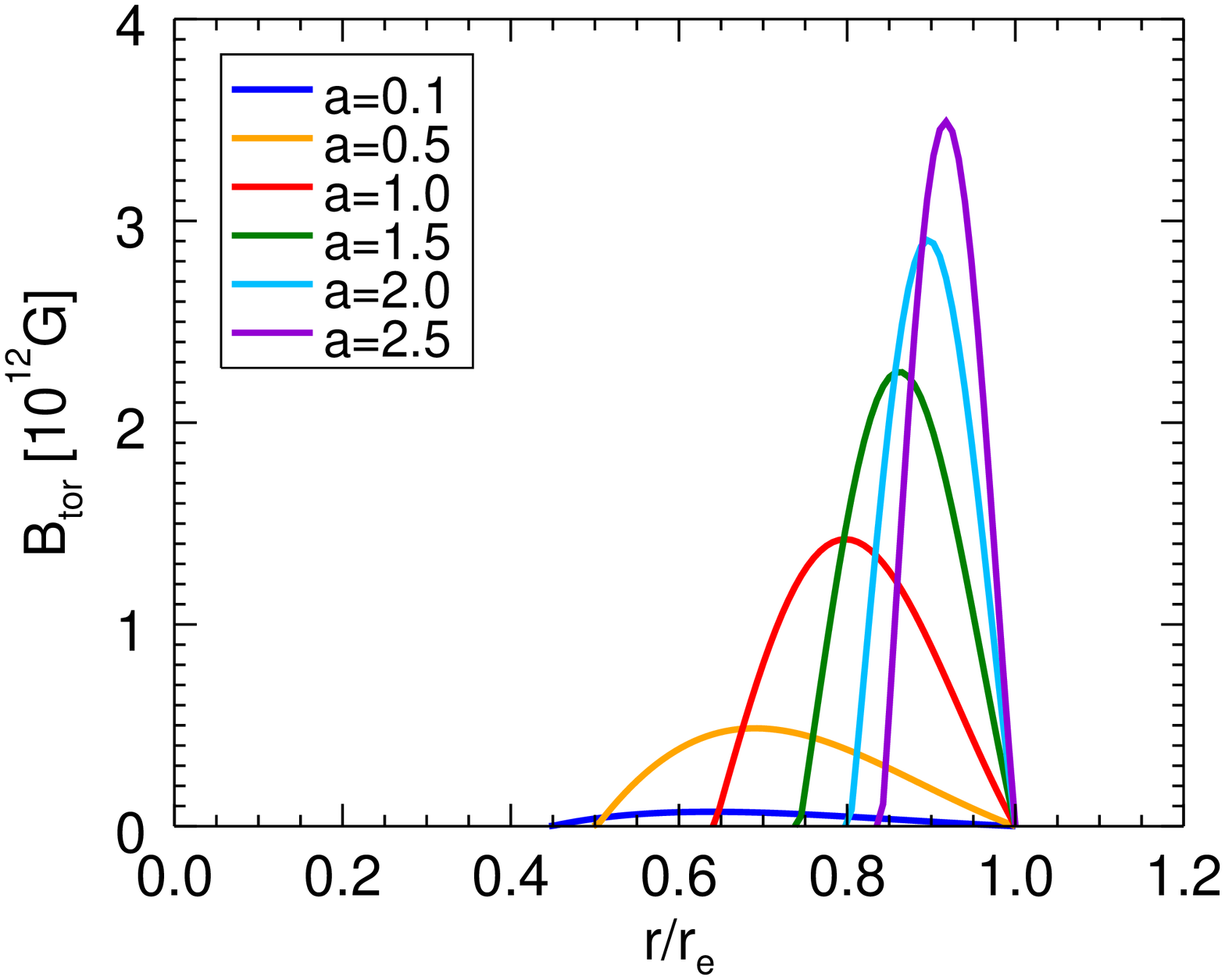}} \\
	\caption{Upper panels:
	profiles of the strength of the poloidal and toroidal 
        components of the magnetic field, along the
        equator, in the weak field limit,
    for our fiducial NS model with $M=1.551 M_\odot$ and $k_{\rm pol}=10^{-6}$. $r_e$ is the equatorial radius.
        Upper left panel is a model with
	$a=0.1$. Upper right panel is a case with $a=2.5$. Lower
        panels: the strength of the poloidal (left) and toroidal
        (right) magnetic field at the equator for various values of
        the parameter $a$. The shrinkage of the torus region, as well
        as the contribution to the poloidal field by extra currents associated to the
        toroidal field is evident.	}
    \label{fig:lowprof}
    \end{figure*}

In Fig.\ref{fig:magenlowB}, the ratio of magnetic energy carried by
the toroidal component of the field, over the total magnetic energy
is shown as a function of the   twisted torus magnetization constant
$a$. It is evident that it is not possible to reach configurations that are
toroidally dominated. In Fig.\ref{fig:lowprof}, we show the
equatorial profile of the poloidal and toroidal components of the
magnetic field, for various values of the parameter $a$, as was done
in Fig.\ref{fig:TTprof} for the case of a much stronger magnetic
field. It is interesting to notice that, as was found in previous
studies, the region occupied by the toroidal field tends to shrink
toward the surface of the star (about 70\% from $a=0.1$ to
$a=2.5$). The effect is the same as seen 
in previous works, that used a different current's
distribution.  In the same plot, done keeping the poloidal magnetization constant
$k_{\rm pol}$ fixed, it is also possible to see the contribution of
the current associated to the toroidal field, to the net dipole moment
(the value of the polar field increases with $a$). Again, as was found
in the case of a strong field, these
peripheral current contribute only marginally (about 20\% for $a=2.5$)
to the net dipole moment. 

\begin{figure*}
	\centering
	{\includegraphics[width=.35\textwidth]{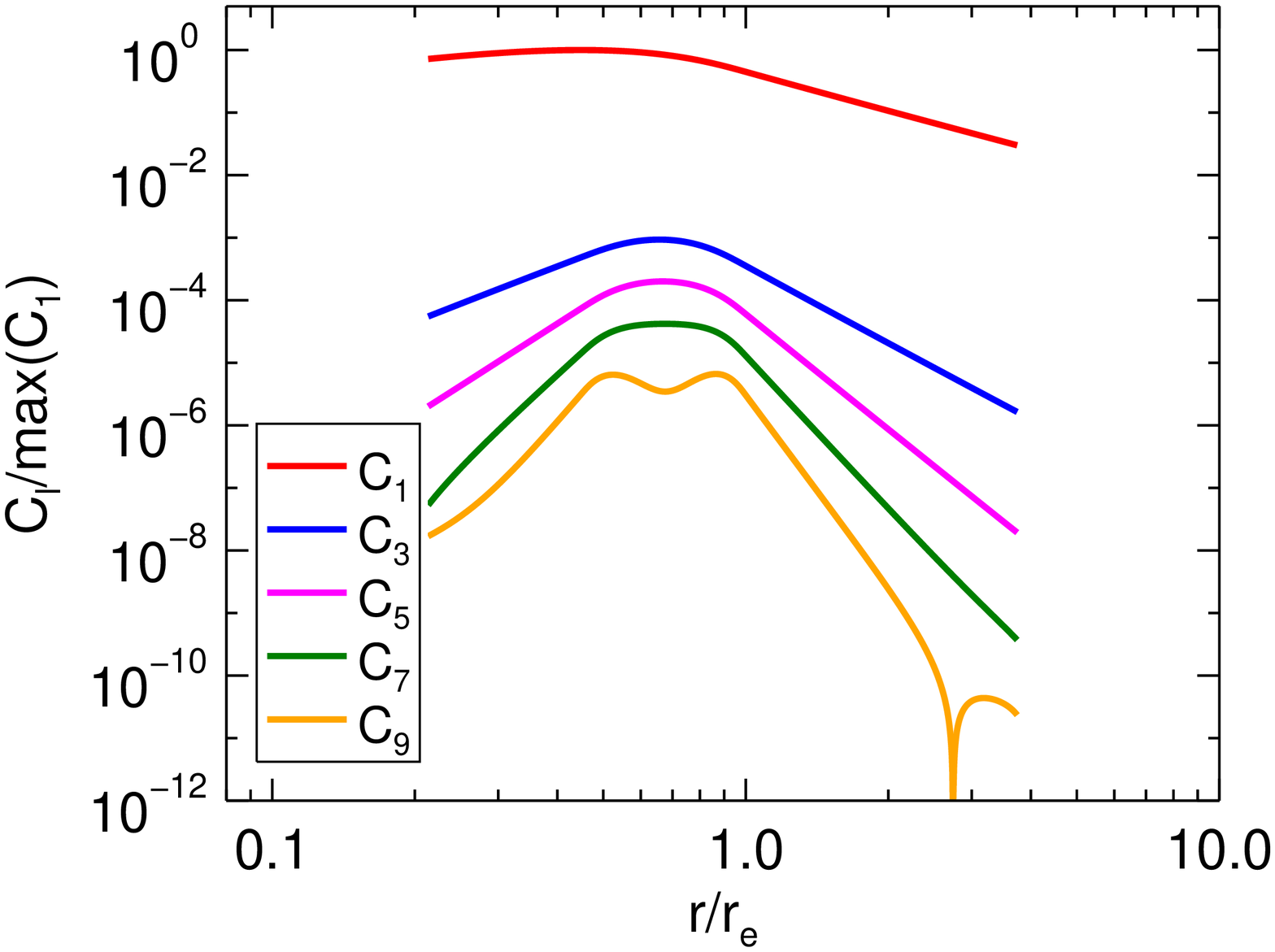}} 
	{\includegraphics[width=.35\textwidth]{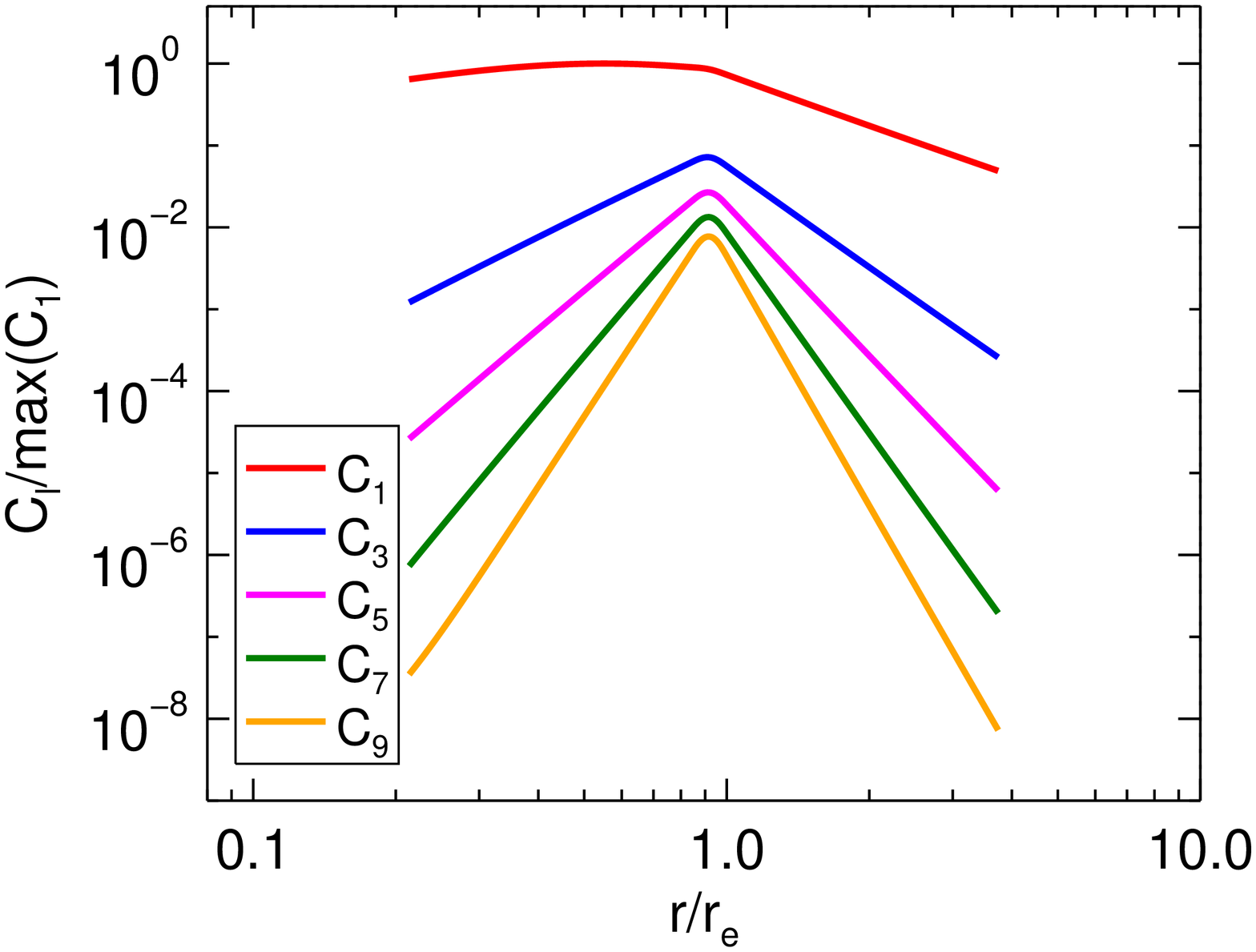}}\\
	\caption{Radial
	profiles of the norm of the $C_l(r)$ terms in the harmonic decomposition
        of the vector potential [see Eq.~\ref{eq:harmonics}], in the
        weak field limit. The
        values are normalized to the maximum of the $C_1(r)$, for
        convenience. Left panel:
	$a=0.1$. Right panel: $a=2.5$.	}
    \label{fig:lowcoef}
    \end{figure*}

Finally, in this low magnetic field limits, it is possible to
investigate the multipolar content of the magnetic field, and how does
it change with respect to the parameter $a$ (i.e. to the ratio of
toroidal magnetic field energy over total magnetic energy). A simple
way to compare the various multipole terms is to look at the relative
strength of the $C_l$ terms in the expansion of the vector potential
Eq.~(\ref{eq:harmonics}), with $C_1$ indicating the dipole term. This is
not possible for stronger fields, because the metric terms are no
longer just a function of $r$, and the $C_l$ will also contain a
geometrical contribution from the metric, which we cannot separate
(spherical harmonics are not eigenfunctions of the angular part of the
Laplacian in a generically curved spacetime). In Fig.~\ref{fig:lowcoef} we
show the values of various $C_l$ terms (normalized to the dipole one)
as a function of radius. As expected, our magnetic configurations are
always dominated by the dipole term. The various multipoles reach a
maximum at the location of the torus, and then drop outside of the
star as $r^{-(l+1)}$. In the case $a=0.1$ the various multipoles are
more than 3-4 orders of magnitude smaller than the dipole term, and in
general each multipole of order $l$ is about one order of magnitude smaller than
the preceding one of order $l-1$ (for smaller values of $a$ the various
multipoles are so small that they are essentially compatible with being
due to numerical noise). In the case $a=2.5$ the multipolar
content of the magnetic field is much higher: the quadrupole term
$l=3$ is only a factor 10 (at peak) smaller than the dipole term, and
in general the ratio between two successive multipoles is only of the
order of a few. 

It this low magnetic field limit, when the metric terms are essentially
independent of the magnetic field strength, we have verified that in order to
get converged solutions of the Grad-Shafranov equation,
Eq.~(\ref{eq:gs}), we need to truncate our decomposition of the vector
potential into spherical harmonics, Eq.~(\ref{eq:harmonics}), at a $l_{\rm max}$ such
that all the neglected multipoles have at least an amplitude
$C_{l>l_{\rm max}}/C_1 < 10^{-5}$. Please note that, while the overall
accuracy of our models is $\sim 10^{-3}$, the accuracy of the elliptic
solver of the Grad-Shafranov equation, is $\sim 10^{-7}$. In fact multipoles with amplitude less than $10^{-7}$ times the leading
dipole term, are dominated by numerical noise (see for example the
behaviour of the $C_{l=9}$ terms in the left panel of Fig.~\ref{fig:lowcoef}).

\section{Global Physical Quantities}

\label{appendix}

To characterise the equilibrium models obtained with our numerical scheme 
we have computed a wide set of global physical quantities that allow
us to provide a parametrization, as complete as possible. Here we give
their definition for  the case of static magnetized configurations,
described within the CFC approximation. 

The most relevant are: the gravitational mass 
\be
M:=\int\left( e + 3p + B^2 \right) \alpha \psi^6 \sin\theta \, dr d\theta d\phi ,
\ee
the baryonic mass
\be
M_0:= \int\rho \psi^6 r^2 \sin\theta \, dr d\theta d\phi,
\ee
the proper mass
\be
M_p:=\int e \psi^6 r^2 \sin \theta \, dr d\theta d\phi,
\ee
the total magnetic energy
\be
\mathscr{H}:=\frac{1}{2}\int B^2 \psi^6 r^2 \sin\theta \, dr d\theta d\phi,
\ee
the magnetic energy in the toroidal component
\be
\mathscr{H}_{\rm tor}:=\frac{1}{2}\int B^\phi B_\phi \psi^6 r^2 \sin\theta \, dr d\theta d\phi,
\ee
the magnetic energy in the poloidal component
\be
\mathscr{H}_{\rm pol}:=\frac{1}{2}\int (B^r B_r + B^\theta B_\theta )\psi^6 r^2 \sin\theta \, dr d\theta d\phi,
\ee
and the binding energy
\be
\mathscr{W}:=M-M_p-\mathscr{H},
\ee
where the integrals are defined over the all three-dimensional space.

In order to characterise the geometrical properties of the magnetic
field,  other quantities must be introduced.
When the magnetic configuration possesses a toroidal component we can evaluate the
 flux of the toroidal magnetic field through a meridional half-plane which, analogously to KY08, 
is given by
\be
\Phi:= \int_0^\pi d\theta \int_0^\infty \sqrt{B^\phi B_\phi} \psi^4 r \, dr.
\ee

In the presence of a poloidal magnetic field we can estimate the magnetic dipole
moment $\mu$ of the star. This is usually defined (see BB95) by the leading term of the
asymptotic behaviour of the magnetic field components at
$r\rightarrow\infty$, where the space-time metric is flat. However this definition, in our opinion, is not
well suited for a numerical scheme. At $r\rightarrow\infty$ the magnetic
field vanishes, and it is not numerically safe, due to interpolation
and round-off errors, to compute a finite
quantity as the ratio of two vanishing ones. On the other hand, if
computed at a finite distance, this definition might introduce errors
due to the metric curvature.  Since our numerical scheme does not use a
compactified domain, and extends only over a few stellar radii outside
a NS, we have derived a definition of magnetic dipole moment that
takes into account the curvature of space-time. This allows us to
measure the dipole moment at finite radii, and we have verified that
the value does not depend on the radius, as expected.
From a multipole expansion of Eq.~(\ref{eq:gs}),
assuming that outside the star the line element is well approximated by the Schwarzschild solution and
selecting the dipole term ($l=1$), one can find a simple relation
that connects the dipole moment $\mu$ to the  $\phi-$component of the vector potential 
$\tilde{A}_\phi$, the gravitational mass $M$ and the radial coordinate $r$, namely
\be
\label{eq:mdm}
\tilde{A}_\phi=\mu\left( 1+\frac{M}{4r} \right)\frac{\sin \theta}{r^2}.
\ee

In the case of mixed field configurations another important global
topological quantity
is the magnetic helicity. Following \cite{Ciolfi_Ferrari+09a} the total
magnetic helicity $H_{\rm m}$ can be defined as
\be
H_{\rm m}:=\int H^0_{\rm m}\alpha \psi^6 r^2 \sin\theta \, dr d\theta d\phi,
\ee
where $H^0_{\rm m}$ is the time component of the helicity four-current
\be
H_{\rm m}^\alpha:=-\frac{1}{2}\epsilon^{\alpha\beta\mu\nu}A_{\beta} F_{\mu\nu}.
\ee
In our case the definition reduces simply to
\be
H_{\rm m}=\int (B^i A_i)\psi^6 r^2 \sin\theta \, dr d\theta d\phi,
\ee
where, using the gauge freedom of the vector potential,
we can impose $A_r=0$ and express $A_\theta$ in function of $A_\phi$ as
\be
A_\theta=\frac{-1}{\sin \theta} \int_\infty^r \frac{\psi^2}{\alpha}\mathcal{I}(A_\phi)dr'.
\ee

Finally there are global quantities related exclusively to the shape
and deformation of the star. These 
are the equatorial radius $r_e$, the polar radius $r_p$, the circumferential
radius
\be
R_{\rm circ}:= \psi^2(r_e,\pi/2)r_e.
\ee
and the mean deformation that, following KY08, is defined by
\be
\bar e := \frac{I_{zz}-I_{xx}}{I_{zz}},
\ee
where $I_{zz}$ and $I_{xx}$ are the
moment of inertia respectively in the parallel
and orthogonal direction to the axis of symmetry
\be
I_{zz}:= \int e r^4 \sin^3\theta dr d\theta d\phi 
\ee
\be
I_{xx}:=\frac{1}{2}\int e r^4 \sin\theta (1+\cos^2\theta) dr d\theta d\phi.
\ee
As was just pointed out in FR12 this definition of  $\bar e$ is strictly Newtonian
and may be not suitable for estimating the gravitational-wave emission of a rotating 
distorted star.

\end{document}